\documentclass[11pt]{article}

\usepackage{style}
\usepackage{authblk}
\usepackage{setspace}

\title{
\bfseries Semiparametric Triple Difference Estimators}
\author[1]{Sina Akbari}
\author[1]{Negar Kiyavash}
\author[2]{AmirEmad Ghassami\vspace{1em}}
\affil[1]{EPFL, Switzerland}
\affil[2]{Department of Mathematics and Statistics, Boston University, USA}
\date{}

\begin{document}

\maketitle
\noindent

\begin{center}
    \vspace{-10mm}
    \small
    {First version:} \href{https://arxiv.org/abs/2502.19788v1}{February 27, 2025}, 
    {Current version:} \href{https://arxiv.org/abs/2502.19788v3}{September 15, 2025}
\end{center}
\vspace{4mm}

\begin{abstract}
    The triple difference causal inference framework is an extension of the well-known difference-in-differences framework.
    It relaxes the parallel trends assumption of the difference-in-differences framework through leveraging data from an auxiliary domain.
    Despite being commonly applied in empirical research, 
    the triple difference framework has received relatively limited attention in the statistics literature. Specifically, investigating the intricacies of identification and the design of robust and efficient estimators for this framework has remained largely unexplored.
    This work aims to address these gaps in the literature.
    From the identification standpoint, we present outcome regression and weighting methods to identify the average treatment effect on the treated in both panel data and repeated cross-section settings.
    For the latter, we relax the commonly made assumption of time-invariant composition of units.
    From the estimation perspective, we develop semiparametric estimators for the triple difference framework in both panel data and repeated cross-sections settings.
    These estimators are based on the cross-fitting technique, and flexible machine learning tools can be used to estimate the nuisance components.
    We characterize conditions under which our proposed estimators are efficient, doubly robust, root-n consistent and asymptotically normal.
    As an application of our proposed methodology, we examined the effect of mandated maternity benefits on the hourly wages of women of childbearing age and found that these mandates result in a $2.6\%$ drop in hourly wages.
\end{abstract}
    
\section{Introduction}

The triple difference framework for causal inference extends the well-known difference-in-differences (DiD) framework \citep{ashenfelter1984using,card1990impact,
card1994minimum,heckman1997matching, card2000minimum,abadie2005semiparametric} by incorporating data from an auxiliary domain.
Specifically, it considers a setting with two domains: a \emph{target domain}, in which the causal parameter of interest is defined, 
and a \emph{reference domain}.
The triple difference framework is designed for settings where the DiD assumptions do not necessarily hold in the target domain.
It characterizes a set of requirements on the relationship between the two domains under which identifiability is attainable by \textit{fusing} data from the two domains \citep{olden2022triple,zhuang2024way,berck2016note}. 

The triple difference framework relaxes the \emph{parallel trends assumption}, which is fundamental to the canonical DiD, and thus allows for a more flexible identification scheme.
Given its flexibility and practical relevance, the triple difference framework has been widely adopted in economics and policy evaluation---see \citep{olden2022triple} for a survey.
However, formal identification and estimation theory for triple difference has received little attention.
Although an identification formula for triple difference appears in \citep{wooldridge2020introductory} and \citep{frohlich2019impact}, others have settled on incorporating the high-level ideas without technical details \citep{lechner2011estimation, angrist2009mostly}.
Recently, \citet{olden2022triple} conducted the first formal study of identification in the triple difference framework.
They outlined the necessary assumptions for the identification of the average treatment effect on the treated (ATT) in the target domain based on outcome regression.
Beyond that, however, key aspects such as weighting-based identification, and the design of robust and efficient
estimators for ATT or other causal quantities 
have remained largely unexplored.
This work aims to fill these gaps.
We consider both identification and estimation aspects of the problem, as described below.

From the identification point of view, we contribute to the triple difference literature by (i) formalizing the outcome regression-based identification for the repeated cross-sections setting (described in \Cref{sec:setup});
and (ii) establishing weighting methods to identify the ATT, which were missing in the literature, for both panel data (also known as repeated outcomes) and repeated cross-sections settings.
In the repeated cross-sections setting, we avoid the widely adopted assumption of \emph{no compositional changes}, i.e., the assumption that the composition of the units is time-invariant. 
Such an assumption requires us to sample observations from the same population, which is unrealistic in most scenarios \citep{hong2013measuring, sant2025difference}.
We show that this assumption, although commonly made in the literature, is not required for identification.

From the estimation point of view, we develop semiparametric estimators for the triple difference framework
in both panel data and repeated cross-sections settings.
We derive influence function-based estimators that attain the semiparametric efficiency bound.
We demonstrate that they are doubly robust, in the sense that
they remain consistent as long as either the estimator for outcome regression or the one for treatment assignment is consistent, but not necessarily both.
Furthermore, we characterize the conditions under which our proposed estimators are $\sqrt{n}$-consistent and asymptotically normal
even if the nuisance function estimators do not converge at $\sqrt{n}$ rate.
In the repeated cross-sections setting, we propose estimators for when compositional changes assumption holds as well as when it is violated.
Additionally, we discuss the implications of this assumption in terms of the robustness properties that our estimators can guarantee.

\paragraph{Related literature.}
The triple difference framework is an extension of the canonical DiD framework.
A large body of work has contributed to refining the latter.    
\citet{abadie2005semiparametric} developed weighting estimators and introduced semiparametric approaches to estimation of ATT in the DiD framework.
More recently, \citet{sant2020doubly} proposed a set of doubly robust estimators, which are robust for inference under parametric assumptions.
Others have proposed methods to combine DiD with synthetic controls \citep{arkhangelsky2021synthetic}, explored heterogeneous treatment effects \citep{nie2019nonparametric}, provided efficient estimation methods for discrete-valued outcomes \citep{li2019double},
analyzed DiD with continuous treatments \citep{callaway2024difference}, and extended DiD to
nonlinear models \citep{athey2006identification, torous2024optimal}.
Yet, similar developments for triple difference remain limited \citep{zhuang2024way, akbari2024triple}.
Our work extends identification and estimation techniques for the triple difference framework building on the aforementioned literature.

The triple difference framework can be naturally interpreted as a data fusion framework \citep{bareinboim2016causal, degtiar2023review,colnet2024causal,yang2024causal}, where the goal is to identify causal effects in a target domain using additional information from an auxiliary or reference domain. 
Despite the conceptual similarity, the data fusion literature has largely focused on conditional independence-type  assumptions on potential outcome variables and has paid little attention to parallel trends-type assumptions. 
As such, existing data fusion frameworks do not directly address the type of identification problems encountered in the triple difference setting.

As mentioned earlier, in the literature of canonical DiD, when repeated cross-sectional data are available, it is commonly assumed that the covariates and the treatment assignment are time-invariant, known as the {no compositional changes} assumption \citep{heckman1997matching, abadie2005semiparametric, sant2020doubly, callaway2021difference}.
To the best of our knowledge, \citet{hong2013measuring} was the first to analyze DiD without imposing this assumption, at the expense of replacing the parallel trends requirement with 
a selection on observables condition.
\citet{zimmert2018efficient} showed that ATT can be identified under a set of weaker independence assumptions and, only recently, \citet{sant2025difference} proved that identification is possible without making such an assumption altogether.
We generalize the viewpoint of \citet{sant2025difference} to the triple difference setting.

\paragraph{Organization.} The rest of this paper is organized as follows:
we formally introduce the triple difference setup and our parameters of interest in \Cref{sec:setup}.
We provide the identification results in \Cref{sec:id}, and in \Cref{sec:est}, we describe our proposed estimators.
In all sections, we present the results for the panel data setting first, followed by those for the repeated cross-sections setting;
proofs of all results are provided in
the Appendix.
In \Cref{sec:sim}, we evaluate our proposed estimators on synthetic data.
In \Cref{sec:application}, we apply our proposed methodology to study the effect of mandated maternity benefits on the hourly wages of women of childbearing age.

\section{Problem Setting}\label{sec:setup}
We assume that data are collected from two domains indicated by $D\in\{0,1\}$, where $D=0$ and $D=1$ represent the reference and target domains, respectively.
Units in each domain are divided into two groups, indicated by $G\in\{0,1\}$.
Members of the groups $G=0$ and $G=1$ are ineligible and eligible to receive a specific treatment, respectively.
For instance, if only certain people, such as those in lower quantiles of income, are eligible for a government financial assistance, the lower quantiles of income correspond to $G=1$.
Let $A\in\{0,1\}$ denote a binary treatment (or policy) which is assigned only to the eligible units in the target domain.
We denote by $X$ a set of observed pre-treatment covariates of the units.
In what follows, we briefly describe the panel data and repeated cross-sections settings.
\subsection{Panel Data Setting}
In this setting, data are collected from the same population at two time points indicated by the index $t\in\{0,1\}$.
Let $Y_0, Y_1\in\mathbb{R}$ represent the observed outcomes at time $t=0$ and $t=1$, respectively.
Moreover, for $t\in\{0,1\}$, let
$Y_t^{(0)}$, $Y_t^{(1)}\in\mathbb{R}$
denote the potential outcomes of $Y_t$, if (potentially contrary to the fact) the treatment was set to $A=0$ (control) and $A=1$ (treatment), respectively.
The treatment is administered after the outcomes are measured at time $t=0$, only in the target domain, and to all eligible units, represented by $(D=1,G=1)$.
We observe i.i.d. samples $\{O_i=(G,D,X,Y_0,Y_1)_i\}_{i=1}^n$.
Our parameter of interest is 
\[\tau_\mathrm{pd}\coloneqq \ex{}\left[Y_1^{(1)}-\Y{1}\:\big|\: G=1, D=1\right].\]
Note that the parameter $\tau_\mathrm{pd}$ represents the average treatment effect at time $t=1$ in the population corresponding to $(D=1,G=1)$, which is the treated population. Therefore, the parameter $\tau_\mathrm{pd}$ is the ATT in the target domain.

\subsection{Repeated Cross-Sections Setting}
In this setting, cross-sectional data are available at two time points.
We denote by $T\in\{0,1\}$ the (random) collection time of samples, where $T=0$ and $T=1$ correspond to pre- and post-treatment periods, respectively.
Let $Y$ denote the observed outcome variable, and $Y^{(0)}, Y^{(1)}\in\mathbb{R}$ represent the potential outcomes under control and treatment, respectively.
The treatment is administered to all units indicated by $(G=1,D=1,T=1)$, that is, the eligible group in the target domain at time $T=1$.
We observe i.i.d. samples $\{O_i=(G,D,T,X,Y)_i\}_{i=1}^n$.
Note that unlike most work in the literature, we do not require the no compositional changes assumption, which requires $p(X,G,D\mid T=0)=p(X,G,D\mid T=1)$.
The parameter of interest is defined as 
\[
    \tau_\mathrm{rc}\coloneqq \ex{}\left[Y^{(1)}-\Y{}\:\big|\: G=1,D=1,T=1\right],
\]
which is the ATT in the target domain.

\subsection{Comparison Between the Two Settings}

In the panel data setting, we track the same individuals (or units) across the two time periods. 
This allows researchers to observe changes within individuals over time. In contrast, in the  repeated cross-sections setting, we are given different samples at the two time points. Each cross-section represents a snapshot of one of the populations at a given moment, but the individuals in each sample are not necessarily the same.
Therefore, the repeated cross-sections setting can introduce additional complications when composition of the units changes across the two populations.
Specifically, the researcher should disentangle whether observed differences between time periods are due to actual changes within individuals or shifts in the composition of the population. 
We will address the complications due to compositional changes in the task of identification and estimation in Sections \ref{sec:id} and \ref{sec:est}, respectively.

\section{Identification}\label{sec:id}
In this section, we present our identification results, first for the panel data setting, followed by the repeated cross-sections setting.
\subsection{Identification in the Panel Data Setting}

We require \emph{consistency} and \emph{no anticipation effects} assumptions in our setting.
The former
links the potential outcomes to the observed outcomes, while the latter requires that receiving treatment has no effect before its actual implementation. 
Both of these assumptions are commonly made in the literature of the DiD framework.

\begin{assumption}[Consistency]\label{as:consistency}
    $Y_t=DGY_t^{(1)} + (1-DG)Y_t^{(0)}$.
\end{assumption}
\begin{assumption}[No anticipation effects]\label{as:noanticip}
    $\Y{0}=Y_0^{(1)}$.
\end{assumption}
\begin{remark}
    While \Cref{as:noanticip} is stated at the individual level for clarity and ease of interpretation, our results only require this condition to hold in expectation. Specifically, we require that $\ex{}\left[\Y{0}\:\big|\: X, G=1, D=1\right]=\ex{}\left[Y_0^{(1)}\:\big|\: X, G=1,D=1\right]$, meaning that, in the subpopulation with $(G=1,D=1)$, for each stratum of the observed covariates $X$, the expected values of $\Y{0}$ and $Y_0^{(1)}$ are equal.
\end{remark}

We also require \emph{positivity}, also known as overlap, in our setting, which is another commonly made assumption in the literature of causal inference.
This assumption
ensures that (i) at least some units are treated in the target domain, to make the parameter of interest well-defined, and (ii) in each stratum of $X$, there is a non-zero probability of being assigned to each group.

\begin{assumption}[Positivity]\label{as:pos3}
    For some $\epsilon\geq0$,
    \begin{enumerate}[label=(\roman*)]
        \item $p(G=1, D=1)>\epsilon$,
        \item $p\Big(\min\big\{p(G=0,D=1\mid X), p(G=1,D=0\mid X), p(G=0,D=0\mid X)\big\}>\epsilon\:\big|\: E\Big)=1$, 
        where $E=\big\{p(G=1,D=1\mid X)>0\big\}$.
    \end{enumerate}
\end{assumption}

We say \emph{strict positivity} holds if $\epsilon>0$.
As we shall see, positivity is sufficient for identification, whereas inference requires strict positivity.
Finally, we make the following assumption.
\begin{assumption}[Conditional parallel difference in trends]\label{as:id3}
For all $x$ such that $p(X=x)p(G,D\mid X=x)>0$, 
    \begin{equation}\label{eq:cet}
    \begin{split}
        \ex{}&\left[
            \Y{1} - \Y{0}\:\big|\: X=x,G=1, D=1
        \right]
        -
        \ex{}\left[
            \Y{1} - \Y{0}\:\big|\: X=x,G=0, D=1
        \right]
        \\&\hspace{1em}=
        \ex{}\left[
            \Y{1} - \Y{0}\:\big|\: X=x,G=1, D=0
        \right]
        -
        \ex{}\left[
            \Y{1} - \Y{0}\:\big|\: X=x,G=0, D=0
        \right].
        \quad
    \end{split}
    \end{equation}
\end{assumption}
\Cref{as:id3} is a relaxation of the \emph{conditional parallel trends} assumption in the canonical DiD framework, in the sense that when data from one domain are available, say $D=1$, conditional parallel trends asserts that the left hand side of Equation \eqref{eq:cet} is equal to $0$.
Rather than assuming that the left hand side term is zero, here, we assume that it is equal to its counterpart in the reference domain $D=0$.

We next present our nonparametric identification result for the ATT based on outcome regression (OR).
\begin{restatable}{theorem}{prporp}\label{prp:orpanel} Under Assumptions \ref{as:consistency}, \ref{as:noanticip}, and \ref{as:id3},
        for all $x$ such that $p(X=x)p(G,D\mid X=x)>0$, the conditional ATT is identified as
        \[
            \ex{}\left[Y_1^{(1)}-\Y{0}\:\big|\: X=x, G=1, D=1\right] =  \mu_{1,1,\Delta}(x) -\mu_{0,1,\Delta}(x)-\mu_{1,0,\Delta}(x)+\mu_{0,0,\Delta}(x),
        \]
        where $\mu_{g,d,\Delta}(X)=\ex{}[Y_1-Y_0\mid X, G=g, D=d]$.
        
           If additionally \Cref{as:pos3} holds, the ATT is identified as
            \(
                \tau_\mathrm{pd} = \psi_\mathrm{pd}
                \),
                where
                \begin{equation*}
                \begin{split}
                \psi_\mathrm{pd}\coloneqq  \ex{}[Y_1-Y_0\mid G=1,D=1]-\ex{}[\mu_{0,1,\Delta}(X)+\mu_{1,0,\Delta}(X)-\mu_{0,0,\Delta}(X)\mid G=1,D=1].
            \end{split}\end{equation*}
\end{restatable}
The first statement in \Cref{prp:orpanel} provides the identification result for the conditional ATT only for $x$ where $p(G,D\mid X=x)$ is positive.
\Cref{as:pos3} ensures that this condition always holds for all $x$ observed in the subpopulation of treated units, $(G=1,D=1)$, resulting in the identification of ATT as given by the second statement in \Cref{prp:orpanel}.
Note that $\psi_\mathrm{pd}$ is a functional of the observational law.
In Section \ref{sec:est}, we will consider $\psi_\mathrm{pd}$ as our parameter of interest.


We next present an identification result for ATT in the triple difference framework based on inverse propensity weighting. 
As in the case of OR-based identification, we first identify the conditional ATT in the target domain, and subsequently use it to identify the ATT.
\begin{restatable}{theorem}{prpipwp}\label{prp:ipwpanel}
    Under Assumptions \ref{as:consistency}, \ref{as:noanticip}, and \ref{as:id3},
        for all $x$ such that $p(X=x)p(G,D\mid X=x)>0$, the conditional ATT is identified as
    \[
        \ex{}\left[
            Y_1^{(1)} - \Y{1}\:\big|\: X=x, G=1, D=1
        \right]
        =
        \ex{}\left[
            \rho_0(X,G,D)\cdot\big(Y_1-Y_0\big) \mid X=x
        \right],
    \]
    where
    \begin{equation}
    \label{eq:rho}
        \rho_0(X,G,D) \coloneqq \sum_{g,d\in\{0,1\}}
        \frac{(1-g-G)(1-d-D)}{\pi_{g,d}(X)}, \quad\text{and }
        \pi_{g,d}(X)\coloneqq p(G=g,D=d\mid X).
    \end{equation}
        If additionally  \Cref{as:pos3} holds, the ATT is identified as
        \begin{equation}
        \begin{split}
        \label{eq:coripwp}
            \tau_\mathrm{pd}
            =\ex{}\left[\frac{\pi_{1,1}(X)}{\ex{}[G\cdot D]}\cdot\rho_0(X,G,D)\cdot(Y_1-Y_0)\right].
        \end{split}
        \end{equation}
\end{restatable}

\subsection{Identification in the Repeated Cross-Sections Setting}
As in the case of panel data setting,
we begin by presenting the necessary assumptions for identification.
Specifically, \Cref{as:consistencyrc} establishes a link between the potential outcomes and the observed outcomes, \Cref{as:noanticiprc} rules out anticipation effects, \Cref{as:posrc} ensures that the parameter of interest is well-defined, and \Cref{as:idrc} ensures the identifiability of the ATT.
These assumptions are the counterparts of Assumptions \ref{as:consistency} through \ref{as:id3} in the repeated cross-sections setting.
\begin{assumption}[Consistency]\label{as:consistencyrc}
    $Y=GDTY^{(1)} + (1-GDT)\Y{}$.
\end{assumption}
\begin{assumption}[No anticipation effects]\label{as:noanticiprc}
    If $T=0$, then $\Y{} = Y^{(1)}$.
\end{assumption}
\begin{remark}
    Similar to the panel data setting, our results only require the weaker assumption of $\ex{}\left[\Y{}\mid X, G=1, D=1, T=0\right]=\ex{}\left[Y^{(1)}\mid X, G=1, D=1, T=0\right]$.
\end{remark}
\begin{assumption}[Positivity]\label{as:posrc}
    For some $\epsilon\geq0$,
    \begin{enumerate}[label=(\roman*)]
        \item $p(G=1, D=1, T=1)>\epsilon$,
        \item 
        \[
        \begin{split}
            p\Big(\min\big\{&p(G=0,D=1,T=1\mid X), p(G=1,D=0,T=1\mid X),\\ p(&G=0,D=0,T=1\mid X),
            p(G=1,D=1,T=0\mid X),
            p(G=0,D=1,T=0\mid X),\\ &\hspace{1em}p(G=1,D=0,T=0\mid X), p(G=0,D=0,T=0\mid X)\big\}
            >\epsilon\mid E\Big)=1,
        \end{split}
        \]
        where $E=\big\{p(G=1,D=1,T=1\mid X)>0\big\}$.
    \end{enumerate}
\end{assumption} 
We will say strict positivity holds when $\epsilon>0$.
\begin{assumption}[Conditional parallel difference in trends]\label{as:idrc}
For all $x$ such that $p(X=x)p(G,D,T\mid X=x)>0$, 
    \[ 
    \begin{split}
        &{\color{white}=} \left(\ex{}\left[
            \Y{}\:\big|\: X=x,G=1, D=1, T=1
        \right] - \ex{}\left[\Y{}\:\big|\: X=x,G=1, D=1, T=0
        \right]\right)\\
        &\hspace{2em}-\left(\ex{}\left[
            \Y{}\:\big|\: X=x,G=0, D=1, T=1
        \right] - \ex{}\left[\Y{}\:\big|\: X=x,G=0, D=1, T=0
        \right]\right)
        \\&\hspace{2em}=\left(\ex{}\left[
            \Y{}\:\big|\: X=x,G=1, D=0, T=1
        \right] - \ex{}\left[\Y{}\:\big|\: X=x,G=1, D=0, T=0
        \right]\right)\\
        &\hspace{4em}-\left(\ex{}\left[
            \Y{}\:\big|\: X=x,G=0, D=0, T=1
        \right] - \ex{}\left[\Y{}\:\big|\: X=x,G=0, D=0, T=0
        \right]\right).
        \quad
    \end{split}
    \]
\end{assumption}

Similar to the previous subsection, we first present the OR-based identification result.
\begin{restatable}{theorem}{prporrc}\label{prp:orrc}
    Under Assumptions \ref{as:consistencyrc}, \ref{as:noanticiprc} and \ref{as:idrc},
    for all $x$ such that $p(X=x)p(G,D,T\mid X=x)>0$, the conditional ATT is identified as
        \[
            \ex{}\left[Y^{(1)}-\Y{}\:\big|\: X=x, G=1, D=1,T=1\right] =  \mu_{1,1,\Delta}(x) -\mu_{0,1,\Delta}(x)-\mu_{1,0,\Delta}(x)+\mu_{0,0,\Delta}(x),
        \]
        where 
        $\mu_{g,d,t}(X)= \ex{}[Y\mid X, G=g, D=d, T=t]$,
    and $\mu_{g,d,\Delta}(X)=\mu_{g,d,1}(X)-\mu_{g,d,0}(X)$.
        If additionally \Cref{as:posrc} holds, the ATT is identified as $\tau_\textrm{rc}=\psi_\mathrm{rc}$, 
            where
            \begin{equation}\label{eq:idorrc}
            \begin{split}
                \psi_\mathrm{rc}\coloneqq
                \ex{}&[Y\mid G=1, D=1, T=1]-\\ &\ex{}\left[\mu_{1,1,0}(X)+\mu_{0,1,\Delta}(X)+\mu_{1,0,\Delta}(X)-\mu_{0,0,\Delta}(X)\mid G=1, D=1, T=1\right].
            \end{split}\end{equation}
            
\end{restatable}


As in the previous case, $\psi_\mathrm{rc}$ is a functional of the observational law, and will be considered as the parameter of interest in \Cref{sec:est}.
We next turn our focus to weighting-based identification for the repeated cross-sections setting.
\begin{restatable}{theorem}{prpipwrc}\label{prp:ipwrc}
    Under Assumptions \ref{as:consistencyrc}, \ref{as:noanticiprc} and \ref{as:idrc}, 
    for all $x$ such that $p(X=x)p(G,D,T\mid X=x)>0$, the conditional ATT is identified as 
    \[\ex{}\left[Y^{(1)}-\Y{}\:\big|\: X=x, G=1,D=1,T=1\right]=\ex{}[\phi_0(X,G,D,T)\cdot Y\mid X=x],\]
    where 
    \begin{equation}\label{eq:phi0}
        \phi_0 (X,G,D,T)
        =-\sum_{g,d,t\in\{0,1\}}\frac{(1-g-G)(1-d-D)(1-t-T)}{\pi_{g,d,t}(X)},
    \end{equation}
    and $\pi_{g,d,t}(X)\coloneqq p(G=g,D=d,T=t\mid X)$.
        If additionally \Cref{as:posrc} holds, the ATT is identified as
        \begin{equation*}
            \begin{split}
                \tau_\mathrm{rc}=\ex{}\left[\frac{\pi_{1,1,1}(X)}{\ex{}[G\cdot D\cdot T]}\cdot\phi_0(X,G,D,T)\cdot Y\right].
            \end{split}
        \end{equation*}
\end{restatable}


\subsubsection{Identification under No Compositional Changes}\label{sec:idncc}
Importantly, \Cref{prp:ipwrc} allows for compositional changes.
That is, the composition of the units is allowed to change arbitrarily across time points.
However, the commonly made no compositional changes assumption simplifies the identification.
Below, we formally present this assumption and discuss its implications for identification of the ATT.
\begin{assumption}[No compositional changes]\label{as:nocompch}
    The distribution of covariates and group indicators $X,G,D$ is time-invariant, that is, $p(X,G,D\mid T=0)=p(X,G,D\mid T=1)$,
    and $p(T=1)$ is a known parameter.
\end{assumption}

Under Assumption \ref{as:nocompch},
$\phi_0(\cdot)$ in \Cref{eq:phi0} can be further simplified to:
\[\begin{split}
    \phi_0(X,G,D,T) 
    &=\left(\frac{T}{p(T=1)}-\frac{1-T}{p(T=0)}\right)\sum_{g,d\in\{0,1\}}\frac{(1-g-G)(1-s-D)}{p(G=g,D=d \mid X)}
    \\&=\frac{T-\ex{}[T]}{\ex{}[T](1-\ex{}[T])}\cdot\rho_0(X,G,D),
\end{split}
\]
where $\rho_0(\cdot)$ is defined in Equation \eqref{eq:rho}.
That is, the identification will now require the less complex nuisance function $\pi_{g,d}(\cdot)$, as opposed to $\pi_{g,d,t}(\cdot)$.
Similarly, the identification functional of the ATT estimand then simplifies to 
    \[
        \tau_\mathrm{rc}=\ex{}\left[\frac{p(G=1,D=1\mid X)}{\ex{}[G\cdot D]}\cdot\rho_0(X,G,D)\cdot \frac{T-\ex{}[T]}{\ex{}[T](1-\ex{}[T])}\cdot Y\right].
    \]
Note how the latter resembles the identification functional for the panel data setting (\Cref{eq:coripwp}) after replacing $(Y_1-Y_0)$ by $\frac{T-\ex{}[T]}{\ex{}[T](1-\ex{}[T])}\cdot Y$.
This aligns well with analogous results in the canonical DiD literature where no compositional changes assumption is made---see, for instance, \citep{abadie2005semiparametric}.

\section{Estimation}\label{sec:est}
In this section, we first present two estimation strategies for each setting based on the identification results of \Cref{sec:id}.
We discuss the potential issues that these estimation strategies can face. Then we propose efficient and robust estimators based on the influence functions of our parameters of interest to address these issues.
Similar to the previous sections, we present our results first for the panel data setting, followed by those for the repeated cross-sections setting.

\subsection{Estimation in the Panel Data Setting}

Based on the definition of the parameter $\psi_{pd}$, 
one can estimate it using the following plug-in estimator:
\[
    \hat{\psi}^\mathrm{or}_\mathrm{pd} =  \frac{\ex{n}\left[\big(Y_1-Y_0-\hat{\mu}_{0,1,\Delta}(X)-\hat{\mu}_{1,0,\Delta}(X)+\hat{\mu}_{0,0,\Delta}(X)\big)\cdot \ind{G=1,D=1}\right]}{\ex{n}[\ind{G=1,D=1}]},
\]
where $\hat{\mu}_{g,d,\Delta}(\cdot)$ is an estimator of the outcome regression function $\mu_{g,d,\Delta}(\cdot)$, and $\ex{n}$ represents empirical mean.
Alternatively, one can estimate the parameter of interest based on the identification functional given by \Cref{prp:ipwpanel}.
Specifically, letting $\hat{\pi}_{g,d}(\cdot)$ be an estimator of $\pi_{g,d}(\cdot)$, $\psi_\mathrm{pd}$ can be estimated using the following weighting estimator:
\begin{equation*}
\begin{split}
    \hat{\psi}^\mathrm{ipw}_\mathrm{pd}
    =\ex{n}\big[\hat{e}_\mathrm{pd}\cdot \hat{\pi}_{1,1}(X)\cdot\hat{\rho}(X,G,D)\cdot(Y_1-Y_0)\big],
\end{split}
\end{equation*}
where $\hat{e}_\mathrm{pd}$ is an estimate of $1/\ex{}[G\cdot D]$, and 
\begin{equation*}
    \hat{\rho}(X,G,D) = \sum_{g,d\in\{0,1\}}
    \frac{(1-g-G)(1-d-D)}{\hat{\pi}_{g,d}(X)}.
\end{equation*}

Both $\hat{\psi}_\mathrm{pd}^\mathrm{or}$ and $\hat{\psi}_\mathrm{pd}^\mathrm{ipw}$ provide consistent estimates of the ATT under correct model specification.
The OR estimator ($\hat{\psi}_\mathrm{pd}^\mathrm{or}$) relies on correctly modeling the outcome regressions, while the IPW estimator ($\hat{\psi}_\mathrm{pd}^\mathrm{ipw}$) depends on the correct specification of the propensity score models.
If either model is misspecified, the respective estimator may be biased. 
To avoid model misspecifications, one can use nonparametric nuisance estimators.
However, 
the convergence rate of such estimators are 
often slower than $\sqrt{n}$.

To address the aforementioned issues, we next propose an estimation strategy
based on the influence function of $\psi_\mathrm{pd}$.
Our estimator is 
(i) robust to model misspecifications (as described in \Cref{thm:drp}), (ii) achieves the semiparametric efficiency bound, and (iii) can achieve parametric convergence rates 
even if the nuisance function estimators do not converge at $\sqrt{n}$ rate.
We start by deriving the efficient influence function of $\psi_\mathrm{pd}$ in the following result.

\begin{restatable}{theorem}{thmif}\label{thm:if}
    In the nonparametric model, the efficient influence function of $\psi_\mathrm{pd}$ is given by
    \begin{equation}
    \label{eq:ifp}
        \psi^{1}_\mathrm{pd}(O) = \frac{1}{\ex{}[G\cdot D]}
            \cdot
            \sum_{g,d\in\{0,1\}}(-1)^{(g+d+1)}w_{g,d}(X,G,D)\cdot
            \big(
                Y_1-Y_0-\mu_{g,d,\Delta}(X)
            \big) - \frac{G\cdot D}{\ex{}[G\cdot D]}\cdot\psi_\mathrm{pd},
    \end{equation}
    where $\mu_{g,d,\Delta}(X)=\ex{}[Y_1-Y_0\mid X, G=g, D=d]$,
\(
    w_{g,d}(X,G,D) = G\cdot D-\mathbbm{1}\{G=g, D=d\}\cdot \pi_{r,g,d}(X)\), and $\pi_{r,g,d}(X)=p(G=1,D=1\mid X)/p(G=g,D=d\mid X)$.
\end{restatable}

Based on the influence function in \Cref{eq:ifp}, 
we propose the following procedure to estimate the parameter $\psi_\mathrm{pd}$.
We use the cross-fitting approach \citep{chernozhukov2018double} to separate the estimation of nuisance parameters from that of the parameter of interest.
In particular, we partition the samples into $L$ equally-sized folds of size $m$ indexed by $\{1,\dots,L\}$.
For each $\ell\in\{1,\dots,L\}$, let $\hat{\mu}^\ell_{g,d,\Delta}(X)$ and $\hat{\pi}^\ell_{r,g,d}(X)$ be the estimators of $\ex{}[Y_1-Y_0\mid X, G=g, D=d]$ and $\pi_{r,g,d}(X)$, respectively, using the data in all but the $\ell$-th 
fold.
Let 
$\ee^\ell_m[\cdot]$ represent the empirical mean in the $\ell$-th fold of data.
Furthermore, let $\hat{e}_\mathrm{pd}^\ell$ be the estimator of $\frac{1}{\ex{}[G\cdot D]}$ using the data in the $\ell$-th fold, defined as \begin{equation}\label{eq:defhate}
\hat{e}_\mathrm{pd}^\ell=\frac{1}{\max\{\frac{1}{m},\ee^{\ell}_m[G\cdot D]\}},\end{equation}
where we take the maximum in the denominator so that $\hat{e}_\mathrm{pd}^\ell$ is well-defined even if $\ee^{\ell}_m[G\cdot D]=0$.
Our estimator for the parameter of interest is:
\begin{equation*}
    \begin{split}
        \hat{\psi}_\mathrm{pd}^\mathrm{dr} = 
        \frac{1}{L}\sum_{\ell=1}^L\hat{e}_\mathrm{pd}^\ell\cdot \ee^\ell_m\big[
            \eta_\mathrm{pd}(O; \{\hat{\mu}^\ell_{g,d}\}_{g,d}, \{\hat{\pi}^\ell_{r,g,d}\}_{g,d})
        \big],
    \end{split}
\end{equation*}
where
\begin{equation*}
    \eta_\mathrm{pd}(O; \{\hat{\mu}^\ell_{g,d}\}_{g,d}, \{\hat{\pi}^\ell_{r,g,d}\}_{g,d})\coloneqq
            \sum_{g,d\in\{0,1\}}(-1)^{(g+d+1)}\hat{w}^\ell_{g,d}(X,G,D)\cdot
            \big(
                Y_1-Y_0-\hat{\mu}^\ell_{g,d,\Delta}(X)
            \big),
\end{equation*}
and
\begin{equation*}
    \hat{w}^\ell_{g,d}(X,G,D) = G\cdot D-\hat{\pi}^\ell_{r,g,d}(X)\cdot\mathbbm{1}\{G=g, D=d\}.
\end{equation*}
Below, we show that our proposed estimator is robust against misspecifications.


\begin{restatable}[Double robustness]{proposition}{thmdrp}\label{thm:drp}
    Suppose $\mathbb{E}[Y_0^2]$ and $\mathbb{E}[Y_1^2]$ are finite. Additionally, suppose that for each $\ell,g,d$, $\norm{\hat{\mu}^\ell_{g,d,\Delta}-f^\ell_{\mu, g,d}}=o_p(1)$ and $\norm{\hat{\pi}^\ell_{r,g,d}-f^\ell_{\pi, g,d}}=o_p(1)$,
    where $f^\ell_{\mu, g,d}$ and  $f^\ell_{\pi, g,d}$ are measurable functions and are in $L^2(p)$, meaning that their second moments with respect to $p(X)$ are finite.
    Under strict positivity (see \Cref{as:pos3}), 
    $\hat{\psi}_\mathrm{pd}^\mathrm{dr}$ is a consistent estimator of $\psi_\mathrm{pd}$, if for every $\ell$  and all $g,d\in\{0,1\}$, either of the following conditions (but not necessarily both) holds:
    \begin{enumerate}[label=(\roman*)]
        \item $f^\ell_{\mu, g,d}(X)= \mu_{g,d,\Delta}(X)$ a.s.
        \item $f^\ell_{\pi,g,d}(X)=\pi_{r,g,d}(X)$
        a.s.
    \end{enumerate}
\end{restatable}
Based on \Cref{thm:drp}, using the estimator $\hat{\psi}^\mathrm{dr}_\mathrm{pd}$ provides the researcher with two opportunities for consistent estimation of the ATT.
Specifically, consistency holds as long as for all $\ell,g,d$, either the outcome estimator $\hat{\mu}_{g,d,\Delta}(X)$, or the propensity score ratio estimator $\hat{\pi}_{r,g,d}(X)$ is $L^2(p)$-consistent for the true nuisance function, but not necessarily both. 

Next, we show that under certain regularity and convergence rate conditions, our estimator will be $\sqrt{n}$-consistent and asymptotically normal (CAN).
Importantly, the requirement is imposed on the product of pairs of convergence rates, rather than on the individual rates themselves.
Consequently, none of the nuisance functions are required to converge at the $\sqrt{n}$ rate.
\begin{assumption}\label{as:infp}
For every $\ell\in\{1,\dots,L\}$ and every $g,d\in\{0,1\}$,
    \begin{itemize}
        \item $\norm{\hat{\mu}^\ell_{g,d,\Delta} - \mu_{g,d,\Delta}}=o_p(1)$ and $\norm{\hat{\pi}_{r,g,d}^\ell - \pi_{r,g,d}}=o_p(1)$.
        \item $\norm{(\hat{\mu}^\ell_{g,d,\Delta})^2 - \mu_{g,d,\Delta}^2}=o_p(1)$ and $\norm{(\hat{\pi}_{r,g,d}^\ell)^2 - (\pi_{r,g,d})^2}=o_p(1)$.
        \item $\norm{\hat{\mu}^\ell_{g,d,\Delta} - \mu_{g,d,\Delta}}=O_p\big(r_{\mu,g,d}(n)\big)$ and $\norm{\hat{\pi}_{r,g,d}^\ell - \pi_{r,g,d}}=O_p\big(r_{\pi,g,d}(n)\big)$ such that $r_{\mu,g,d}(n)\times\allowbreak r_{\pi,g,d}(n)=o(n^{-1/2})$.
        \item  $\norm{\mu_{g,d,\Delta}^2}$, $\ex{}[Y_0^4]$, and $\ex{}[Y_1^4]$ are finite.
    \end{itemize}
\end{assumption}
\begin{restatable}[CAN]{theorem}{thmcanp}\label{thm:canp}
    Under strict positivity (Assumption \ref{as:pos3}), and \Cref{as:infp}, we have \[         \sqrt{n}(\hat{\psi}_\mathrm{pd}^\mathrm{dr}-\psi_\mathrm{pd})\overset{D}{\to}\mathcal{N}\Big(0,\mathrm{var}\big(             \psi^{1}_p(O)         \big)\Big),     \] where $\overset{D}{\to}$ represents convergence in distribution.
\end{restatable}

As a corollary of \Cref{thm:canp}, we can use the influence function $\psi^{1}_\mathrm{pd}(O)$ to obtain confidence intervals for the parameter of interest, $\psi_\mathrm{pd}$.
Specifically, for every $\ell\in\{1,\dots,L\}$, we estimate the variance of $\psi^{1}_\mathrm{pd}(\cdot)$ in the $\ell$-th fold as
\[
    \hat{\sigma}^2_\ell = \ee_m^\ell\left[\Big(\hat{e}^{-\ell}_\mathrm{pd}\cdot\eta_\mathrm{pd}(O; \{\hat{\mu}^\ell_{g,d}\}_{g,d}, \{\hat{\pi}^\ell_{r,g,d}\}_{g,d})
    - \hat{e}^{-\ell}_\mathrm{pd}\cdot G\cdot D\cdot \hat{\psi}_\mathrm{pd}^\mathrm{dr}\Big)^2\right],
\]
where $\hat{e}^{-\ell}_\mathrm{pd}$ is the  estimator of $1/\ex{}[G\cdot D]$ using data from all but the $\ell$-th fold of data, defined similarly to \Cref{eq:defhate}.
Then we define $\hat{\sigma}^2=\frac{1}{L}\sum_{\ell=1}^L\hat{\sigma}^2_\ell$.
Using this estimated variance, the $100(1-\alpha)\%$ confidence interval of $\psi_\mathrm{pd}$ can be obtained as 
\[
    \hat{\psi}_\mathrm{pd}^\mathrm{dr} \pm z_{1-\alpha/2}\frac{\hat{\sigma}}{\sqrt{n}},
\]
where $z_{1-\alpha/2}$ is the $(1-\alpha/2)$-quantile of the standard normal distribution.
\subsection{Estimation in the Repeated Cross-Sections Setting}
Similar to the panel data setting, based on the definition of $\psi_\mathrm{rc}$, 
it can be estimated using the following plug-in estimator:
\begin{align*}
    \hat{\psi}^\mathrm{or}_\textrm{rc} 
                =
                \frac{
                \ex{n}\left[
                    \big(Y - \hat{\mu}_{1,1,0}(X)-\hat{\mu}_{0,1,\Delta}(X)-\hat{\mu}_{1,0,\Delta}(X)+\hat{\mu}_{0,0,\Delta}(X)\big)
                    \cdot \ind{G=1,D=1,T=1}
                \right]
                }{\ex{n}[\ind{G=1,D=1,T=1}]},
\end{align*}
where $\hat{\mu}_{1,1,0}(\cdot)$ and $\hat{\mu}_{g,d,\Delta}(\cdot)$ are estimators of $\mu_{1,1,0}(\cdot)$, and $(\mu_{g,d,1}-\mu_{g,d,0})(\cdot)$, respectively.
Alternatively, \Cref{prp:ipwrc} suggests the following weighting-based estimator:
\[
    \hat{\psi}^\mathrm{ipw}_\mathrm{rc}=\ex{}\big[\hat{e}_\mathrm{rc}\cdot\hat{\pi}_{1,1,1}(X)\cdot\hat{\phi}(X,G,D,T)\cdot Y\big],
\]
where $\hat{e}_\mathrm{rc}$ is an estimate of $1/\ex{}[G\cdot D\cdot T]$, $\hat{\pi}_{g,d,t}(\cdot)$ is an estimator of the propensity score function $\pi_{g,d,t}(\cdot)$, and 
\[
    \hat{\phi} (X,G,D,T)
        =-\sum_{g,d,t\in\{0,1\}}\frac{(1-g-G)(1-d-D)(1-t-T)}{\hat{\pi}_{g,d,t}(X)}.
\]

As in the panel data setting, these two estimators ($\hat{\psi}^\mathrm{or}_\mathrm{rc}$ and $\hat{\psi}^\mathrm{ipw}_\mathrm{rc}$) can be biased if the corresponding models are misspecified.
One can use nonparametric nuisance function estimators to avoid misspecifications, but such estimators often have slow convergence rates.
We therefore propose an influence function-based estimator which is (i) robust to misspecifications (see \Cref{thm:drrc}), (ii) achieves the semiparametric efficiency bound, and (iii) can attain parametric convergence rates even if the nuisance functions converge slower than the $\sqrt{n}$ rate.
We begin by deriving the efficient influence function of $\psi_\mathrm{rc}$ in the following result.
\begin{restatable}{theorem}{thmifrc}\label{thm:ifrc}
    In the nonparametric model, the efficient influence function of $\psi_\mathrm{rc}$ is given by
    \begin{equation}\label{eq:ifrc}
        \begin{split}
            \psi^1_\mathrm{rc,1}(O)=\frac{1}{\ex{}[G\cdot D\cdot T]}
            \cdot
            \sum_{g,d,t\in\{0,1\}}(-1)^{(g+d+t)}\omega_{g,d,t}(X,G,D,T)\cdot\big(Y-&\mu_{g,d,t}(X)\big)\\&-\frac{G\cdot D\cdot T}{\ex{}[G\cdot D\cdot T]}\cdot\psi_\mathrm{rc},
        \end{split}
    \end{equation}
    where $\mu_{g,d,t}(X)= \ex{}[Y\mid X, G=g, D=d, T=t]$,
\(
    \omega_{g,d,t}(X,G,D,T) = G\cdot D\cdot T - 
    \pi_{r,g,d,t}(X)
    \cdot \mathbbm{1}\{G=g,D=d,T=t\}
\),
and
$\pi_{r,g,d,t}(X)=p(G=1, D=1, T=1\mid X)/p(G=g, D=d, T=t\mid X)$.
\end{restatable}

The influence function in \Cref{eq:ifrc}
suggests the following estimation strategy based on cross-fitting \citep{chernozhukov2018double}.
We partition the data into $L$ folds of size $m$, indexed by $\{1,\dots,L\}$.
For each $\ell\in\{1,\dots,L\}$, we let
$\hat{\mu}^\ell_{g,d,t}(X)$ and $\hat{\pi}^\ell_{r,g,d,t}(X)$ be estimators of $\ex{}[Y\mid X, G=g, D=d, T=t]$ and $\pi_{r,g,d,t}(X)$, respectively, using the data in all but $\ell$-th fold.
Additionally, let $\hat{e}^\ell_\mathrm{rc}$ be the estimator of $(1/\ex{}[G\cdot D\cdot T])$ using the data in the $\ell$-th fold, defined as
\begin{equation}\label{eq:erc}
    \hat{e}^\ell_\mathrm{rc} = \frac{1}{\max\{\frac{1}{m},\ee_m^{\ell}[G\cdot D\cdot T]\}}.
\end{equation}
Our proposed estimator for $\psi_\mathrm{rc}$ is:
\[
    \hat{\psi}_\mathrm{rc,1}^\mathrm{dr}=
    \frac{1}{L}\sum_{\ell=1}^L\hat{e}^\ell_\mathrm{rc}\cdot\ee_m^\ell
    \Big[
        \eta_\mathrm{rc,1}(O; \{\hat{\mu}^\ell_{g,d,t}\}_{g,d,t}, \{\hat{\pi}^\ell_{r,g,d,t}\}_{g,d,t})
    \Big],
\]
where
\[
    \eta_\mathrm{rc,1}(O;\{\hat{\mu}^\ell_{g,d,t}\}_{g,d,t}, \{\hat{\pi}^\ell_{r,g,d,t}\}_{g,d,t}) \coloneqq  
        \sum_{g,d,t\in\{0,1\}}(-1)^{(g+d+t)}\hat{\omega}^\ell_{g,d,t}(X,G,D,T)\cdot\big(Y-\hat{\mu}^\ell_{g,d,t}(X)\big),
\]
and,
\[
    \hat{\omega}_{g,d,t}(X,G,D,T) = G\cdot D\cdot T - \hat{\pi}_{r,g,d,t}(X)\cdot\mathbbm{1}\{G=g,D=d,T=t\}.
\]
Below, we show that $\hat{\psi}_\mathrm{rc,1}^\mathrm{dr}$ is robust against misspecifications.


\begin{restatable}[Double robustness]{proposition}{thmdrrc}\label{thm:drrc}
    Suppose $\mathbb{E}[Y^2]$ is finite. Additionally, suppose that for each $\ell,g,d,t$, $\norm{\hat{\mu}^\ell_{g,d,t}-f^\ell_{\mu, g,d,t}}=o_p(1)$ and $\norm{\hat{\pi}^\ell_{r,g,d,t}-f^\ell_{\pi, g,d,t}}=o_p(1)$,
    where $f^\ell_{\mu, g,d,t}$, $f^\ell_{\pi, g,d,t}$ are measurable functions in $L^2(p)$.
    Under strict positivity (see \Cref{as:posrc}), 
    $\hat{\psi}_\mathrm{rc,1}^\mathrm{dr}$ is a consistent estimator of $\psi_\mathrm{rc}$, if for every $\ell$  and all $g,d,t\in\{0,1\}$, either of the following conditions (but not necessarily both) holds:
    \begin{enumerate}[label=(\roman*)]
        \item $f^\ell_{\mu, g,d,t}(X)=\mu_{g,d,t}(X)$ a.s.
        \item $f^\ell_{\pi,g,d,t}(X)=\pi_{r,g,d,t}(X)$ a.s.
    \end{enumerate}
\end{restatable}
\Cref{thm:drrc} demonstrates that $\hat{\psi}^\mathrm{dr}_\mathrm{rc,1}$ is consistent as long as for all $\ell,g,d,t$, either the outcome regression estimator $\hat{\mu}_{g,d,t}$ or the propensity score ratio estimator $\hat{\pi}_{r,g,d,t}$ (but not necessarily both) is $L^2(p)$-consistent for the true nuisance function.
Next, we show that under certain regularity and convergence rate conditions, our estimator will be
$\sqrt{n}$-consistent and asymptotically normal (CAN). 
Similarly to the panel data setting, the requirement is imposed on the
product of pairs of convergence rates, rather than on the individual rates themselves. 
Consequently,
none of the nuisance functions are required to converge at the $\sqrt{n}$ rate.
\begin{assumption}\label{as:infrc}
For every $\ell\in\{1,\dots,L\}$ and every $g,d,t\in\{0,1\}$,
    \begin{itemize}
        \item $\norm{\hat{\mu}^\ell_{g,d,t} - \mu_{g,d,t}}=o_p(1)$ and $\norm{\hat{\pi}_{r,g,d,t}^\ell - \pi_{r,g,d,t}}=o_p(1)$.
        \item $\norm{(\hat{\mu}^\ell_{g,d,t})^2 - \mu_{g,d,t}^2}=o_p(1)$ and $\norm{(\hat{\pi}_{r,g,d,t}^\ell)^2 - (\pi_{r,g,d,t})^2}=o_p(1)$.
        \item $\norm{\hat{\mu}^\ell_{g,d,t} - \mu_{g,d,t}}=O_p\big(r_{\mu,g,d,t}(n)\big)$ and $\norm{\hat{\pi}_{r,g,d,t}^\ell - \pi_{r,g,d,t}}=O_p\big(r_{\pi,g,d,t}(n)\big)$ such that $r_{\mu,g,d,t}(n)\times\allowbreak r_{\pi,g,d,t}(n)=o(n^{-1/2})$.
        \item  $\norm{\mu_{g,d,t}^2}$, and $\ex{}[Y^4]$ are finite.
    \end{itemize}
\end{assumption}

\begin{restatable}[CAN]{theorem}{thmcanrc}\label{thm:canrc}
    Under strict positivity (\Cref{as:posrc}), and \Cref{as:infrc}, we have
    \[         
        \sqrt{n}(\hat{\psi}_\mathrm{rc,1}^\mathrm{dr}-\psi_\mathrm{rc})\overset{D}{\to}\mathcal{N}\Big(0,\mathrm{var}\big(             \psi^1_\mathrm{rc,1}(O)     \big)\Big).     
    \] 
\end{restatable}
As a corollary of \Cref{thm:canrc}, one can use the influence function $\psi^1_\mathrm{rc,1}(\cdot)$ to obtain confidence intervals for $\psi_\mathrm{rc}$.
Specifically, for every $\ell\in\{1,\dots,L\}$, we estimate the variance of $\psi^1_\mathrm{rc,1}(O)$ in the $\ell$-th fold as
\begin{equation*}
\begin{split}
    \hat{\sigma}^2_\ell = \ee_m^\ell\left[\Big(\hat{e}^{-\ell}_\mathrm{rc}\cdot\eta_\mathrm{rc,1}(O; \{\hat{\mu}^\ell_{g,d,t}\}_{g,d,t}, \{\hat{\pi}^\ell_{r,g,d,t}\}_{g,d,t})\big]
    - \hat{e}^{-\ell}_\mathrm{rc}\cdot G\cdot D\cdot T\cdot \hat{\psi}^\mathrm{dr}_\mathrm{rc,1}\Big)^2\right],
\end{split}
\end{equation*}
where $\hat{e}^{-\ell}_\mathrm{rc}$ is the estimator of $1/\ex{}[G\cdot D\cdot T]$ using data in all but $\ell$-th fold, defined similarly to \Cref{eq:erc}.
Then we define the estimated variance $\hat{\sigma}^2=\frac{1}{L}\sum_{\ell=1}^L\hat{\sigma}^2_\ell$, and, 
the $100(1-\alpha)\%$ confidence interval of $\psi_\mathrm{rc}$ can be obtained as 
\[
    \hat{\psi}_\mathrm{rc,1}^\mathrm{dr} \pm z_{1-\alpha/2}\frac{\hat{\sigma}}{\sqrt{n}},
\]
where $z_{1-\alpha/2}$ is the $(1-\alpha/2)$-quantile of the normal distribution.
\subsubsection{Estimation under No Compositional Changes}
Both \Cref{thm:drrc} and \Cref{thm:canrc} allow for compositional changes.
In \Cref{sec:idncc}, we showed how \Cref{as:nocompch} simplifies weighting-based identification.
Here, we show that this assumption also has significant implications for estimation.
To this end, we first provide the efficient influence function of $\psi_\mathrm{rc}$ under \Cref{as:nocompch}.


\begin{restatable}{theorem}{thmifncc}\label{thm:ifncc}
    In the model which places no restrictions on the obseved data other than that $T$ is independent of $(X,G,D)$, the efficient influence function of $\psi_{rc}$ is given by
    \begin{equation}\label{eq:ifncc}
    \begin{split}
        \psi^1_{rc,2}(O) = \frac{1}{\ex{}[G\cdot D]}
            \cdot\!
            \sum_{g,d,t\in\{0,1\}}\!\!(-1)^{(g+d+t)}\alpha_{g,d,t}(X,G,D,T)\cdot \big( Y &- \mu_{g,d,t}(X)\big) - \frac{G\cdot D}{\ex{}[G\cdot D]}\cdot\psi_{rc},
    \end{split}
    \end{equation}
    where $\mu_{g,d,t}(X)= \ex{}[Y\mid X, G=g, D=d, T=t]$, 
    \(
        \alpha_{g,d,t}(X,G,D,T) = G\cdot D - 
        \big(
        \mathbbm{1}\{T=t\}
        /
        p(T=t)
        \big)
        \cdot \mathbbm{1}\{G=g,D=d\}\cdot \pi_{r,g,d}(X)
    \),
    and $\pi_{r,g,d}(X)=p(G=1,D=1\mid X)/p(G=g,D=d\mid X)$.
\end{restatable}

Let $\hat{\mu}^\ell_{g,d,t}(X)$, and $\hat{\pi}^\ell_{r,g,d}(X)$
be estimators of $\ex{}[Y\mid X, G=g, D=d, T=t]$, and $\pi_{r,g,d}(X)$
respectively, using the data in all but $\ell$-th fold.
Also, define 
\[
    \hat{e}^\ell_\mathrm{rc,2}=\frac{1}{\max\{\frac{1}{m},\ee_m^{\ell}[G\cdot D]\}}.
\]
Here, based on the influence function in \Cref{eq:ifncc}, we propose the following estimator for $\psi_\mathrm{rc}$ under \Cref{as:nocompch}:
\[
    \begin{split}
    \hat{\psi}_{\mathrm{rc},2}^\mathrm{dr}=
    \frac{1}{L} \sum_{\ell=1}^L
    \hat{e}_\mathrm{rc,2}^\ell\cdot
    \ee_m^\ell
    \Big[
        \eta_\mathrm{rc,2}(O; 
        \{\hat{\mu}^\ell_{g,d,t}\}_{g,d,t}, \{\hat{\pi}^\ell_{r,g,d}\}_{g,d})
    \Big],
    \end{split}
\]
where 
\[
\begin{split}
    \eta_\mathrm{rc,2}&(O;
    \{\hat{\mu}^\ell_{g,d,t}\}_{g,d}, \{\hat{\pi}^\ell_{r,g,d}\}_{g,d}) \coloneqq
    \sum_{g,d,t\in\{0,1\}}(-1)^{(g+d+t)}\hat{\alpha}^\ell_{g,d,t}(X,G,D,T)\cdot 
    \big(
        Y
        - \hat{\mu}^\ell_{g,d,t}(X)
    \big),
\end{split}
\]
and
\begin{equation*}
    \hat{\alpha}^\ell_{g,d,t}(X,G,D,T) = G\cdot D-
    \frac{1}{p(T=t)}
    \cdot\hat{\pi}^\ell_{r,g,d}(X)\cdot\mathbbm{1}\{G=g, D=d,T=t\}.
\end{equation*}
The following result demonstrates the robustness of $\hat{\psi}_\mathrm{rc,2}^\mathrm{dr}$ under \Cref{as:nocompch} against misspecifications.
\begin{restatable}[Double robustness]{proposition}{thmindep}\label{thm:drindep}
    Suppose $\mathbb{E}[Y^2]$ is finite. Additionally, suppose that for each $\ell,g,d,t$, $\norm{\hat{\mu}^\ell_{g,d,t}-f^\ell_{\mu, g,d,t}}=o_p(1)$ and $\norm{\hat{\pi}^\ell_{r,g,d}-f^\ell_{\pi, g,d}}=o_p(1)$,
    where $f^\ell_{\mu, g,d,t}$, $f^\ell_{\pi, g,d}$ are measurable functions in $L^2(p)$.
    Under \Cref{as:nocompch}, and strict positivity of $p(G,D\mid X)$ (see \Cref{as:pos3}),
    $\hat{\psi}_{\mathrm{rc,2}}^\mathrm{dr}$ is a consistent estimator of $\psi_\mathrm{rc}$ if for every $\ell$ and all $g,d\in\{0,1\}$, either of the following conditions (but not necessarily both) holds:
    \begin{enumerate}[label=(\roman*)]
        \item $f^\ell_{\mu, g,d,1}(X)-f^\ell_{\mu, g,d,0}(X)=\mu_{g,d,1}(X)-\mu_{g,d,0}(X)$ a.s.
     \item $f^\ell_{\pi, g,d}(X)=\pi_{r,g,d}(X)$ a.s.
     \end{enumerate}
\end{restatable}

As evident from \Cref{thm:drindep}, under \Cref{as:nocompch}, a stronger robustness can be achieved.
In particular, $\hat{\psi}_{\mathrm{rc},2}^\mathrm{dr}$ is consistent under similar assumptions to the panel data setting; it suffices to have access to consistent estimators of either $\mu_{g,d,1}-\mu_{g,d,0}$, or $\pi_{r,g,d}$.
However, without this assumption, one needs consistent estimators of both $\mu_{g,d,1}$ and $\mu_{g,d,0}$ (as opposed to only their contrast), or consistent estimators of both $\pi_{r,g,d,1}$ and $\pi_{r,g,d,0}$ (as opposed to only $\pi_{r,g,d}$). 
Under \Cref{as:nocompch}, the required assumptions for $\sqrt{n}$-consistency and asymptotic normality of $\hat{\psi}^\mathrm{dr}_\mathrm{rc,2}$ are weaker than those for $\hat{\psi}^\mathrm{dr}_\mathrm{rc,1}$.
In particular, convergence rate conditions are imposed only on $(\hat{\mu}_{g,d,1}-\hat{\mu}_{g,d,0})$ and $\hat{\pi}_{r,g,d}$, rather than on $\hat{\mu}_{g,d,1},\hat{\mu}_{g,d,0},\hat{\pi}_{r,g,d,1}$, and $\hat{\pi}_{r,g,d,0}$ (which was required in \Cref{as:infrc}).

\begin{assumption}\label{as:infrc2}
For every $\ell\in\{1,\dots,L\}$ and every $g,d,t\in\{0,1\}$,
    \begin{itemize}
        \item $\norm{\hat{\mu}^\ell_{g,d,t} - \mu_{g,d,t}}=o_p(1)$ and $\norm{\hat{\pi}_{r,g,d}^\ell - \pi_{r,g,d}}=o_p(1)$.
        \item $\norm{(\hat{\mu}^\ell_{g,d,t})^2 - \mu_{g,d,t}^2}=o_p(1)$ and $\norm{(\hat{\pi}_{r,g,d}^\ell)^2 - ({\pi_{r,g,d}})^2}=o_p(1)$.
        \item $\norm{(\hat{\mu}^\ell_{g,d,1}-\hat{\mu}^\ell_{g,d,0}) - (\mu_{g,d,1}-\mu_{g,d,0})}=O_p\big(r_{\mu,g,d}(n)\big)$ and $\norm{\hat{\pi}_{r,g,d}^\ell - \pi_{r,g,d}}=O_p\big(r_{\pi,g,d,}(n)\big)$ such that $r_{\mu,g,d}(n)\times\allowbreak r_{\pi,g,d}(n)=o(n^{-1/2})$.
        \item  $\norm{\mu_{g,d,t}^2}$, and $\ex{}[Y^4]$ are finite.
    \end{itemize}
\end{assumption}

\begin{restatable}[CAN]{theorem}{thmcanrctwo}\label{thm:canrc2}
    Under strict positivity of $p(G,D\mid X)$ (\Cref{as:pos3}), and Assumptions \ref{as:nocompch} and \ref{as:infrc2}, we have
    \[         
        \sqrt{n}(\hat{\psi}_\mathrm{rc,2}^\mathrm{dr}-\psi_\mathrm{rc})\overset{D}{\to}\mathcal{N}\Big(0,\mathrm{var}\big(             \psi^1_\mathrm{rc,2}(O)\big)\Big).     
    \] 
\end{restatable}
Once again, a corollary of \Cref{thm:canrc2} is that confidence intervals for the parameter of interest can be obtained by estimating the variance of the influence function $\psi^1_\mathrm{rc,2}(O)$.

\section{Simulation Studies}\label{sec:sim}
In this section, we present simulation studies to assess the performance of our proposed methodology.
We adapted the data-generating mechanism of \cite{kang2007demystifying} to the triple-difference setting, with the key modification of introducing an unobserved confounder $U$, drawn from a standard normal distribution.
Four observed covariates $(X_1, X_2, X_3, X_4)$ were constructed from nonlinear transformations of latent normal variables, while treatment eligibility and domain assignments $(G,D)$ were jointly generated from a multinomial logit model that depends on both observed covariates and $U$.
Outcomes were generated from nonlinear functions of $(X_1,\dots,X_4)$ and $U$, with potential outcomes defined to satisfy the conditional parallel difference in trends assumption (\Cref{as:id3} and \Cref{as:idrc} in the panel data and repeated cross-sections settings, respectively).
In the repeated cross-sections setting, the time indicator $T$ was sampled from an expit model with a parameter that depends on both the covariates and the treatment and domain indicators, thereby violating no-compositional changes (\Cref{as:nocompch}).
A detailed description of our data generating process is included in \Cref{apx:sim}.

\begin{figure}[t]
    \centering
    \includegraphics{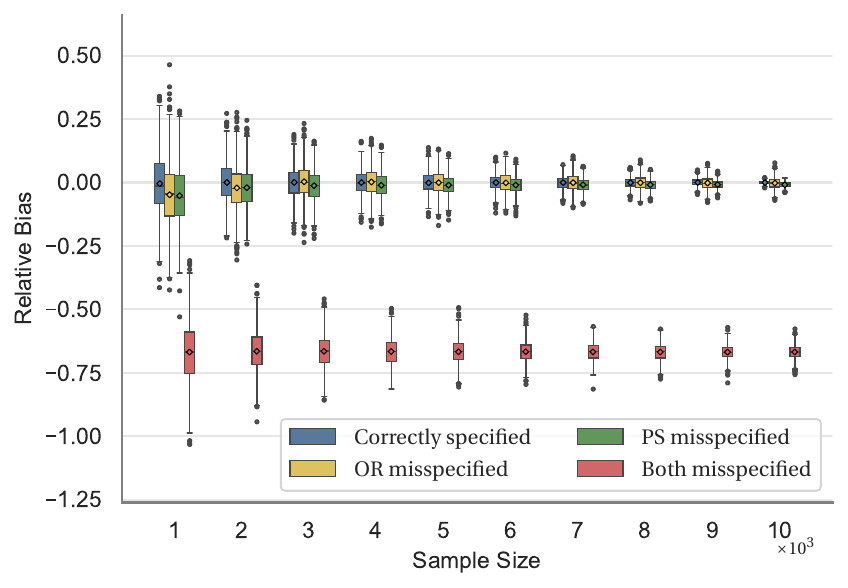}
    \caption{Relative bias vs sample size in the panel data setting.
    This pattern confirms the double-robustness property of $\hat{\psi}_\mathrm{pd}^\mathrm{dr}$.}
    \label{fig:biasp}
\end{figure}
\begin{table}[ht!]
\centering
\caption{Average bias and RMSE of $\hat{\psi}_\mathrm{pd}^\mathrm{dr}$ across sample sizes, based on 1000 Monte Carlo replications.
The true parameter value is equal to $10$.
CS, PSM, ORM, and BM represent the specification of nuisance functions: Correctly Specified, Propensity Score Misspecified, Outcome Regression Misspecified, and Both Misspecified, respectively.
}
\label{tab:panel}
\setlength{\tabcolsep}{4.1pt} 
\begin{tabular}{lrrrrrrrrrrr}
\toprule
 & \multicolumn{11}{c}{\textbf{Sample size}} \\
\cmidrule(lr){3-12}
 & & 1000 & 2000 & 3000 & 4000 & 5000 & 6000 & 7000 & 8000 & 9000 & 10000 \\
\midrule
\multirow{2}{*}{CS} 
  & Avg. Bias     & -0.045 & 0.002 & 0.004 & -0.002 & -0.013 & -0.006 & -0.009 & -0.006 & -0.001 & -0.006 \\
  & Avg. RMSE & 1.153 & 0.805 & 0.607 & 0.481 & 0.395 & 0.314 & 0.260 & 0.202 & 0.143 & 0.065 \\
\midrule
\multirow{2}{*}{PSM} 
  & Avg. Bias     & -0.521 & -0.204 & -0.127 & -0.111 & -0.107 & -0.102 & -0.092 & -0.092 & -0.083 & -0.093 \\
  & Avg. RMSE & 1.269 & 0.825 & 0.615 & 0.493 & 0.407 & 0.336 & 0.280 & 0.224 & 0.174 & 0.131 \\
\midrule
\multirow{2}{*}{ORM} 
  & Avg. Bias     & -0.481 & -0.214 & 0.029 & 0.021 & -0.006 & -0.014 & -0.011 & -0.008 & -0.021 & -0.023 \\
  & Avg. RMSE & 1.313 & 0.887 & 0.681 & 0.545 & 0.471 & 0.392 & 0.336 & 0.273 & 0.242 & 0.216 \\
\midrule
\multirow{2}{*}{BM} 
  & Avg. Bias     & -6.684 & -6.650 & -6.649 & -6.661 & -6.671 & -6.669 & -6.677 & -6.681 & -6.684 & -6.679 \\
  & Avg. RMSE & 6.789 & 6.700 & 6.680 & 6.683 & 6.687 & 6.681 & 6.687 & 6.689 & 6.689 & 6.684 \\
\bottomrule
\end{tabular}
\end{table}

In our evaluations, we considered four different variants of estimators $\hat{\psi}^\mathrm{dr}_\mathrm{pd}$ and $\hat{\psi}^\mathrm{dr}_\mathrm{rc,1}$: (i) with both outcome regression (OR) functions and propensity scores (PS) correctly specified, (ii) with correctly specified OR functions but misspecified PS, (iii) with misspecified OR functions but correctly specified PS, and (iv) with all nuisance functions misspecified.
To help ensure that the nuisance functions were adequately captured and the estimator can be regarded as correctly specified, we used fully connected neural networks, designed with sufficient depth and width to flexibly model complex, non-linear nuisance functions.
For OR functions, we used networks with three hidden layers, while for PS we used networks with two hidden layers. 
For the case of misspecified OR function, we used a ridge regression estimator that takes only $X_1$, the first observed covariate, into account.
For the case of misspecified PS, we used a logistic regression estimator that works only with $X_1$, the first observed covariate.
In all four cases, cross-fitting with three folds was employed.
The complete implementation details and reproducing python code are available at \href{https://github.com/SinaAkbarii/triplediff}{https://github.com/SinaAkbarii/triplediff}.

\begin{figure}[t]
    \centering
    \includegraphics{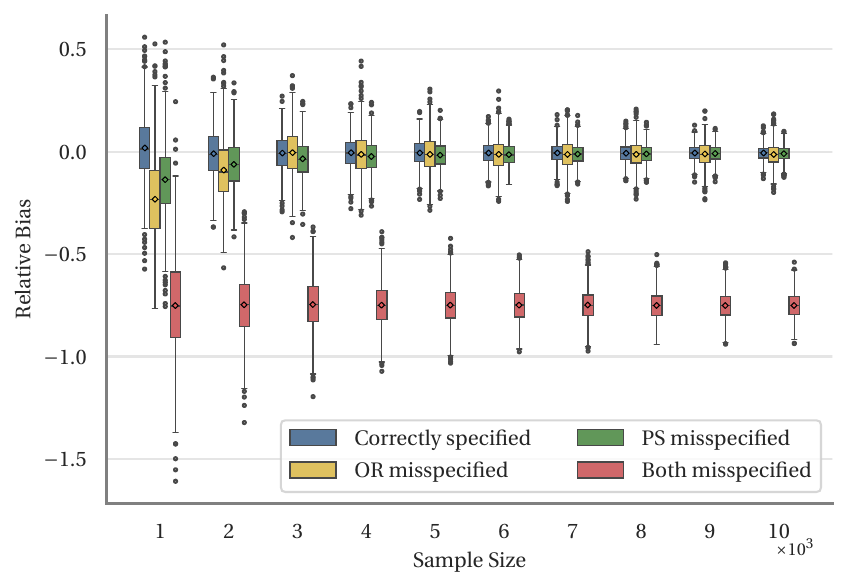}
    \caption{Relative bias vs sample size in the repeated cross-sections setting.
    This pattern confirms the double-robustness property of $\hat{\psi}_\mathrm{rc,1}^\mathrm{dr}$.}
    \label{fig:biasrc}
\end{figure}
\begin{table}[ht!]
\centering
\caption{Average bias and RMSE of $\hat{\psi}_\mathrm{rc,1}^\mathrm{dr}$ across sample sizes, based on 1000 Monte Carlo replications.
The true parameter value is equal to $10$.
CS, PSM, ORM, and BM represent the specification of nuisance functions: Correctly Specified, Propensity Score Misspecified, Outcome Regression Misspecified, and Both Misspecified, respectively.
}
\label{tab:rc}
\setlength{\tabcolsep}{4.1pt} 
\begin{tabular}{lrrrrrrrrrrr}
\toprule
 & \multicolumn{11}{c}{\textbf{Sample size}} \\
\cmidrule(lr){3-12}
 & & 1000 & 2000 & 3000 & 4000 & 5000 & 6000 & 7000 & 8000 & 9000 & 10000 \\
\midrule
\multirow{2}{*}{CS} 
  & Avg. Bias     & 0.180 & -0.099 & -0.070 & -0.053 & -0.063 & -0.066 & -0.063 & -0.076 & -0.065 & -0.067 \\
  & Avg. RMSE & 1.606 & 1.184 & 0.897 & 0.748 & 0.642 & 0.564 & 0.501 & 0.458 & 0.406 & 0.375 \\
\midrule
\multirow{2}{*}{PSM} 
  & Avg. Bias     & -1.368 & -0.622 & -0.350 & -0.237 & -0.170 & -0.147 & -0.119 & -0.111 & -0.094 & -0.100 \\
  & Avg. RMSE & 2.251 & 1.324 & 0.981 & 0.797 & 0.670 & 0.569 & 0.515 & 0.473 & 0.421 & 0.390 \\
\midrule
\multirow{2}{*}{ORM} 
  & Avg. Bias     & -2.323 & -0.900 & -0.046 & -0.135 & -0.136 & -0.135 & -0.137 & -0.135 & -0.113 & -0.137 \\
  & Avg. RMSE & 3.065 & 1.781 & 1.181 & 1.045 & 0.907 & 0.828 & 0.738 & 0.668 & 0.615 & 0.592 \\
\midrule
\multirow{2}{*}{BM} 
  & Avg. Bias     & -7.511 & -7.476 & -7.462 & -7.494 & -7.491 & -7.490 & -7.486 & -7.508 & -7.512 & -7.508 \\
  & Avg. RMSE & 7.875 & 7.643 & 7.572 & 7.568 & 7.550 & 7.535 & 7.525 & 7.543 & 7.543 & 7.535 \\
\bottomrule
\end{tabular}
\end{table}

\Cref{fig:biasp} illustrates the relative bias of each of the variants of $\hat{\psi}_\mathrm{pd}^\mathrm{dr}$ in the panel data setting with sample sizes ranging from $1000$ to $10000$.
Relative bias is defined as $(\hat{\psi}/\psi)-1$, where $\hat{\psi}$ is the estimate and $\psi$ is the true value of the parameter of interest, which was $10$ in our data generating mechanism.
Each boxplot summarizes results from 1,000 Monte Carlo replications.
The results show that when at least one set of the OR functions or PS  is correctly specified, the relative bias converges to zero, confirming the double-robustness property. 
In contrast, when all nuisance functions are misspecified, the estimates converge to a value that differs from the true parameter, causing the relative bias to converge to a negative value rather than vanish.
\Cref{tab:panel} presents the numerical values of average bias and root mean squared error (RMSE).

\Cref{fig:biasrc} and \Cref{tab:rc} show the relative bias, average bias, and RMSE of variants of $\hat{\psi}_\mathrm{rc,1}^\mathrm{dr}$ estimators in the repeated cross-sections setting.
Overall, the patterns closely mirror those observed in the panel data setting. However, for a fixed sample size, these estimators exhibit greater variance than their panel data counterparts. This is expected, since each panel data sample contains both $Y_0$ and $Y_1$ and therefore provides more information, whereas each repeated cross-sections sample includes only one outcome.
To make the comparison clearer, in
\Cref{tab:bias_coverage_mse}, we report the bias, RMSE, and empirical coverage of 95\% confidence intervals of our estimators $\hat{\psi}_\mathrm{pd}^\mathrm{dr}$ and $\hat{\psi}_\mathrm{rc,1}^\mathrm{dr}$, 
when the nuisance functions are correctly specified. 
Moreover, the results of this table support the asymptotic normality of these estimators, as coverage exceeds $95\%$ with increasing sample size.


\begin{table}[ht]
\centering
\caption{Bias, RMSE, and empirical coverage of 95\% confidence intervals across sample sizes, based on 1000 Monte Carlo replications.
The true parameter value is equal to $10$.}
\label{tab:bias_coverage_mse}
\setlength{\tabcolsep}{4.3pt} 
\begin{tabular}{lrrrrrrrrrrr}
\toprule
 & \multicolumn{11}{c}{\textbf{Sample size}} \\
\cmidrule(lr){3-12}
 & & 1000 & 2000 & 3000 & 4000 & 5000 & 6000 & 7000 & 8000 & 9000 & 10000 \\
\midrule
\multirow{3}{*}{$\hat{\psi}_\mathrm{pd}^\mathrm{dr}$} 
  & Avg. Bias     & -0.045 & 0.002 & 0.004 & -0.002 & -0.013 & -0.006 & -0.009 & -0.006 & -0.001 & -0.006 \\
  & Avg. RMSE & 1.153 & 0.805 & 0.607 & 0.481 & 0.395 & 0.314 & 0.260 & 0.202 & 0.143 & 0.065 \\
  &  Coverage     & 94.9 & 95.9 & 98.0 & 98.4 & 99.2 & 99.7 & 100.0 & 100.0 & 100.0 & 100.0 \\
\midrule
\multirow{3}{*}{$\hat{\psi}_\mathrm{rc,1}^\mathrm{dr}$} 
  & Avg. Bias     & 0.180 & -0.099 & -0.070 & -0.053 & -0.063 & -0.066 & -0.063 & -0.076 & -0.065 & -0.067 \\
  & Avg. RMSE & 1.606 & 1.184 & 0.897 & 0.748 & 0.642 & 0.564 & 0.501 & 0.458 & 0.406 & 0.375 \\
  & Coverage      & 93.8 & 94.8 & 96.1 & 96.3 & 96.9 & 97.6 & 97.8 & 97.8 & 98.8 & 99.1 \\
\bottomrule
\end{tabular}
\end{table}


\section{Application: Effect of Mandated Maternity Benefits on Wages}\label{sec:application}
We applied our method to examine the impact of mandated maternity benefits on wages of women of childbearing age in the United States. 
We used the May Current Population Survey (CPS) of \cite{census_cps_may_2023} for the years 1974--1975, representing the pre-treatment period, and 1977--1978, representing the post-treatment period. 
The treatment of interest ($A$) is the adoption of state-level laws in the late 1970s requiring employers to offer health insurance policies covering childbirth costs and maternity benefits.
The outcome of interest ($Y$) is the logarithm of the hourly wages.
All wage measures were adjusted using the Consumer Price Index (CPI) for the corresponding year: 49.3 (1974), 53.8 (1975), 60.6 (1977), and 65.2 (1978). 
The design of our study follows that of \citet{gruber1994maternity}.
The analysis was restricted to eight states: Connecticut, Illinois, Indiana, Massachusetts, New Jersey, New York, North Carolina, and Ohio, where Illinois, New Jersey and New York adopted the policy and were considered as the target domain ($D=1$), whereas the other states did not do so and served as the reference domain ($D=0$).
Eligibility for the maternity mandate is defined by an indicator variable $G=1$ for married women aged 20--40 (childbearing age) and zero otherwise. 
We used the following demographic and occupational covariates ($X$): education, age, sex, race (white/non-white), marital status, union status, and white-collar occupation status. 
We further restricted the sample to individuals aged 20--65 and followed Gruber's exclusion criteria, removing single or divorced women aged 20--40 as well as married men in the same age range.
Individuals with hourly wages below $1$ or above $100$ US dollars were also excluded.

The primary distinction between our analysis and that of \citet{gruber1994maternity} lies in the estimation strategy.
While Gruber considered a simple parametric model with fixed effects for the outcome such that one of the parameters will become the causal parameter of interest,
we used our semiparametric influence function-based estimator $\hat{\psi}_\mathrm{rc,1}^\mathrm{dr}$,
where nuisance functions were estimated using fully connected neural networks as described in \Cref{sec:sim}.
The results of our analysis is presented in \Cref{tab:real_data_results}.
For the standard error and p-value (for testing $H_0$: effect $=0$), we report results based on both the influence function and 
bootstrap through 1000 replications.\footnote{The python implementations to reproduce these results can be found at \url{https://github.com/SinaAkbarii/triplediff}.}
We note that mandated maternity benefits increase labor costs for employers, which are likely passed on to employees in the form of lower wages. 
In particular, it is implausible that mandated maternity benefits would raise wages for married women of childbearing age. 
For this reason, we report a one-sided p-value.
Our analysis concluded a point estimate of 
$\hat{\psi}_\mathrm{rc,1}^\mathrm{dr} = -0.02633$ with standard error $0.02223$ for change in the log of hourly wages. This implies that the mandates resulted in a drop of $2.6\%$ in the hourly wages.
This is slightly lower than Gruber's conclusion, which was a $4.2\%$ drop in hourly wages.
For evaluating the significance, the one-sided p-values are 
$0.118$
and $0.109$,
obtained from the influence function and bootstrap approaches, respectively.
Taken together, these results provide evidence that mandated maternity benefits reduced the wages of married women of childbearing age in adopting states. 

\begin{table}[ht]
\centering
\caption{Point estimate, standard errors, and and one-sided p-values based on influence function (IF) and bootstrap (BS) approaches.}
\label{tab:real_data_results}
\setlength{\tabcolsep}{18pt} 
\begin{tabular}{lccc}
\toprule
 & \multirow{2}{*}{Point estimate}
 & Standard error
 & One-sided p-value\\
 &  
  & (IF/BS)
  & (IF/BS)\\
\midrule
$\hat{\psi}_\mathrm{rc,1}^\mathrm{dr}$ & -0.02633   
&  0.02223 / 0.02208
& 0.118 / 0.109
 \\
\bottomrule
\end{tabular}
\end{table}

\section{Conclusion}
We studied the identification and estimation of the average treatment effect on the treated within the triple difference framework, focusing on both panel data and repeated cross-sections settings.
From the identification standpoint, we presented the first weighting-based identification results, notably, while allowing for compositional changes in the repeated cross-sections setting.
From an estimation standpoint, we proposed semiparametric estimators for the triple difference framework in both panel data and repeated cross-section settings. 
These estimators employ a cross-fitting approach, allowing flexible machine learning methods to estimate the nuisance functions. 
We identified the conditions under which our estimators are efficient, doubly robust, 
$\sqrt{n}$-consistent, and asymptotically normal.
Our work lays a foundation for further analysis of the triple difference framework.
Given the practical relevance and growing use of this framework in empirical research, there is ample scope for future studies to extend such analyses, especially by exploring settings with multiple time periods, dynamic treatment effects, and continuous treatments.
As an application of our proposed methodology, we assessed the effect of mandated maternity benefits on the hourly wages of women of childbearing age and found that these mandates result in a $2.6\%$ drop in hourly wages.
Our estimated 2.6\% reduction in hourly wages is somewhat smaller than the 4.2\% reported by \cite{gruber1994maternity}; however, while Gruber relied on a simple parametric model, our approach employs nonparametric machine learning methods, which are more flexible and reliable.
Therefore, the effect of mandated maternity benefit policies might be slightly smaller than previously estimated.
\bibliographystyle{plainnat}
\bibliography{biblio}

\begin{thebibliography}{34}
\providecommand{\natexlab}[1]{#1}
\providecommand{\url}[1]{\texttt{#1}}
\expandafter\ifx\csname urlstyle\endcsname\relax
  \providecommand{\doi}[1]{doi: #1}\else
  \providecommand{\doi}{doi: \begingroup \urlstyle{rm}\Url}\fi

\bibitem[Abadie(2005)]{abadie2005semiparametric}
Alberto Abadie.
\newblock Semiparametric difference-in-differences estimators.
\newblock \emph{The review of economic studies}, 72\penalty0 (1):\penalty0 1--19, 2005.

\bibitem[Akbari and Kiyavash(2024)]{akbari2024triple}
Sina Akbari and Negar Kiyavash.
\newblock Triple changes estimator for targeted policies.
\newblock In \emph{Forty-first International Conference on Machine Learning}, 2024.

\bibitem[Angrist and Pischke(2009)]{angrist2009mostly}
Joshua~D Angrist and J{\"o}rn-Steffen Pischke.
\newblock \emph{Mostly harmless econometrics: An empiricist's companion}.
\newblock Princeton university press, 2009.

\bibitem[Arkhangelsky et~al.(2021)Arkhangelsky, Athey, Hirshberg, Imbens, and Wager]{arkhangelsky2021synthetic}
Dmitry Arkhangelsky, Susan Athey, David~A Hirshberg, Guido~W Imbens, and Stefan Wager.
\newblock Synthetic difference-in-differences.
\newblock \emph{American Economic Review}, 111\penalty0 (12):\penalty0 4088--4118, 2021.

\bibitem[Ashenfelter and Card(1984)]{ashenfelter1984using}
Orley~C Ashenfelter and David Card.
\newblock Using the longitudinal structure of earnings to estimate the effect of training programs, 1984.

\bibitem[Athey and Imbens(2006)]{athey2006identification}
Susan Athey and Guido~W Imbens.
\newblock Identification and inference in nonlinear difference-in-differences models.
\newblock \emph{Econometrica}, 74\penalty0 (2):\penalty0 431--497, 2006.

\bibitem[Bareinboim and Pearl(2016)]{bareinboim2016causal}
Elias Bareinboim and Judea Pearl.
\newblock Causal inference and the data-fusion problem.
\newblock \emph{Proceedings of the National Academy of Sciences}, 113\penalty0 (27):\penalty0 7345--7352, 2016.

\bibitem[Berck and Villas-Boas(2016)]{berck2016note}
Peter Berck and Sofia~B Villas-Boas.
\newblock A note on the triple difference in economic models.
\newblock \emph{Applied Economics Letters}, 23\penalty0 (4):\penalty0 239--242, 2016.

\bibitem[Callaway and Sant’Anna(2021)]{callaway2021difference}
Brantly Callaway and Pedro~HC Sant’Anna.
\newblock Difference-in-differences with multiple time periods.
\newblock \emph{Journal of econometrics}, 225\penalty0 (2):\penalty0 200--230, 2021.

\bibitem[Callaway et~al.(2024)Callaway, Goodman-Bacon, and Sant'Anna]{callaway2024difference}
Brantly Callaway, Andrew Goodman-Bacon, and Pedro~HC Sant'Anna.
\newblock Difference-in-differences with a continuous treatment.
\newblock Technical report, National Bureau of Economic Research, 2024.

\bibitem[Card(1990)]{card1990impact}
David Card.
\newblock The impact of the mariel boatlift on the miami labor market.
\newblock \emph{Ilr Review}, 43\penalty0 (2):\penalty0 245--257, 1990.

\bibitem[Card and Krueger(1994)]{card1994minimum}
David Card and Alan~B Krueger.
\newblock Minimum wages and employment: A case study of the fast food industry in new jersey and pennsylvania.
\newblock \emph{American Economic Review}, 84:\penalty0 772--793, 1994.

\bibitem[Card and Krueger(2000)]{card2000minimum}
David Card and Alan~B Krueger.
\newblock Minimum wages and employment: a case study of the fast-food industry in new jersey and pennsylvania: reply.
\newblock \emph{American Economic Review}, 90:\penalty0 1397--1420, 2000.

\bibitem[Chernozhukov et~al.(2018)Chernozhukov, Chetverikov, Demirer, Duflo, Hansen, Newey, and Robins]{chernozhukov2018double}
Victor Chernozhukov, Denis Chetverikov, Mert Demirer, Esther Duflo, Christian Hansen, Whitney Newey, and James Robins.
\newblock Double/debiased machine learning for treatment and structural parameters, 2018.

\bibitem[Colnet et~al.(2024)Colnet, Mayer, Chen, Dieng, Li, Varoquaux, Vert, Josse, and Yang]{colnet2024causal}
B{\'e}n{\'e}dicte Colnet, Imke Mayer, Guanhua Chen, Awa Dieng, Ruohong Li, Ga{\"e}l Varoquaux, Jean-Philippe Vert, Julie Josse, and Shu Yang.
\newblock Causal inference methods for combining randomized trials and observational studies: a review.
\newblock \emph{Statistical science}, 39\penalty0 (1):\penalty0 165--191, 2024.

\bibitem[Degtiar and Rose(2023)]{degtiar2023review}
Irina Degtiar and Sherri Rose.
\newblock A review of generalizability and transportability.
\newblock \emph{Annual Review of Statistics and Its Application}, 10\penalty0 (1):\penalty0 501--524, 2023.

\bibitem[Fr{\"o}hlich and Sperlich(2019)]{frohlich2019impact}
Markus Fr{\"o}hlich and Stefan Sperlich.
\newblock \emph{Impact evaluation}.
\newblock Cambridge University Press, 2019.

\bibitem[Gruber(1994)]{gruber1994maternity}
Jonathan Gruber.
\newblock The incidence of mandated maternity benefits.
\newblock \emph{The American Economic Review}, 84\penalty0 (3):\penalty0 622--641, 1994.

\bibitem[Heckman et~al.(1997)Heckman, Ichimura, and Todd]{heckman1997matching}
James~J Heckman, Hidehiko Ichimura, and Petra~E Todd.
\newblock Matching as an econometric evaluation estimator: Evidence from evaluating a job training programme.
\newblock \emph{The review of economic studies}, 64\penalty0 (4):\penalty0 605--654, 1997.

\bibitem[Hong(2013)]{hong2013measuring}
Seung-Hyun Hong.
\newblock Measuring the effect of napster on recorded music sales: difference-in-differences estimates under compositional changes.
\newblock \emph{Journal of Applied Econometrics}, 28\penalty0 (2):\penalty0 297--324, 2013.

\bibitem[Kang and Schafer(2007)]{kang2007demystifying}
Joseph D.~Y. Kang and Joseph~L. Schafer.
\newblock Demystifying double robustness: A comparison of alternative strategies for estimating a population mean from incomplete data.
\newblock \emph{Statistical Science}, 22\penalty0 (4):\penalty0 523--539, 2007.
\newblock \doi{10.1214/07-STS227}.

\bibitem[Lechner et~al.(2011)]{lechner2011estimation}
Michael Lechner et~al.
\newblock The estimation of causal effects by difference-in-difference methods.
\newblock \emph{Foundations and Trends{\textregistered} in Econometrics}, 4\penalty0 (3):\penalty0 165--224, 2011.

\bibitem[Li and Li(2019)]{li2019double}
Fan Li and Fan Li.
\newblock Double-robust estimation in difference-in-differences with an application to traffic safety evaluation.
\newblock \emph{arXiv e-prints}, pages arXiv--1901, 2019.

\bibitem[Nie et~al.(2019)Nie, Lu, and Wager]{nie2019nonparametric}
Xinkun Nie, Chen Lu, and Stefan Wager.
\newblock Nonparametric heterogeneous treatment effect estimation in repeated cross sectional designs.
\newblock \emph{arXiv preprint arXiv:1905.11622}, 2019.

\bibitem[Olden and M{\o}en(2022)]{olden2022triple}
Andreas Olden and Jarle M{\o}en.
\newblock The triple difference estimator.
\newblock \emph{The Econometrics Journal}, 25\penalty0 (3):\penalty0 531--553, 2022.

\bibitem[Sant'Anna and Xu(2025)]{sant2025difference}
Pedro~HC Sant'Anna and Qi~Xu.
\newblock Difference-in-differences with compositional changes.
\newblock \emph{arXiv preprint arXiv:2304.13925}, 2025.

\bibitem[Sant’Anna and Zhao(2020)]{sant2020doubly}
Pedro~HC Sant’Anna and Jun Zhao.
\newblock Doubly robust difference-in-differences estimators.
\newblock \emph{Journal of econometrics}, 219\penalty0 (1):\penalty0 101--122, 2020.

\bibitem[{the U.S. Census Bureau}()]{census_cps_may_2023}
{the U.S. Census Bureau}.
\newblock Current population survey.
\newblock URL \url{https://www.census.gov/programs-surveys/cps.html}.

\bibitem[Torous et~al.(2024)Torous, Gunsilius, and Rigollet]{torous2024optimal}
William Torous, Florian Gunsilius, and Philippe Rigollet.
\newblock An optimal transport approach to estimating causal effects via nonlinear difference-in-differences.
\newblock \emph{Journal of Causal Inference}, 12\penalty0 (1):\penalty0 20230004, 2024.

\bibitem[Van~der Vaart(2000)]{van2000asymptotic}
Aad~W Van~der Vaart.
\newblock \emph{Asymptotic statistics}, volume~3.
\newblock Cambridge university press, 2000.

\bibitem[Wooldridge(2020)]{wooldridge2020introductory}
Jeffrey~M. Wooldridge.
\newblock \emph{Introductory Econometrics: A Modern Approach}.
\newblock Cengage Learning, Boston, MA, 7th edition, 2020.
\newblock ISBN 978-1-337-55886-0.

\bibitem[Yang et~al.(2024)Yang, Lee, K{\"o}hler, and Ghassami]{yang2024causal}
Zou Yang, Seung~Hee Lee, Julia~R K{\"o}hler, and AmirEmad Ghassami.
\newblock Causal data fusion for panel data without pre-intervention period.
\newblock \emph{arXiv preprint arXiv:2410.16391}, 2024.

\bibitem[Zhuang(2024)]{zhuang2024way}
Castiel~Chen Zhuang.
\newblock A way to synthetic triple difference.
\newblock \emph{arXiv preprint arXiv:2409.12353}, 2024.

\bibitem[Zimmert(2018)]{zimmert2018efficient}
Michael Zimmert.
\newblock Efficient difference-in-differences estimation with high-dimensional common trend confounding.
\newblock \emph{arXiv preprint arXiv:1809.01643}, 2018.

\end{thebibliography}
\clearpage
\appendix
\begin{center}
    \centering
    \bfseries
    \Large
    Appendices
\end{center}
The appendix is organized as follows.
In \Cref{apx:if} we derive the efficient influence functions of our parameters of interest, providing the proofs of Theorems \ref{thm:if}, \ref{thm:ifrc}, and \ref{thm:ifncc}.
In \Cref{apx:proofs} we present the proofs of the results appearing in the main text.
\Cref{apx:sim} provides further details of the data generating mechanism used in the simulations of \Cref{sec:sim}.
\section{Efficient Influence Functions}\label{apx:if}
The following techniques are used throughout proofs:
\begin{enumerate}
    \item[T1.]
    \[
        \frac{\partial }{\partial\epsilon}\ex{\epsilon}[h(X)]=\int h(x)\frac{\partial }{\partial\epsilon}p_\epsilon(x)dx = \frac{\partial }{\partial\epsilon}\int h(x)s_\epsilon(x)p_\epsilon(x)dx = \ex{\epsilon}[h(X)s_\epsilon(X)].
    \]
    \item[T2.]
    \[\begin{split}
        \ex{}[h(X)\mid B=b] = \int h(x)p(x\mid b)dx &= \int h(x)\frac{p(b\mid x)}{p(b)}p(x)dx \\&= \frac{1}{p(B=b)}\ex{}[p(B=b\mid X)h(X)].
    \end{split}
    \]
    \item[T3.]
    \[
        \ex{}[\frac{\ind{B=b}}{p(B=b\mid X)}h(X)] = \ex{}\big[\ex{}[\frac{\ind{B=b}}{p(B=b\mid X)}h(X)\mid X]\big]=\ex{}[h(X)].
    \]
    \item[T4.] (follows from the previous one) 
    \[
        \ex{}[h(X,b)] = \ex{}[\frac{\ind{B=b}}{p(B=b\mid X)}h(X, b)] = \ex{}[\frac{\ind{B=b}}{p(B=b\mid X)}h(X, B)]
    \]
    \item[T5.] (follows from the previous one)
    \[\begin{split}
        \ex{}\big[\ex{}[h(X,Y,G,D)\mid X,G=g,D=d]\big] &=  \ex{}\big[\frac{\ind{G=g,D=d}}{p(G=g,D=d\mid X)}\ex{}[h(X,Y,G,D)\mid X,G,D]\big]\\
        &=\ex{}[\frac{\ind{G=g,D=d}}{p(G=g,D=d\mid X)}h(X,Y,G,D)].
    \end{split}\]
    \item[T6.] if $\ex{\epsilon}[h(X)]$ is a constant, then
    \[
        \ex{\epsilon}[h(X)s_\epsilon(X)] \overset{T1}{=} \frac{\partial}{\partial\epsilon}\ex{\epsilon}[h(X)] = 0.
    \]
    \item[T7.] (follows from the previous one) for any $h(\cdot)$
    \[
        \ex{}[h(X)s(Y\mid X)] = \ex{}\big[h(X)\ex{}[s(Y\mid X)\mid X]\big]= \ex{}[h(X)\cdot0]=0.
    \]
    \item[T8.] (follows from the previous one) for any $h(\cdot)$
    \[
        \ex{}[s(X)h(X)] = \ex{}[s(X,Y)h(X)].
    \]
    \item[T9.] for any $h(\cdot)$,
    \[
        \ex{}\big[(h(X)-\ex{}[h(X)\mid Y])s( Y)\big]=  0.
    \]
    \item[T10.] 
    \[\begin{split}
        \ex{}[h(X)s(X\mid Y)] \overset{T7}{=}\ex{}\big[(h(X)-\ex{}[h(X)\mid Y])s(X\mid Y)\big]\overset{T9}{=}  \ex{}\big[(h(X)-\ex{}[h(X)\mid Y])s(X,Y)\big].      
    \end{split}
    \]
\end{enumerate}

\thmif*
\begin{proof}
    The parameter $\psi_\mathrm{pd}$ is defined as
\begin{equation}
    \begin{split}
        \psi_\mathrm{pd} = &\ex{}\Big[
            \ex{}[Y_1-Y_0\mid X, G=1, D=1]
        \\&-
            \ex{}[Y_1-Y_0\mid X, G=0, D=1]
        \\&-
            \ex{}[Y_1-Y_0\mid X, G=1, D=0]
        \\&+
            \ex{}[Y_1-Y_0\mid X, G=0, D=0]
        \:\big\vert\: G=1, D=1
        \Big]
        \\&=\ex{}\Big[
            \frac{p(G=1,D=1\mid X)}{\ex{}[G\cdot D]}\cdot
            \ex{}[Y_1-Y_0\mid X, G=1, D=1]
        \\&-\frac{p(G=1,D=1\mid X)}{\ex{}[G\cdot D]}\cdot
            \ex{}[Y_1-Y_0\mid X, G=0, D=1]
        \\&-\frac{p(G=1,D=1\mid X)}{\ex{}[G\cdot D]}\cdot
            \ex{}[Y_1-Y_0\mid X, G=1, D=0]
        \\&+\frac{p(G=1,D=1\mid X)}{\ex{}[G\cdot D]}\cdot
            \ex{}[Y_1-Y_0\mid X, G=0, D=0]
        \Big]
        \\&
        =\sum_{g,d\in\{0,1\}}(-1)^{g+d}\ex{}\Big[\frac{p(G=1,D=1\mid X)}{\ex{}[G\cdot D]}\cdot
            \ex{}[Y_1-Y_0\mid X, G=g, D=d]\Big]
        .
    \end{split}
\end{equation}

Define $\mu_{g,d,\Delta}(X)\coloneqq\ex{}[Y_1-Y_0\mid X,G=g,D=d]$.
Consider a regular parametric sub-model $p_\epsilon\big(Y_0, Y_1, X, G,D\big)$ which coincides with $p\big(Y_0, Y_1, X, G,D\big)$ when $\epsilon=0$.
We denote the score function corresponding to $p_\epsilon$ by $s_\epsilon$.
We also use $\ex{\epsilon}[
\cdot
]$ to represent expectation with respect to $p_\epsilon$.
Define $\psi_{g,d}(\epsilon)\coloneqq\ex{\epsilon}\big[\frac{p(G=1,D=1\mid X)}{\ex{\epsilon}[G\cdot D]}\cdot
            \ex{\epsilon}[Y_1-Y_0\mid X, G=g, D=d]\big]$, and from above we know $\psi_\mathrm{pd}(\epsilon)=\sum_{g,d\in\{0,1\}}(-1)^{g+d}\psi_{g,d}(\epsilon)$.
Similarly,
\begin{equation}\label{eq:sumifp}
    \frac{\partial}{\partial\epsilon}\psi_\mathrm{pd}(\epsilon)=\sum_{g,d\in\{0,1\}}(-1)^{g+d}\frac{\partial}{\partial\epsilon}\psi_{g,d}(\epsilon),
\end{equation}
where
\begin{equation*}\label{eq:if}
    \begin{split}
        \frac{\partial}{\partial\epsilon}&\psi_{g,d}(\epsilon)
        =
        \frac{\partial}{\partial\epsilon}
        \frac{\int\int\int 
        p_\epsilon(G=1,D=1\mid x)\big(y_1-y_0\big)
        p_\epsilon\big(y_1,y_0\mid x, G=g, D=d\big)
        p_\epsilon(x)
        dy_0dy_1dx}
        {\int
        p_\epsilon(G=1,D=1\mid x)p_\epsilon(x)
        dx}.
    \end{split}
\end{equation*}
We now evaluate $\frac{\partial}{\partial\epsilon}\psi_{g,d}(\epsilon)$ at $\epsilon=0$.
\begingroup
\allowdisplaybreaks
\begin{align*}
    \frac{\partial}{\partial\epsilon}&\psi_{g,d}(\epsilon)
    \big\vert_{\epsilon=0}
    \\&\overset{(T1)}{=}\frac{1}{\ex{}[G\cdot D]}\ex{}\Big[
        p(G=1,D=1\mid X)\ex{}\big[
            \big(Y_1-Y_0\big)s\big(Y_0,Y_1\mid X, G=g, D=d\big)\mid X,\\&\hspace{32em} G=g, D=d
        \big]
    \Big]
    \\&+\frac{1}{\ex{}[G\cdot D]}\ex{}\Big[
        p(G=1,D=1\mid X)s(X)\ex{}\big[
            \big(Y_1-Y_0\big)\mid X, G=g, D=d
        \big]
    \Big]
    \\&+\frac{1}{\ex{}[G\cdot D]}\ex{}\Big[
        p(G=1,D=1\mid X)s(G=1,D=1\mid X)\ex{}\big[
            \big(Y_1-Y_0\big)\mid X, G=g, D=d
        \big]
    \Big]
    \\&-\frac{\psi_{g,d}}{\ex{}[G\cdot D]}\ex{}[
        p(G=1,D=1\mid X)s(G=1, D=1\mid X)
    ]
    \\&-\frac{\psi_{g,d}}{\ex{}[G\cdot D]}\ex{}[
        p(G=1,D=1\mid X)s(X)
    ]
    \\&\overset{(T5,T8)}{=}\frac{1}{\ex{}[G\cdot D]}\ex{}\Big[
        \frac{\mathbbm{1}\{G=g,D=d\}}{p(G=g,D=d\mid X)}p(G=1,D=1\mid X)
            \big(Y_1-Y_0\big)s\big(Y_0,Y_1\mid X, G,D\big)
    \Big]
    \\&+\frac{1}{\ex{}[G\cdot D]}\ex{}\Big[
        s\big(Y_0,Y_1,X,G,D\big)p(G=1,D=1\mid X)\ex{}\big[
            Y_1-Y_0\mid X, G=g, D=d
        \big]
    \Big]
    \\&+\frac{1}{\ex{}[G\cdot D]}\ex{}\Big[
        \ex{}[G\cdot D\mid X]s(G=1,D=1\mid X)\ex{}\big[
            Y_1-Y_0\mid X, G=g, D=d
        \big]
    \Big]
    \\&-\frac{\psi_{g,d}}{\ex{}[G\cdot D]}\ex{}[
        \ex{}[G\cdot D\mid X]s(G=1, D=1\mid X)
    ]
    \\&-\frac{\psi_{g,d}}{\ex{}[G\cdot D]}\ex{}[
        s\big(Y_0,Y_1,X,G,D\big)p(G=1,D=1\mid X)
    ]
    \\&\overset{(T7)}{=}\frac{1}{\ex{}[G\cdot D]}\ex{}\Big[
        \frac{\mathbbm{1}\{G=g,D=d\}}{p(G=g,D=d\mid X)}p(G=1,D=1\mid X)
            \big(Y_1-Y_0-\mu_{g,d,\Delta}(X)\big)\\&\hspace{27em}s\big(Y_0,Y_1\mid X, G,D\big)
    \Big]
    \\&+\frac{1}{\ex{}[G\cdot D]}\ex{}\Big[
        s\big(Y_0,Y_1,X,G,D\big)p(G=1,D=1\mid X)\ex{}\big[
            Y_1-Y_0\mid X, G=g, D=d
        \big]
    \Big]
    \\&+\frac{1}{\ex{}[G\cdot D]}\ex{}\Big[
        G\cdot D\cdot s(G,D\mid X)\ex{}\big[
            Y_1-Y_0\mid X, G=g, D=d
        \big]
    \Big]
    \\&-\frac{\psi_{g,d}}{\ex{}[G\cdot D]}\ex{}[
        G\cdot D\cdot s(G,D\mid X)
    ]
    \\&-\frac{\psi_{g,d}}{\ex{}[G\cdot D]}\ex{}[
        s\big(Y_0,Y_1,X,G,D\big)p(G=1,D=1\mid X)
    ]
    \\&\overset{(T6,T7,T8)}{=}\frac{1}{\ex{}[G\cdot D]}\ex{}\Big[
        \frac{\mathbbm{1}\{G=g,D=d\}}{p(G=g,D=d\mid X)}p(G=1,D=1\mid X)
            \big(Y_1-Y_0-\mu_{g,d,\Delta}(X)\big)\\&\hspace{30em}s\big(Y_0,Y_1, X, G,D\big)
    \Big]
    \\&+\frac{1}{\ex{}[G\cdot D]}\ex{}\Big[
        s\big(Y_0,Y_1,X,G,D\big)p(G=1,D=1\mid X)\ex{}\big[
            Y_1-Y_0\mid X, G=g, D=d
        \big]
    \Big]
    \\&+\frac{1}{\ex{}[G\cdot D]}\ex{}\Big[
        \big(G\cdot D-p(G=1,D=1\mid X)\big)s\big(Y_0,Y_1,X,G,D\big)\\&\hspace{24em}\ex{}\big[
            Y_1-Y_0\mid X, G=g, D=d
        \big]
    \Big]
    \\&-\frac{\psi_{g,d}}{\ex{}[G\cdot D]}\ex{}[
        \big(G\cdot D-p(G=1,D=1\mid X)\big) s\big(Y_0,Y_1,X,G,D\big)
    ]
    \\&-\frac{\psi_{g,d}}{\ex{}[G\cdot D]}\ex{}[
        s\big(Y_0,Y_1,X,G,D\big)p(G=1,D=1\mid X)
    ],
\end{align*}
\endgroup
which is equal to
\[
\begin{split}
    &\ex{}\Big[
        s\big(Y_0,Y_1,X,G,D\big)\cdot 
        \frac{1}{\ex{}[G\cdot D]}
        \cdot
        \big[\\&
        \frac{\mathbbm{1}\{G=g,D=d\}}{p(G=g,D=d\mid X)}p(G=1,D=1\mid X)
            \big(Y_1-Y_0-\mu_{g,d,\Delta}(X)\big)
        \\&+p(G=1,D=1\mid X)\mu_{g,d,\Delta}(X)
    \\&+
        \big(G\cdot D-p(G=1,D=1\mid X)\big)\mu_{g,d,\Delta}(X)
    \\&-\psi_{g,d}\cdot
        \big(G\cdot D-p(G=1,D=1\mid X)\big) 
    \\&-\psi_{g,d}\cdot
        p(G=1,D=1\mid X)
    \big]\Big],
\end{split}
\]
and with simplification,
\[\begin{split}
    \frac{\partial}{\partial\epsilon}\psi_{g,d}(\epsilon)\big\vert_{\epsilon=0}&=\ex{}\Big[
        s\big(Y_0,Y_1,X,G,D\big)\cdot \frac{1}{\ex{}[G\cdot D]}
        \cdot\big[\\&
        \frac{\mathbbm{1}\{G=g,D=d\}}{p(G=g,D=d\mid X)}p(G=1,D=1\mid X)
            \big(Y_1-Y_0-\mu_{g,d,\Delta}(X)\big)\\&
            +G\cdot D\cdot(\mu_{g,d,\Delta}-\psi_{g,d})\big]\Big].
\end{split}\]
Combining the latter with Equation \eqref{eq:sumifp},
\begingroup
\allowdisplaybreaks
\begin{align*}
    \frac{\partial}{\partial\epsilon}\psi_\mathrm{pd}(\epsilon)\big\vert_{\epsilon=0}&=
    \\&\ex{}\Big[
        s\big(Y_0,Y_1,X,G,D\big)\cdot
        \frac{1}{\ex{}[G\cdot D]}
        \cdot
        \big[\\&
        G\cdot D\cdot
            \big(Y_1-Y_0-\mu_{1,1,\Delta}(X)\big)
        \\&-p(G=1,D=1\mid X)\cdot\frac{(1-G)\cdot D}{p(G=0,D=1\mid X)}\cdot
            \big(Y_1-Y_0-\mu_{0,1,\Delta}(X)\big)
        \\&-p(G=1,D=1\mid X)\cdot\frac{G\cdot(1-D)}{p(G=1,D=0\mid X)}\cdot
            \big(Y_1-Y_0-\mu_{1,0,\Delta}(X)\big)
        \\&+p(G=1,D=1\mid X)\cdot\frac{(1-G)\cdot(1-D)}{p(G=0,D=0\mid X)}\cdot
            \big(Y_1-Y_0-\mu_{0,0,\Delta}(X)\big)
        \\&+G\cdot D\cdot\big(\mu_{1,1,\Delta}(X)-\mu_{0,1,\Delta}(X)-
        \mu_{1,0,\Delta}(X)
        +\mu_{0,0,\Delta}(X)
        -\psi_\mathrm{pd}\big)\big]\Big].
\end{align*}
\endgroup

An influence function is therefore:
\[
\begin{split}
        \psi^1_\mathrm{pd}(O)=&\frac{1}{\ex{}[G\cdot D]}
        \cdot
        \Big(
        G\cdot D\cdot
            \big(Y_1-Y_0-\mu_{0,1,\Delta}(X)-
        \mu_{1,0,\Delta}(X)
        +\mu_{0,0,\Delta}(X)
        -\psi_\mathrm{pd}\big)
        \\&-p(G=1,D=1\mid X)\cdot\frac{(1-G)\cdot D}{p(G=0,D=1\mid X)}\cdot
            \big(Y_1-Y_0-\mu_{0,1,\Delta}(X)\big)
        \\&-p(G=1,D=1\mid X)\cdot\frac{G\cdot(1-D)}{p(G=1,D=0\mid X)}\cdot
            \big(Y_1-Y_0-\mu_{1,0,\Delta}(X)\big)
        \\&+p(G=1,D=1\mid X)\cdot\frac{(1-G)\cdot(1-D)}{p(G=0,D=0\mid X)}\cdot
            \big(Y_1-Y_0-\mu_{0,0,\Delta}(X)\big)\Big).
\end{split}
\]
Since the model is
nonparametric, there is a unique influence function, and it is efficient in the model.
\end{proof}

\thmifrc*
\begin{proof}
    The parameter $\psi_\mathrm{rc}$ can be written as
\begin{equation}
    \begin{split}
        &\psi_\mathrm{rc} = \ex{}\Big[
            \ex{}[Y\mid X, G=1, D=1, T=1]
            -\ex{}[Y\mid X, G=1, D=1, T=0]
        \\&-
            \ex{}[Y\mid X, G=0, D=1, T=1]
            +\ex{}[Y\mid X, G=0, D=1, T=0]
        \\&-
            \ex{}[Y\mid X, G=1, D=0, T=1]
            +\ex{}[Y\mid X, G=1, D=0, T=0]
        \\&+
            \ex{}[Y\mid X, G=0, D=0, T=1]
            -\ex{}[Y\mid X, G=0, D=0, T=0]
        \:\big\vert\: G=1, D=1, T=1
        \Big]
        \\&=\ex{}\Big[\sum_{g,d,t\in\{0,1\}}(-1)^{(g+d+t+1)}
            \frac{p(G=1,D=1, T=1\mid X)}{\ex{}[G\cdot D\cdot T]}\cdot
            \ex{}[Y\mid X, G=g, D=d, T=t]
        \Big]
        .
    \end{split}
\end{equation}

Define $\mu_{g,d,t}(X)\coloneqq \ex{}[Y\mid G=g, D=d,T=t]$.
Consider a regular parametric sub-model $p_\epsilon\big(Y,T, X, G, D \big)$ which  coincides with $p\big(Y, T, X, G,D\big)$ when $\epsilon=0$.
We denote the score function corresponding to $p_\epsilon$ by $s_\epsilon$, and expectation with respect to $p_\epsilon$ by $\ex{\epsilon}[\cdot]$.

\[
    \begin{split}
        \frac{\partial}{\partial\epsilon}\psi_\mathrm{rc}(\epsilon)=
        \sum_{g,d,t\in\{0,1\}}(-1)^{(g+d+t+1)}&
        \frac{\partial}{\partial\epsilon}
        \frac{\int\int y
        p_\epsilon(G=1,D=1,T=1\mid x)
        p_\epsilon(y\mid x, g,d,t)
        p_\epsilon(x)
        dydx}
        {\int
        p_\epsilon(G=1,D=1,T=1\mid x)p_\epsilon(x)
        dx}.
    \end{split}
\]
Consider one of the terms on the right-hand side, evaluated at $\epsilon=0$:
\begingroup
\allowdisplaybreaks
\begin{align*}
    \frac{\partial}{\partial\epsilon}&\psi_{g,d,t}(\epsilon)\big\vert_{\epsilon=0}=\frac{\partial}{\partial\epsilon}
    \frac{\int\int y
    p_\epsilon(G=1,D=1,T=1\mid x)
    p_\epsilon(y\mid x, G=g, D=d, T=t)
    p_\epsilon(x)
    dydx}
    {\int
    p_\epsilon(G=1,D=1,T=1\mid x)
    p_\epsilon(x)
    dx}\big\vert_{\epsilon=0}
    \\&=\frac{1}{\ex{}[G\cdot D\cdot T]}\ex{}\Big[
        p(G=1,D=1,T=1\mid X)\ex{}\big[
            Ys(Y\mid X, G=g, D=d, T=t)\mid X, g,d,t
        \big]
    \Big]
    \\&+\frac{1}{\ex{}[G\cdot D\cdot T]}\ex{}\Big[
        p(G=1,D=1,T=1\mid X)s(X)\ex{}\big[
            Y\mid X, G=g, D=d, T=t
        \big]
    \Big]
    \\&+\frac{1}{\ex{}[G\cdot D\cdot T]}\ex{}\Big[
        p(G=1,D=1,T=1\mid X)s(G=1,D=1,T=1\mid X)\\&\hspace{25em}\ex{}\big[
            Y\mid X, G=g, D=d, T=t
        \big]
    \Big]
    \\&-\frac{\psi_{g,d,t}}{\ex{}[G\cdot D\cdot T]}\ex{}[
        p(G=1,D=1,T=1\mid X)s(G=1, D=1, T=1\mid X)
    ]
    \\&-\frac{\psi_{g,d,t}}{\ex{}[G\cdot D\cdot T]}\ex{}[
        p(G=1,D=1,T=1\mid X)s(X)
    ]
    \\&=\frac{1}{\ex{}[G\cdot D\cdot T]}\ex{}\Big[
        \frac{\mathbbm{1}\{G=g,D=d,T=t\}}{p(G=g,D=d,T=t\mid X)}p(G=1,D=1,T=1\mid X)
            Ys(Y\mid X, G,D,T)
    \Big]
    \\&\overset{(T8)}{+}\frac{1}{\ex{}[G\cdot D\cdot T]}\ex{}\Big[
        s(Y,X,G,D,T)p(G=1,D=1,T=1\mid X)\ex{}\big[
            Y\mid X, G=g, D=d, T=t
        \big]
    \Big]
    \\&+\frac{1}{\ex{}[G\cdot D\cdot T]}\ex{}\Big[
        \ex{}[G\cdot D\cdot T\mid X]s(G=1,D=1,T=1\mid X)\ex{}\big[
            Y\mid X, G=g, D=d,T=t
        \big]
    \Big]
    \\&-\frac{\psi_{g,d,t}}{\ex{}[G\cdot D\cdot T]}\ex{}\big[
        \ex{}[G\cdot D\cdot T\mid X]s(G=1, D=1, T=1\mid X)
    \big]
    \\&\overset{(T8)}{-}\frac{\psi_{g,d,t}}{\ex{}[G\cdot D\cdot T]}\ex{}[
        s(Y,X,G,D,T)p(G=1,D=1,T=1\mid X)
    ]
    \\&\overset{(T7)}{=}\frac{1}{\ex{}[G\cdot D\cdot T]}\ex{}\Big[
        \frac{\mathbbm{1}\{G=g,D=d,T=t\}}{p(G=g,D=d,T=t\mid X)}p(G=1,D=1,T=1\mid X)
            \big(Y-\mu_{g,d,t}(X)\big)\\&\hspace{32em}s(Y\mid X, G,D,T)
    \Big]
    \\&+\frac{1}{\ex{}[G\cdot D\cdot T]}\ex{}\Big[
        s(Y,X,G,D,T)p(G=1,D=1,T=1\mid X)\ex{}\big[
            Y\mid X, G=g, D=d,T=t
        \big]
    \Big]
    \\&+\frac{1}{\ex{}[G\cdot D\cdot T]}\ex{}\Big[
        G\cdot D\cdot T\cdot s(G,D,T\mid X)\ex{}\big[
            Y\mid X, G=g, D=d,T=t
        \big]
    \Big]
    \\&-\frac{\psi_{g,d,t}}{\ex{}[G\cdot D\cdot T]}\ex{}[
        G\cdot D\cdot T\cdot s(G,D,T\mid X)
    ]
    \\&-\frac{\psi_{g,d,t}}{\ex{}[G\cdot D\cdot T]}\ex{}[
        s(Y,X,G,D,T)p(G=1,D=1,T=1\mid X)
    ]
    \\&\overset{(T10)}{=}\frac{1}{\ex{}[G\cdot D\cdot T]}\ex{}\Big[
        \frac{\mathbbm{1}\{G=g,D=d,T=t\}}{p(G=g,D=d,T=t\mid X)}p(G=1,D=1,T=1\mid X)
            \\&\hspace{25em}\big(Y-\mu_{g,d,t}(X)\big)s(Y, X, G,D, T)
    \Big]
    \\&+\frac{1}{\ex{}[G\cdot D\cdot T]}\ex{}\Big[
        s(Y,X,G,D,T)p(G=1,D=1,T=1\mid X)\ex{}\big[
            Y\mid X, G=g, D=d, T=t
        \big]
    \Big]
    \\&\overset{(T10,T7)}{+}\frac{1}{\ex{}[G\cdot D\cdot T]}\ex{}\Big[
        \big(G\cdot D\cdot T-p(G=1,D=1,T=1\mid X)\big)s(Y,X,G,D,T)\\&\hspace{25em}\ex{}\big[
            Y\mid X, G=g, D=d, T=t
        \big]
    \Big]
    \\&\overset{(T10,T7)}{-}\frac{\psi_{g,d,t}}{\ex{}[G\cdot D\cdot T]}\ex{}[
        \big(G\cdot D\cdot T-p(G=1,D=1,T=1\mid X)\big) s(Y,X,G,D, T)
    ]
    \\&-\frac{\psi_{g,d,t}}{\ex{}[G\cdot D\cdot T]}\ex{}[
        s(Y,X,G,D,T)p(G=1,D=1,T=1\mid X)
    ],
\end{align*}
\endgroup
which is equal to
\begingroup
\allowdisplaybreaks
\begin{align*}
    \ex{}\Big[
        s&(Y,X,G,D,T)\cdot 
        \frac{1}{\ex{}[G\cdot D\cdot T]}
        \cdot
        \big[\\&
        \frac{\mathbbm{1}\{G=g,D=d,T=t\}}{p(G=g,D=d,T=t\mid X)}p(G=1,D=1,T=1\mid X)
            \big(Y-\mu_{g,d,t}(X)\big)
        \\&+p(G=1,D=1,T=1\mid X)\mu_{g,d,t}(X)
    \\&+
        \big(G\cdot D\cdot T-p(G=1,D=1,T=1\mid X)\big)\mu_{g,d,t}(X)
    \\&-\psi_{g,d,t}
        \big(G\cdot D\cdot T-p(G=1,D=1,T=1\mid X)\big) 
    \\&-\psi_{g,d,t}\cdot
        p(G=1,D=1,T=1\mid X)
    \big]\Big]\\
    =\ex{}&\Big[
        s(Y,X,G,D,T)\cdot 
        \frac{1}{\ex{}[G\cdot D\cdot T]}
        \cdot
        \big[\\&
        \frac{\mathbbm{1}\{G=g,D=d,T=t\}}{p(G=g,D=d,T=t\mid X)}\cdot p(G=1,D=1,T=1\mid X)
            \big(Y-\mu_{g,d,t}(X)\big)
    \\&+
        G\cdot D\cdot T\big(\mu_{g,d,t}(X)-\psi_{g,d,t}\big)
    \big]\Big].
\end{align*}
\endgroup
As a result, 
\begingroup
\allowdisplaybreaks
\begin{align*}
    \frac{\partial}{\partial\epsilon}\psi_\mathrm{rc}(\epsilon)\big\vert_{\epsilon=0}&=
    \\&\ex{}\Big[
        s\big(Y_0,Y_1,X,G,D\big)\cdot
        \frac{1}{\ex{}[G\cdot D\cdot T]}
        \cdot
        \big[\\&
        G\cdot D\cdot T\big(Y-\mu_{1,1,1}(X)\big) 
        \\&- p(G=1,D=1,T=1\mid X)\cdot\frac{G\cdot D\cdot(1-T)}{p(G=1,D=1,T=0\mid X)}\big(Y-\mu_{1,1,0}(X)\big)
        \\&- p(G=1,D=1,T=1\mid X)\cdot\frac{(1-G)\cdot D\cdot T}{p(G=0,D=1,T=1\mid X)}\big(Y-\mu_{0,1,1}(X)\big)
        \\&+ p(G=1,D=1,T=1\mid X)\cdot\frac{(1-G)\cdot D\cdot(1-T)}{p(G=0,D=1,T=0\mid X)}\big(Y-\mu_{0,1,0}(X)\big)
        \\&- p(G=1,D=1,T=1\mid X)\cdot\frac{G\cdot (1-D)\cdot T}{p(G=1,D=0,T=1\mid X)}\big(Y-\mu_{1,0,1}(X)\big)
        \\&+ p(G=1,D=1,T=1\mid X)\cdot\frac{G\cdot (1-D)\cdot (1-T)}{p(G=1,D=0,T=0\mid X)}\big(Y-\mu_{1,0,0}(X)\big)
        \\&+ p(G=1,D=1,T=1\mid X)\cdot\frac{(1-G)\cdot (1-D)\cdot T}{p(G=0,D=0,T=1\mid X)}\big(Y-\mu_{0,0,1}(X)\big)
        \\&- p(G=1,D=1,T=1\mid X)\cdot\frac{(1-G)\cdot (1-D)\cdot (1-T)}{p(G=0,D=0,T=0\mid X)}\big(Y-\mu_{0,0,0}(X)\big)
        \\&+G\cdot D\cdot T\cdot\big(\mu_{1,1,1}(X)-\mu_{1,1,0}(X)-\mu_{0,1,1}(X)+\mu_{0,1,0}(X)
        \\&-
        \mu_{1,0,1}(X)+
        \mu_{1,0,0}(X)
        +\mu_{0,0,1}(X)-\mu_{0,0,0}(X)
        -\psi_\mathrm{rc}\big)\big]\Big],
\end{align*}
\endgroup
which can be organized as
\begingroup
\allowdisplaybreaks
\begin{align*}
    \frac{\partial}{\partial\epsilon}&\psi_\mathrm{rc}(\epsilon)\big\vert_{\epsilon=0}=
    \\&
        \ex{}\Big[
        s\big(Y_0,Y_1,X,G,D\big)
        \cdot\frac{1}{\ex{}[G\cdot D\cdot T]}\cdot
        \big[
        \\&\big(G\cdot D\cdot T- p(G=1,D=1,T=1\mid X)\cdot\frac{G\cdot D\cdot(1-T)}{p(G=1,D=1,T=0\mid X)}\big)\big(Y-\mu_{1,1,0}(X)\big)
        \\&\big(G\cdot D\cdot T - p(G=1,D=1,T=1\mid X)\cdot\frac{(1-G)\cdot D\cdot T}{p(G=0,D=1,T=1\mid X)}\big)\big(Y-\mu_{0,1,1}(X)\big)
        \\&-\big(G\cdot D\cdot T -p(G=1,D=1,T=1\mid X) \cdot\frac{(1-G)\cdot D\cdot(1-T)}{p(G=0,D=1,T=0\mid X)}\big)\big(Y-\mu_{0,1,0}(X)\big)
        \\&\big(G\cdot D\cdot T - p(G=1,D=1,T=1\mid X)\cdot\frac{G\cdot (1-D)\cdot T}{p(G=1,D=0,T=1\mid X)}\big)\big(Y-\mu_{1,0,1}(X)\big)
        \\&-\big(G\cdot D\cdot T - p(G=1,D=1,T=1\mid X)\cdot\frac{G\cdot (1-D)\cdot (1-T)}{p(G=1,D=0,T=0\mid X)}\big)\big(Y-\mu_{1,0,0}(X)\big)
        \\&-\big(G\cdot D\cdot T - p(G=1,D=1,T=1\mid X)\cdot\frac{(1-G)\cdot (1-D)\cdot T}{p(G=0,D=0,T=1\mid X)}\big)\big(Y-\mu_{0,0,1}(X)\big)
        \\&\big(G\cdot D\cdot T - p(G=1,D=1,T=1\mid X)\cdot\frac{(1-G)\cdot (1-D)\cdot (1-T)}{p(G=0,D=0,T=0\mid X)}\big)\big(Y-\mu_{0,0,0}(X)\big)
        \\&
        -G\cdot D\cdot T\cdot\psi_\mathrm{rc}\big]\Big].
\end{align*}
\endgroup

An influence function is therefore:
\[
\begin{split}
    &\psi^1_\mathrm{rc,1}(O) = \frac{1}{\ex{}[G\cdot D\cdot T]}
        \cdot
        \sum_{g,d,t\in\{0,1\}}(-1)^{(g+d+t)}\omega_{g,d,t}(X,G,D,T)\cdot\big(Y-\mu_{g,d,t}(X)\big)
        \\&\hspace{30em}-\frac{G\cdot D\cdot T}{\ex{}[G\cdot D\cdot T]}\cdot\psi_\mathrm{rc},
\end{split}
\]
where
\[
    \omega_{g,d,t}(X,G,D,T) = G\cdot D\cdot T - p(G=1,D=1,T=1\mid X)\cdot\frac{\mathbbm{1}\{G=g,D=d,T=t\}}{p(G=g, D=d, T=t\mid X)}.
\]
Since the model is
nonparametric, there is a unique influence function, and it is efficient in the model.
\end{proof}

\thmifncc*
\begin{proof}
Under \Cref{as:nocompch}, $\psi_\mathrm{rc}$
can be written as
\begin{equation}
    \begin{split}
        \psi_\mathrm{rc}=&\ex{}\Big[\sum_{g,d,t\in\{0,1\}}(-1)^{(g+d+t+1)}
            \frac{p(G=1,D=1, T=1\mid X)}{\ex{}[G\cdot D\cdot T]}\cdot
            \ex{}[Y\mid X, G=g, D=d, T=t]
        \Big]
        \\&=\ex{}\Big[\sum_{g,d,t\in\{0,1\}}(-1)^{(g+d+t+1)}
            \frac{p(G=1,D=1\mid X)}{\ex{}[G\cdot D]}\cdot
            \ex{}[Y\mid X, G=g, D=d, T=t]
        \Big]
        .
    \end{split}
\end{equation}

Define $\mu_{g,d,t}(X)\coloneqq \ex{}[Y\mid X, G=g, D=d,T=t]$.

Consider a regular parametric sub-model $p_\epsilon\big(Y,X, G,D,\big)$ which  coincides with $p\big(Y,X, G,D,T\big)$ when $\epsilon=0$.
We denote the score function corresponding to $p_\epsilon$ by $s_\epsilon$.
Under \Cref{as:nocompch}, $p_\epsilon(Y,X,G,D,T)$ can be decomposed as 
\begingroup
\allowdisplaybreaks
\begin{align*}
p_\epsilon(Y,X,G,D,T) =
\sum_{g,d,t\in\{0,1\}}&
\ind{G=g,D=d,T=t}\cdot
p_\epsilon(Y\mid X, G=g, D=d, T=t)\cdot\\&
p_\epsilon(G,D\mid X)\cdot
p_\epsilon(X)\cdot
p_\epsilon(T),
\end{align*}
\endgroup
and the score function is then given by
\begingroup
\allowdisplaybreaks
\begin{align*}
s_\epsilon(Y,X,G,D,T) =
\sum_{g,d,t\in\{0,1\}}&
\ind{G=g,D=d,T=t}\cdot
s_\epsilon(Y\mid X, G=g, D=d, T=t)+\\&
s_\epsilon(G,D\mid X)+
s_\epsilon(X)+
s_\epsilon(T).
\end{align*}
\endgroup
Therefore, defining for each $(g,d,t)\in\{0,1\}^3$,
\begingroup
\allowdisplaybreaks
\begin{align*}
    \mathcal{T}_{Y|X,g,d,t}=\big\{\ind{G=g,D=d,T=t}&\cdot \beta_{g,d,t}(Y,X):\beta_{g,d,t}\in L^2(p), \\&\ex{}\left[\beta_{g,d,t}(X,Y)\mid X,G=g,D=d,T=t\right]=0 \text{ a.s.}\big\},
\end{align*}
\endgroup
as well as
\begingroup
\allowdisplaybreaks
\begin{align*}
    \mathcal{T}_{G,D\mid X}=&\big\{\nu(X,G,D):\nu\in L^2(p), \ex{}\left[\nu(X,G,D)\mid X\right]=0 \text{ a.s.}\big\},
    \\
    \mathcal{T}_{X}=&\big\{\gamma(X):\gamma\in L^2(p), \ex{}\left[\gamma(X)\right]=0\big\},\\
    \mathcal{T}_{T}=&\big\{u(T):u\in L^2(p), \ex{}\left[u(T)\right]=0\big\},
\end{align*}
\endgroup
the semiparamteric tangent space is the $L^2(p)$
closure of 
\[
    \mathcal{T}_0=\bigcup_{g,d,t}\mathcal{T}_{Y|X,g,d,t}\cup \mathcal{T}_{G,D\mid X} \cup\mathcal{T}_{X}\cup\mathcal{T}_{T}.
\]
Next, we derive the pathwise derivative of $\psi_\mathrm{rc}(p_\epsilon)$, defined as follows.
\[
    \begin{split}
        \frac{\partial}{\partial\epsilon}\psi_\mathrm{rc}(\epsilon)=
        \sum_{g,d,t\in\{0,1\}}(-1)^{(g+d+t+1)}&
        \frac{\partial}{\partial\epsilon}
        \frac{\int\int y
        p_\epsilon(G=1,D=1\mid x)
        p_\epsilon(y\mid x, g,d,t)
        p_\epsilon(x)
        dydx}
        {\int
        p_\epsilon(G=1,D=1\mid x)p_\epsilon(x)
        dx}.
    \end{split}
\]
Consider one of the terms on the right-hand side:
\[
\begin{split}
    \frac{\partial}{\partial\epsilon}&\psi_{g,d,t}(\epsilon)=\frac{\partial}{\partial\epsilon}
    \frac{\int\int y
    p_\epsilon(G=1,D=1\mid x)
    p_\epsilon(y\mid x, G=g, D=d, T=t)
    p_\epsilon(x)
    dydx}
    {\int
    p_\epsilon(G=1,D=1\mid x)
    p_\epsilon(x)
    dx}.
\end{split}
\]
Following the same steps as the previous case,
\[
\begin{split}
    \frac{\partial}{\partial\epsilon}\psi_{g,d,t}(\epsilon)\big\vert_{\epsilon=0}
    =\ex{}\Big[
        s&(Y,X,G,D,T)\cdot 
        \frac{1}{\ex{}[G\cdot D]}
        \cdot
        \big[\\&
        \frac{\mathbbm{1}\{G=g,D=d,T=t\}}{p(G=g,D=d\mid X)\cdot p(T=t)}p(G=1,D=1\mid X)
            \big(Y-\mu_{g,d,t}(X)\big)
    \\&+
        G\cdot D\big(\mu_{g,d,t}(X)-\psi_{g,d,t}\big)
    \big]\Big].
\end{split}
\]

As a result, 
\begingroup
\allowdisplaybreaks
\begin{align*}
    \frac{\partial}{\partial\epsilon}\psi_\mathrm{rc}(\epsilon)\big\vert_{\epsilon=0}&=
    \\&\ex{}\Big[
        s\big(Y_0,Y_1,X,G,D\big)\cdot
        \frac{1}{\ex{}[G\cdot D]}
        \cdot
        \big[\\&
        G\cdot D\cdot \frac{T}{p(T=1)}\big(Y-\mu_{1,1,1}(X)\big) 
        \\&- G\cdot D\cdot\frac{1-T}{1-p(T=1)}\big(Y-\mu_{1,1,0}(X)\big)
        \\&- p(G=1,D=1\mid X)\cdot \frac{T}{p(T=1)}\cdot\frac{(1-G)\cdot D}{p(G=0,D=1\mid X)}\big(Y-\mu_{0,1,1}(X)\big)
        \\&+ p(G=1,D=1\mid X)\cdot\frac{1-T}{1-p(T=1)}\cdot\frac{(1-G)\cdot D}{p(G=0,D=1\mid X)}\big(Y-\mu_{0,1,0}(X)\big)
        \\&- p(G=1,D=1\mid X)\cdot \frac{T}{p(T=1)}\cdot\frac{G\cdot (1-D)}{p(G=1,D=0\mid X)}\big(Y-\mu_{1,0,1}(X)\big)
        \\&+ p(G=1,D=1\mid X)\cdot \frac{1-T}{1-p(T=1)}\cdot\frac{G\cdot (1-D)}{p(G=1,D=0\mid X)}\big(Y-\mu_{1,0,0}(X)\big)
        \\&+ p(G=1,D=1\mid X)\cdot \frac{T}{p(T=1)}\cdot\frac{(1-G)\cdot (1-D)}{p(G=0,D=0\mid X)}\big(Y-\mu_{0,0,1}(X)\big)
        \\&- p(G=1,D=1\mid X)\cdot \frac{1-T}{1-p(T=1)}\cdot\frac{(1-G)\cdot (1-D)}{p(G=0,D=0\mid X)}\big(Y-\mu_{0,0,0}(X)\big)
        \\&+G\cdot D\cdot\big(\mu_{1,1,1}(X)-\mu_{1,1,0}(X)-\mu_{0,1,1}(X)+\mu_{0,1,0}(X)
        \\&-
        \mu_{1,0,1}(X)+
        \mu_{1,0,0}(X)
        +\mu_{0,0,1}(X)-\mu_{0,0,0}(X)
        -\psi_\mathrm{rc}\big)\big]\Big],
\end{align*}
\endgroup
which can be reorganized as
\begingroup
\allowdisplaybreaks
\begin{align*}
    \frac{\partial}{\partial\epsilon}&\psi_\mathrm{rc}(\epsilon)\big\vert_{\epsilon=0}=
    \\&
        \ex{}\Big[
        s\big(Y_0,Y_1,X,G,D\big)
        \cdot\frac{1}{\ex{}[G\cdot D]}\cdot
        \big[
        \\&-\left(G\cdot D- \frac{G\cdot D\cdot T}{p(T=1)}\cdot\frac{p(G=1,D=1\mid X)}{p(G=1,D=1\mid X)}\right)\big(Y-\mu_{1,1,1}(X)\big)
        \\&+\left(G\cdot D- \frac{G\cdot D\cdot (1-T)}{p(T=0)}\cdot\frac{p(G=1,D=1\mid X)}{p(G=1,D=1\mid X)}\right)\big(Y-\mu_{1,1,0}(X)\big)
        \\&+\left(G\cdot D- \frac{(1-G)\cdot D\cdot T}{p(T=1)}\cdot\frac{p(G=1,D=1\mid X)}{p(G=0,D=1\mid X)}\right)\big(Y-\mu_{0,1,1}(X)\big)
        \\&-\left(G\cdot D- \frac{(1-G)\cdot D\cdot (1-T)}{p(T=0)}\cdot\frac{p(G=1,D=1\mid X)}{p(G=0,D=1\mid X)}\right)\big(Y-\mu_{0,1,0}(X)\big)
        \\&+\left(G\cdot D- \frac{G\cdot (1-D)\cdot T}{p(T=1)}\cdot\frac{p(G=1,D=1\mid X)}{p(G=1,D=0\mid X)}\right)\big(Y-\mu_{1,0,1}(X)\big)
        \\&-\left(G\cdot D- \frac{G\cdot (1-D)\cdot (1-T)}{p(T=0)}\cdot\frac{p(G=1,D=1\mid X)}{p(G=1,D=0\mid X)}\right)\big(Y-\mu_{1,0,0}(X)\big)
        \\&-\left(G\cdot D- \frac{(1-G)\cdot (1-D)\cdot T}{p(T=1)}\cdot\frac{p(G=1,D=1\mid X)}{p(G=0,D=0\mid X)}\right)\big(Y-\mu_{0,0,1}(X)\big)
        \\&+\left(G\cdot D- \frac{(1-G)\cdot (1-D)\cdot (1-T)}{p(T=0)}\cdot\frac{p(G=1,D=1\mid X)}{p(G=0,D=0\mid X)}\right)\big(Y-\mu_{0,0,0}(X)\big)
        \\&
        -G\cdot D\cdot\psi_\mathrm{rc}\big]\Big].
\end{align*}
\endgroup

An influence function is therefore:
\[
\begin{split}
&\psi^1_\mathrm{rc,2}(O) = 
    \frac{1}{\ex{}[G\cdot D]}
        \cdot
        \sum_{g,d,t\in\{0,1\}}(-1)^{(g+d+t)}\alpha_{g,d,t}(X,G,D,T)\cdot\big(Y-\mu_{g,d,t}(X)\big)
        -\frac{G\cdot D}{\ex{}[G\cdot D]}\cdot\psi_\mathrm{rc},
\end{split}
\]
where
\[
    \alpha_{g,d,t}(X,G,D,T) = G\cdot D - \frac{\mathbbm{1}\{T=t\}}{p(T=t)}\cdot \mathbbm{1}\{G=g,D=d\}\cdot\frac{p(G=1,D=1\mid X)}{p(G=g, D=d\mid X)}.
\]

To complete the proof, we show that $\psi^1_\mathrm{rc,2}$ lies in the tangent space.
Define 
\begingroup
\allowdisplaybreaks
\begin{align*}\beta_{g,d,t}(Y,X)&=\frac{(-1)^{(g+d+t+1)}}{\ex{}[G\cdot D]}\cdot \frac{p(G=1,D=1\mid X)}{p(G=g,D=d\mid X)p(T=t)}\cdot \left(Y-\mu_{g,d,t}(X)\right),
\\
\nu(X,G,D) &= \frac{G\cdot D}{\ex{}[G\cdot D]}\cdot\left(\sum_{g,d,t\in\{0,1\}}(-1)^{(g+d+t+1)}\mu_{g,d,t}(X)
-\psi_\mathrm{rc}\right).
\end{align*}
\endgroup
Clearly, 
\[
\psi^1_\mathrm{rc,2}(O) = \sum_{g,d,t\in\{0,1\}}\beta_{g,d,t}(Y,X)+\nu(X,G,D).
\]
It suffices to show that $\ex{}\left[\beta_{g,d,t}(X,Y)\mid X,G=g,D=d,T=t\right]=0$, and, $\left[\nu(X,G,D)\mid X\right]=0$.
The former follows by definition of $\mu_{g,d,t}$:
\begingroup
\allowdisplaybreaks
\begin{align*}
    \ex{}&\left[\beta_{g,d,t}(X,Y)\mid X,G=g,D=d,T=t\right]
    \\&=\frac{(-1)^{(g+d+t+1)}}{\ex{}[G\cdot D]}\cdot \frac{p(G=1,D=1\mid X)}{p(G=g,D=d\mid X)p(T=t)}\cdot \left(\ex{}[Y\mid X,G=g,D=d,T=t]-\mu_{g,d,t}(X)\right)
    \\&=\frac{(-1)^{(g+d+t+1)}}{\ex{}[G\cdot D]}\cdot \frac{p(G=1,D=1\mid X)}{p(G=g,D=d\mid X)p(T=t)}\cdot \left(\mu_{g,d,t}(X)-\mu_{g,d,t}(X)\right)=0,
\end{align*}
\endgroup
and the latter is
\begingroup
\allowdisplaybreaks
\begin{align*}
\ex{}
&\left[
    \frac{G\cdot D}{\ex{}[G\cdot D]}\cdot\left(\sum_{g,d,t\in\{0,1\}}(-1)^{(g+d+t+1)}\mu_{g,d,t}(X)
    -\psi_\mathrm{rc}\right)
\right]
\\
&=
\ex{}
\left[
    \sum_{g,d,t\in\{0,1\}}(-1)^{(g+d+t+1)}\mu_{g,d,t}(X)
    \:\Big|\: G=1,D=1
\right]
-\psi_\mathrm{rc}
\\
&\overset{(a)}{=}
\ex{}
\left[
    \sum_{g,d,t\in\{0,1\}}(-1)^{(g+d+t+1)}\mu_{g,d,t}(X)
    \:\Big|\: G=1,D=1,T=1
\right]
-\psi_\mathrm{rc}
\\
&\overset{(b)}{=}
\ex{}[\mu_{1,1,1}(X)\mid G=1,D=1,T=1]-
\ex{}
\big[
    \mu_{1,1,0}(X)
    +\mu_{0,1,\Delta}(X)
    \\&\hspace{4em}+\mu_{1,0,\Delta}(X)
    -\mu_{0,0,\Delta}(X)
    \mid G=1,D=1,T=1
\big]
-\psi_\mathrm{rc}
\\
&\overset{(c)}{=}
\ex{}[Y\mid G=1,D=1,T=1]-
\ex{}
\big[
    \mu_{1,1,0}(X)
    +\mu_{0,1,\Delta}(X)
    \\&\hspace{4em}+\mu_{1,0,\Delta}(X)
    -\mu_{0,0,\Delta}(X)
    \mid G=1,D=1,T=1
\big]
-\psi_\mathrm{rc}
\\
&\overset{(d)}{=}
\psi_\mathrm{rc}
-\psi_\mathrm{rc}=0,
\end{align*}
\endgroup
where $(a)$ holds because of \Cref{as:nocompch}, for
$(b)$ we recall that $\mu_{g,d,\Delta}(X)=\mu_{g,d,1}(X)-\mu_{g,d,0}(X)$, $(c)$ is due to the law of iterated expectations,
and $(d)$ follows from the definition of $\psi_\mathrm{rc}$ in \Cref{eq:idorrc}.
\end{proof}

\pagebreak
\section{Omitted Proofs}\label{apx:proofs}
\prporp*
\begin{proof}
Part $(i)$.
    \[
    \begin{split}
        &
        \ex{}[Y_1^{(1)} - \Y{1}\mid X, G=1, D=1]
        \\&=
        \ex{}[Y_1^{(1)}\mid X, G=1, D=1] - \big(
            \ex{}[\Y{0}\mid X, G=1, D=1]
            + \ex{}[\Y{1}-\Y{0}\mid X, G=1, D=1]
        \big)
        \\&\overset{(a)}{=}\ex{}[Y_1^{(1)}\mid X, G=1, D=1] - \Big(
            \ex{}[\Y{0}\mid X, G=1, D=1]
        \\&+
            \ex{}[\Y{1}-\Y{0}\mid X, G=0, D=1]
            +
            \ex{}[\Y{1}-\Y{0}\mid X, G=1, D=0]
        \\&-
            \ex{}[\Y{1}-\Y{0}\mid X, G=0, D=0]
        \Big)
        \\&\overset{(b)}{=}\ex{}[Y_1^{(1)}\mid X, G=1, D=1] - \Big(
            \ex{}[Y_0^{(1)}\mid X, G=1, D=1]
        \\&+
            \ex{}[\Y{1}-\Y{0}\mid X, G=0, D=1]
            +
            \ex{}[\Y{1}-\Y{0}\mid X, G=1, D=0]
        \\&-
            \ex{}[\Y{1}-\Y{0}\mid X, G=0, D=0]
        \Big)
        \\&\overset{(c)}{=}\ex{}[Y_1\mid X, G=1, D=1] - \big(
            \ex{}[Y_0\mid X, G=1, D=1]
        \\&+
            \ex{}[Y_1-Y_0\mid X, G=0, D=1]
            +
            \ex{}[Y_1-Y_0\mid X, G=1, D=0]
        \\&-
            \ex{}[Y_1-Y_0\mid X, G=0, D=0]
        \big)\\&=
        \mu_{1,1,\Delta}(X)-
        \mu_{0,1,\Delta}(X)-
        \mu_{1,0,\Delta}(X)+
        \mu_{0,0,\Delta}(X)
        ,
    \end{split}
\]
where (a), (b) and (c) follow from Assumptions \ref{as:id3}, \ref{as:noanticip} and \ref{as:consistency}, respectively.

Part $(ii)$.

\begin{equation*}\label{eq:id}
    \begin{split}
        &\ex{}[Y_1^{(1)} - \Y{1}\mid G=1, D=1]
        \\&=
        \ex{}\big[
            \ex{}[Y_1^{(1)} - \Y{1}\mid X, G=1, D=1]\mid G=1, D=1
        \big]
        \\&\overset{(a)}{=}
        \ex{}[\mu_{1,1,\Delta}(X)-
        \mu_{0,1,\Delta}(X)-
        \mu_{1,0,\Delta}(X)+
        \mu_{0,0,\Delta}(X)\mid G=1, D=1]
        \\&=\ex{}[Y_1-Y_0\mid G=1, D=1]
        \\&-
        \ex{}[
        \mu_{0,1,\Delta}(X)+
        \mu_{1,0,\Delta}(X)-
        \mu_{0,0,\Delta}(X)\mid G=1, D=1]
        ,
    \end{split}
\end{equation*}
where (a) follows from \Cref{as:pos3} and part $(i)$.
\end{proof}
\prpipwp*

\begin{proof}
Part $(i)$.
    \begin{equation*}
        \begin{split}
            \ex{}\big[
                \rho_0(X,G,D)&\cdot\big(Y_1-Y_0\big)\mid X
            \big]
            \\&=
            \ex{}\big[
                \rho_0(X,G,D)\cdot\big(Y_1-Y_0\big)\mid X, G=1, D=1
            \big] P(G=1,D=1\mid X)
            \\
            &+
            \ex{}\big[
                \rho_0(X,G,D)\cdot\big(Y_1-Y_0\big)\mid X, G=0, D=1
            \big] P(G=0,D=1\mid X)
            \\&+
            \ex{}\big[
                \rho_0(X,G,D)\cdot\big(Y_1-Y_0\big)\mid X, G=1, D=0
            \big] P(G=1,D=0\mid X)
            \\
            &+
            \ex{}\big[
                \rho_0(X,G,D)\cdot\big(Y_1-Y_0\big)\mid X, G=0, D=0
            \big] P(G=0,D=0\mid X)
            \\
            &=
            \ex{}\big[
                Y_1-Y_0\mid X, G=1, D=1
            \big]
            \\
            &-
            \ex{}\big[
                Y_1-Y_0\mid X, G=0, D=1
            \big]
            \\&-
            \ex{}\big[
                Y_1-Y_0\mid X, G=1, D=0
            \big]
            \\
            &+
            \ex{}\big[
                Y_1-Y_0\mid X, G=0, D=0
            \big]
            \\
            &\overset{(a)}{=}
            \ex{}\big[
                Y_1^{(1)}-Y_0^{(1)}\mid X, G=1, D=1
            \big]
            -
            \ex{}\big[
                \Y{1}-\Y{0}\mid X, G=0, D=1
            \big]
            \\&-
            \Big(
            \ex{}\big[
                \Y{1}-\Y{0}\mid X, G=1, D=0
            \big]
            -
            \ex{}\big[
                \Y{1}-\Y{0}\mid X, G=0, D=0
            \big]
            \Big)
            \\
            &\overset{(b)}{=}
            \ex{}\big[
                Y_1^{(1)}-\Y{0}\mid X, G=1, D=1
            \big]
            -
            \ex{}\big[
                \Y{1}-\Y{0}\mid X, G=0, D=1
            \big]
            \\&-
            \Big(
            \ex{}\big[
                \Y{1}-\Y{0}\mid X, G=1, D=0
            \big]
            -
            \ex{}\big[
                \Y{1}-\Y{0}\mid X, G=0, D=0
            \big]
            \Big)
            \\
            &\overset{(b)}{=}
            \ex{}\big[
                Y_1^{(1)}-\Y{0}\mid X, G=1, D=1
            \big]
            -
            \ex{}\big[
                \Y{1}-\Y{0}\mid X, G=0, D=1
            \big]
            \\&-
            \Big(
            \ex{}\big[
                \Y{1}-\Y{0}\mid X, G=1, D=1
            \big]
            -
            \ex{}\big[
                \Y{1}-\Y{0}\mid X, G=0, D=1
            \big]
            \Big)
            \\&=
            \ex{}\big[
                Y_1^{(1)} - \Y{1}\mid X, G=1, D=1
            \big].
        \end{split}
    \end{equation*}
    where (a) is due to \Cref{as:consistency}, and (b) and (c) follow from \Cref{as:noanticip} and \Cref{as:id3}, respectively.

Part $(ii)$.
\begingroup
\allowdisplaybreaks
\begin{align*}
    \ex{}\big[&
        Y_1^{(1)} - \Y{1}\mid G=1, D=1
    \big]
    \\&=
    \ex{}\Big[\ex{}\big[
        Y_1^{(1)} - \Y{1}\mid X, G=1, D=1
    \big]
    \mid G=1, D=1\Big]\\
    &\overset{(a)}{=}\ex{}
    \Big[\ex{}\big[
        \rho_0(X,G,D)\cdot\big(Y_1-Y_0\big) \mid X
    \big]
    \mid G=1, D=1\Big]\\
    &\overset{}{=}\int\ex{}\big[
        \rho_0(X,G,D)\cdot\big(Y_1-Y_0\big) \mid X
    \big]dP(X\mid G=1,D=1)
    \\
    &\overset{}{=}\int\ex{}\big[
        \rho_0(X,G,D)\cdot\big(Y_1-Y_0\big) \mid X
    \big]\frac{p(G=1,D=1\mid X)}{p(G=1,D=1)}dP(X)
    \\
    &=\frac{1}{p(G=1,D=1)}\cdot\ex{}\big[p(G=1,D=1\mid X)\cdot\ex{}[\rho_0(X,G,D)\cdot(Y_1-Y_0)\mid X]\big]
    \\&=\frac{1}{\ex{}[G\cdot D]}\cdot\ex{}[p(G=1,D=1\mid X)\cdot\rho_0(X,G,D)\cdot(Y_1-Y_0)],
\end{align*}
\endgroup
where (a) is due to \Cref{as:pos3} and part $(i)$.
\end{proof}
\prporrc*
\begin{proof}
Part $(i)$.
    \[
    \begin{split}
        &
        \ex{}[Y^{(1)} - \Y{}\mid X, G=1, D=1, T=1]
        \\&=
        \ex{}[Y^{(1)}\mid X, G=1, D=1, T=1] - 
            \ex{}[\Y{}\mid X, G=1, D=1, T=0]
            \\&- \big(\ex{}[\Y{}\mid X, G=1, D=1, T=1]-\ex{}[\Y{}\mid X, G=1, D=1, T=0]
        \big)
        \\&\overset{(a)}{=}\ex{}[Y^{(1)}\mid X, G=1, D=1, T=1] - 
            \ex{}[\Y{}\mid X, G=1, D=1, T=0]
        \\&-\Big(
            \big(\ex{}\big[
            \Y{}\mid X,G=0, D=1, T=1
        \big] - \ex{}\big[\Y{}\mid X,G=0, D=1, T=0
        \big]\big)
        \\&+\big(\ex{}\big[
            \Y{}\mid X,G=1, D=0, T=1
        \big] - \ex{}\big[\Y{}\mid X,G=1, D=0, T=0
        \big]\big)\\
        &-\big(\ex{}\big[
            \Y{}\mid X,G=0, D=0, T=1
        \big] - \ex{}\big[\Y{}\mid X,G=0, D=0, T=0
        \big]\big)
        \Big)
        \\&\overset{(b)}{=}\ex{}[Y\mid X, G=1, D=1, T=1] - 
            \ex{}[Y\mid X, G=1, D=1, T=0]
        \\&-
            \big(\ex{}\big[
            Y\mid X,G=0, D=1, T=1
        \big] - \ex{}\big[Y\mid X,G=0, D=1, T=0
        \big]\big)
        \\&-\big(\ex{}\big[
            Y\mid X,G=1, D=0, T=1
        \big] - \ex{}\big[Y\mid X,G=1, D=0, T=0
        \big]\big)\\
        &+\big(\ex{}\big[
            Y\mid X,G=0, D=0, T=1
        \big] - \ex{}\big[Y\mid X,G=0, D=0, T=0
        \big]\big)
        \\&=
        \mu_{1,1,\Delta}(X)-
        \mu_{0,1,\Delta}(X)-
        \mu_{1,0,\Delta}(X)+
        \mu_{0,0,\Delta}(X)
        ,
    \end{split}
\]
where (a) follows from \Cref{as:idrc} and (b) is due to Assumptions \ref{as:noanticiprc} and \ref{as:consistencyrc}.

Part $(ii)$.
    \[\begin{split}
        \tau_\mathrm{rc} &= \ex{}[Y^{(1)}-\Y{}\mid G=1,D=1,T=1]
        \\&=
        \ex{}\big[
        \ex{}[Y^{(1)}-\Y{}\mid X, G=1,D=1,T=1]
        \mid G=1, D=1, T=1    
        \big],
    \end{split}
    \]
and the rest follows from part $(i)$ and \Cref{as:posrc}.
\end{proof}
\prpipwrc*
\begin{proof}
Part $(i)$.
    \begin{equation*}
        \begin{split}
            \ex{}\big[
                &\phi_0(X,G,D,T)\cdot Y\mid X
            \big]
            \\&=
            \sum_{g,d,t\in\{0,1\}}\ex{}\big[
                \phi_0(X,G,D,T)\cdot Y\mid X, G=g, D=d, T=t
            \big] p(G=g,D=d,T=t\mid X)
            \\&=
            \sum_{g,d,t\in\{0,1\}}\ex{}\big[
                -\sum_{g',d',t'\in\{0,1\}}(1-g'-g)(1-d'-d)(1-t'-t)\cdot Y\mid X, G=g, D=d, T=t
            \big]
            \\&\overset{(a)}{=}
            \sum_{g,d,t\in\{0,1\}}\ex{}\big[
                -(1-2g)(1-2d)(1-2t)\cdot Y\mid X, G=g, D=d, T=t
            \big]
            \\&=
            \sum_{g,d,t\in\{0,1\}}(-1)^{g+d+t+1}\ex{}\big[ Y\mid X, G=g, D=d, T=t
            \big]
            \\
            &=
            \big(\ex{}\big[
                Y\mid X, G=1, D=1, T=1
            \big]
            -
            \ex{}\big[
                Y\mid X, G=1, D=1, T=0
            \big]\big)
            \\
            &-
            \big(\ex{}\big[
                Y\mid X, G=0, D=1, T=1
            \big]
            -
            \ex{}\big[
                Y\mid X, G=0, D=1, T=0
            \big]\big)
            \\
            &-
            \big(\ex{}\big[
                Y\mid X, G=1, D=0, T=1
            \big]
            -
            \ex{}\big[
                Y\mid X, G=1, D=0, T=0
            \big]\big)
            \\&+
            \big(\ex{}\big[
                Y\mid X, G=0, D=0, T=1
            \big]
            -
            \ex{}\big[
                Y\mid X, G=0, D=0, T=0
            \big]\big)
            \\
            &\overset{(b)}{=}
            \big(\ex{}\big[
                Y^{(1)}\mid X, G=1, D=1, T=1
            \big]
            -
            \ex{}\big[
                \Y{}\mid X, G=1, D=1, T=0
            \big]\big)
            \\
            &-
            \big(\ex{}\big[
                \Y{}\mid X, G=0, D=1, T=1
            \big]
            -
            \ex{}\big[
                \Y{}\mid X, G=0, D=1, T=0
            \big]\big)
            \\
            &-
            \big(\ex{}\big[
                \Y{}\mid X, G=1, D=0, T=1
            \big]
            -
            \ex{}\big[
                \Y{}\mid X, G=1, D=0, T=0
            \big]\big)
            \\&+
            \big(\ex{}\big[
                \Y{}\mid X, G=0, D=0, T=1
            \big]
            -
            \ex{}\big[
                \Y{}\mid X, G=0, D=0, T=0
            \big]\big)
            \\
            &\overset{(c)}{=}
            \ex{}\big[
                Y^{(1)}\mid X, G=1, D=1
            ,T=1\big]
            -
            \ex{}\big[
                \Y{}\mid X, G=1, D=1,T=1
            \big],
        \end{split}
    \end{equation*}
    where $(a)$ holds because $(1-g-g')$ is non-zero if and only if $g=g'$, $(b)$ is due to Assumptions \ref{as:noanticiprc} and \ref{as:consistencyrc}, and $(c)$ is due to  \Cref{as:idrc}.

Part $(ii)$.
    \begingroup
    \allowdisplaybreaks
    \begin{align*}
            \ex{}\big[&
                Y^{(1)} - \Y{}\mid G=1, D=1, T=1
            \big]
            \\&
            \ex{}\Big[\ex{}\big[
                Y^{(1)} - \Y{}\mid X, G=1, D=1, T=1
            \big]
            \mid G=1, D=1, T=1\Big]\\
            &\overset{(a)}{=}\ex{}
            \Big[\ex{}\big[
                \phi_0(X,G,D,T)\cdot Y \mid X
            \big]
            \mid G=1, D=1, T=1\Big]\\
            &\overset{}{=}\int\ex{}\big[
                \phi_0(X,G,D,T)\cdot Y \mid X
            \big]dP(X\mid G=1,D=1,T=1)
            \\
            &\overset{}{=}\int\ex{}\big[
                \rho_0(X,G,D,T)\cdot Y \mid X
            \big]\frac{p(G=1,D=1,T=1\mid X)}{p(G=1,D=1,T=1)}dP(X)
            \\
            &=\frac{1}{p(G=1,D=1,T=1)}\cdot\ex{}\big[p(G=1,D=1,T=1\mid X)\cdot\ex{}[\phi_0(X,G,D,T)\cdot Y\mid X]\big]
            \\&=\frac{1}{\ex{}[G\cdot D\cdot T]}\cdot\ex{}[p(G=1,D=1,T=1\mid X)\cdot\phi_0(X,G,D,T)\cdot Y],
\end{align*}
\endgroup
where (a) is due to \Cref{as:posrc} and part $(i)$.
\end{proof}

\begin{lemma}\label{lem:ratioestimator}
    Let $V$ be a Bernoulli random variable with parameter $q$, where $q$ is a positive constant.
    Suppose $\{V_i\}_{i=1}^n$ are i.i.d. samples of $V$.
    Define 
    \[
        e=\frac{1}{\ex{}[V]}, \quad \hat{e}_n=\frac{n}{\max\{1,\sum_{i=1}^n V_i\}}.
    \]
    Then $\hat{e}_n-e =  O_p(n^{-1/2})$, $\ex{}[\hat{e}_n-e]=o(n^{-1/2})$, $\ex{}[|\hat{e}_n-e|]=O(n^{-1/2})$, and $\norm{\hat{e}_n-e}=O(n^{-1/2})$.
\end{lemma}
\begin{proof}
    Define 
    \[
        \Tilde{e}_n=\frac{n}{\sum_{i=1}^n V_i}.
    \]
    $V_i$ has variance $q(1-q)$.
    Applying the central limit theorem, $\sqrt{n}(\frac{1}{n}\sum_iV_i-\ex{}[V])\overset{D}{\to}\mathcal{N}\big(0,q(1-q)\big)$.
    By delta method, 
    \begin{align*}
        \sqrt{n}(\Tilde{e}_n-e)\overset{D}{\to}\mathcal{N}\big(0, \frac{1-q}{q^3}\big).
    \end{align*}
    Moreover, for any $\epsilon>0$,
    \[
        p\left(\vert\sqrt{n}(\hat{e}_n-e)-\sqrt{n}(\Tilde{e}_n-e)\vert>\epsilon\right)\leq p\left(\sum_{i=1}^nV_i=0\right)=(1-q)^n\to0,
    \]
    and using Slutsky's theorem,
    \begin{align*}
        \sqrt{n}(\hat{e}_n-e)\overset{D}{\to}\mathcal{N}\big(0, \frac{1-q}{q^3}\big),
    \end{align*}
    which implies that $\hat{e}_n-e=O_p(n^{-1/2})$, 
    $\ex{}[\hat{e}_n-e]=o(n^{-1/2})$,
    and $\mathrm{var}(\hat{e}_n-e)=O(n^{-1})$.
    Finally, 
    \begin{align*}
        \norm{\hat{e}_n-e}=\sqrt{\ex{}[(\hat{e}_n-e)^2]}
        \leq
        \sqrt{\mathrm{var}(\hat{e}_n-e)}=O(n^{-1/2}).
    \end{align*}
\end{proof}

\thmdrp*

\begin{proof}
Define $e=\frac{1}{\ex{}[G\cdot D]}$ and $\pi_{r,g,d}(X)=\frac{\pi_{1,1}(X)}{\pi_{g,d}(X)}$.
Suppose that the data is split into $L$ folds of size $m$, so that $n=mL$. 
We drop the subscript of $\hat{e}_\mathrm{pd}^\ell$ in the remainder of this proof.
We first show that $\ex{}\left[
\eta_\mathrm{pd}(O;\{f^\ell_{\mu,g,d}\}_{g,d}, \{f^\ell_{\pi,g,d}\}_{g,d})
\right]=\psi_\mathrm{pd}/e$.
\begingroup
\allowdisplaybreaks
\begin{align*}
    &\hspace{-.5em}\ex{}\left[\eta_\mathrm{pd}(O;\{f^\ell_{\mu,g,d}\}_{g,d}, \{f^\ell_{\pi,g,d}\}_{g,d})
    -
    \eta_\mathrm{pd}(O;\{\mu_{g,d,\Delta}\}_{g,d}, \{\pi_{r,g,d}\}_{g,d})
    \right] 
    =
    \\
    &\sum_{g,d\in\{0,1\}}(-1)^{(g+d+1)}
    \ex{}\Big[
        -G\cdot D\cdot(f^\ell_{\mu,g,d}-\mu_{g,d,\Delta})
        -\ind{G=g,D=d}\cdot(Y_1-Y_0)\cdot(f^\ell_{\pi,g,d}-\pi_{r,g,d})
        \\&+\ind{G=g,D=d}\cdot(f^\ell_{\mu,g,d}f^\ell_{\pi,g,d}-\mu_{g,d,\Delta}\pi_{r,g,d})
    \Big]
    \\&=
    -\sum_{g,d\in\{0,1\}}(-1)^{(g+d+1)}
    \ex{}\left[
        \big(G\cdot D - \ind{G=g,D=d}\cdot\pi_{r,g,d}\big)\cdot(f^\ell_{\mu,g,d}-\mu_{g,d,\Delta})
        \right]
        \\&
        -\sum_{g,d\in\{0,1\}}(-1)^{(g+d+1)}\ex{}\left[
        \ind{G=g,D=d}\cdot(Y_1-Y_0-\mu_{g,d,\Delta})\cdot(f^\ell_{\pi,g,d}-\pi_{r,g,d})
        \right]
        \\&
        +\sum_{g,d\in\{0,1\}}(-1)^{(g+d+1)}\ex{}\left[
        \ind{G=g,D=d}\cdot(f^\ell_{\mu,g,d}-\mu_{g,d,\Delta})\cdot(f^\ell_{\pi,g,d}-\pi_{r,g,d})
    \right]
    \\&\overset{(a)}{=}
    -\sum_{g,d\in\{0,1\}}(-1)^{(g+d+1)}
    \ex{}\left[
        \big(\pi_{1,1} - \pi_{g,d}\cdot\pi_{r,g,d}\big)\cdot(f^\ell_{\mu,g,d}-\mu_{g,d,\Delta})
        \right]
        \\&
        -\sum_{g,d\in\{0,1\}}(-1)^{(g+d+1)}\ex{}\left[
        \ind{G=g,D=d}\cdot(Y_1-Y_0-\mu_{g,d,\Delta})\cdot(f^\ell_{\pi,g,d}-\pi_{r,g,d})
        \right]
        \\&
        +\sum_{g,d\in\{0,1\}}(-1)^{(g+d+1)}\ex{}\left[
        \ind{G=g,D=d}\cdot(f^\ell_{\mu,g,d}-\mu_{g,d,\Delta})\cdot(f^\ell_{\pi,g,d}-\pi_{r,g,d})
    \right]
    \\&\overset{(b)}{=}
        -\sum_{g,d\in\{0,1\}}\ex{}\left[
        \ind{G=g,D=d}\cdot(Y_1-Y_0-\mu_{g,d,\Delta})\cdot(f^\ell_{\pi,g,d}-\pi_{r,g,d})
        \right]
    \\&\overset{(c)}{=}
        -\sum_{g,d\in\{0,1\}}(-1)^{(g+d+1)}\ex{}\left[
        (Y_1-Y_0-\mu_{g,d,\Delta})\cdot(f^\ell_{\pi,g,d}-\pi_{r,g,d})
        \big| G=g,D=d
        \right]\cdot p(G=g,D=d)
        \\&\overset{(d)}{=}
        -\sum_{g,d\in\{0,1\}}(-1)^{(g+d+1)}\ex{}\left[
        \ex{}\left[
        Y_1-Y_0-\mu_{g,d,\Delta}|X,G=g,D=d\right]\cdot(f^\ell_{\pi,g,d}-\pi_{r,g,d})
        \big| G=g,D=d
        \right]\\&\cdot p(G=g,D=d)
        \\&\overset{(e)}{=}
        -\sum_{g,d\in\{0,1\}}(-1)^{(g+d+1)}\ex{}\left[
        (\mu_{g,d,\Delta}-\mu_{g,d,\Delta})\cdot(f^\ell_{\pi,g,d}-\pi_{r,g,d})
        \big| G=g,D=d
        \right]\cdot p(G=g,D=d)
        \\&=0,
\end{align*}
where $(a)$ and $(d)$ are due to the law of iterated expectations, $(b)$ follows since $\pi_{1,1}-\pi_{g,d}\pi_{r,g,d}=\pi_{1,1}-\pi_{1,1}=0$, 
and the third term is $0$ because either $f^\ell_{\mu,g,d}-\mu_{g,d,\Delta}=0$ or $f^\ell_{\pi,g,d}-\pi_{r,g,d}=0$,
$(c)$ is an application of the total probability law, and $(e)$ is by definition of $\mu_{g,d,\Delta}$.
We get
\begin{equation}\label{eq:proofdrexpec}
    \ex{}\left[
\eta_\mathrm{pd}(O;\{f^\ell_{\mu,g,d}\}_{g,d}, \{f^\ell_{\pi,g,d}\}_{g,d})
\right]
=
\ex{}\left[
    \eta_\mathrm{pd}(O;\{\mu_{g,d,\Delta}\}_{g,d}, \{\pi_{r,g,d}\}_{g,d})
\right]
=\frac{\psi_\mathrm{pd}}{e}.
\end{equation}
\endgroup
We also show that $\eta_\mathrm{pd}(O;\{f^\ell_{\mu,g,d}\}_{g,d}, \{f^\ell_{\pi,g,d}\}_{g,d})$ is integrable:
\begingroup
\allowdisplaybreaks
\begin{align*}
    \ex{}\big[\big\vert\eta_\mathrm{pd}(O;&\{f^\ell_{\mu,g,d}\}_{g,d}, \{f^\ell_{\pi,g,d}\}_{g,d})\big\vert\big] 
    \\&\overset{(a)}{\leq}
    \sum_{g,d\in\{0,1\}}\Big[
    \pp\big[\vert G\cdot D\cdot(Y_1-Y_0)\vert\big]
    +\pp\big[\vert(Y_1-Y_0)\cdot f^\ell_{\pi,g,d}(X)\vert\big]
    \\&\hspace{3em}+\pp\big[\vert G\cdot D\cdot f^\ell_{\mu,g,d}(X)\vert\big]
    +\pp\big[\vert
    f^\ell_{\mu,g,d}(X)\cdot f^\ell_{\pi,g,d}(X)
    \vert\big]
            \Big]
    \\&\overset{(b)}{\leq}
    \sum_{g,d\in\{0,1\}}\Big[
    (\norm{Y_1}+\norm{Y_0})
    \cdot(1+ \norm{f^\ell_{\pi,g,d}})
    \\&\hspace{7em}+\norm{f^\ell_{\mu,g,d}}
    +\norm{
    f^\ell_{\mu,g,d}}\cdot \norm{f^\ell_{\pi,g,d}}
            \Big]
            \overset{(c)}{<}\infty,
\end{align*}
\endgroup
where $(a)$, $(b)$, and $(c)$ are due to triangle inequality, Cauchy-Schwarz, and the bounded moment assumptions of the proposition.
Moreover,
\begingroup
\allowdisplaybreaks
\begin{align*}
    &\hspace{-8em}\pp\left[\big\vert\eta_\mathrm{pd}(O;\{\hat{\mu}^\ell_{g,d,\Delta}\}_{g,d}, \{\hat{\pi}^\ell_{r,g,d}\}_{g,d})
    -
    \eta_\mathrm{pd}(O;\{f^\ell_{\mu,g,d}\}_{g,d}, \{f^\ell_{\pi,g,d}\}_{g,d})\big\vert\right]
    \\&\overset{(a)}{\leq}
    \sum_{g,d\in\{0,1\}}\Big[
    \pp\big[\vert G\cdot D\cdot\big(\hat{\mu}^\ell_{g,d,\Delta}(X) - f^\ell_{\mu,g,d}(X)\big)\vert\big]
    \\&+\pp\big[\vert(Y_1-Y_0)\cdot\big(
    \hat{\pi}^\ell_{r,g,d}(X)
    - f^\ell_{\pi,g,d}(X)\big)\vert\big]
    \\&+
    \pp\big[\vert
    f^\ell_{\mu,g,d}(X)\big(
    \hat{\pi}^\ell_{r,g,d}(X)
    - f^\ell_{\pi,g,d}(X)\big)
    \vert\big]
    \\&+
    \pp\big[\vert
    f^\ell_{\pi,g,d}(X)\big(
    \hat{\mu}^\ell_{g,d,\Delta}(X)
    - f^\ell_{\mu,g,d}(X)\big)
    \vert\big]
    \\&+
    \pp\big[\vert
    \big(
    \hat{\pi}^\ell_{r,g,d}(X)
    - f^\ell_{\pi,g,d}(X)\big)
    \big(
    \hat{\mu}^\ell_{g,d,\Delta}(X)
    - f^\ell_{\mu,g,d}(X)\big)\vert\big]
            \Big]
    \\
    &\overset{(b)}{\leq}\sum_{g,d\in\{0,1\}}\Big[
    \norm{\big(\hat{\mu}^\ell_{g,d,\Delta} - f^\ell_{\mu,g,d}\big)}
    \\&+\norm{(Y_1-Y_0)}\cdot\norm{\big(
    \hat{\pi}^\ell_{r,g,d}
    - f^\ell_{\pi,g,d}\big)}
    \\&+
    \norm{
    f^\ell_{\mu,g,d}}\cdot\norm{\big(
    \hat{\pi}^\ell_{r,g,d}
    - f^\ell_{\pi,g,d}\big)
    }
    \\&+
    \norm{
    f^\ell_{\pi,g,d}}\cdot\norm{\big(
    \hat{\mu}^\ell_{g,d,\Delta}
    - f^\ell_{\mu,g,d}\big)}
    \\&+
    \norm{
    \big(
    \hat{\pi}^\ell_{r,g,d}
    - f^\ell_{\pi,g,d}\big)}\cdot\norm{
    \big(
    \hat{\mu}^\ell_{g,d,\Delta}
    - f^\ell_{\mu,g,d}\big)}
            \Big]
    \\&\overset{(c)}{=}o_p(1),
\end{align*}
\endgroup
where $(a)$ is due to triangle inequality, $(b)$ is due to Cauchy-Schwartz, and $(c)$ holds because of the assumptions of the proposition.
Therefore, $\eta_\mathrm{pd}(O;\{\hat{\mu}^\ell_{g,d,\Delta}\}_{g,d}, \{\hat{\pi}^\ell_{r,g,d}\}_{g,d})$ converges in $L_1$ to $\eta_\mathrm{pd}(O;\{f^\ell_{\mu,g,d}\}_{g,d}, \{f^\ell_{\pi,g,d}\}_{g,d})$.
Now consider 
\begingroup
\allowdisplaybreaks
\begin{align*}
    &\hspace{-2em}\ee_m^\ell\left[\eta_\mathrm{pd}(O;\{\hat{\mu}^\ell_{g,d,\Delta}\}_{g,d}, \{\hat{\pi}^\ell_{r,g,d}\}_{g,d})\right] - \frac{\psi_\mathrm{pd}}{e}
    \\
    &=
    \ee_m^\ell\left[\eta_\mathrm{pd}(O;\{\hat{\mu}^\ell_{g,d,\Delta}\}_{g,d}, \{\hat{\pi}^\ell_{r,g,d}\}_{g,d})
    -
    \eta_\mathrm{pd}(O;\{f^\ell_{\mu,g,d}\}_{g,d}, \{f^\ell_{\pi,g,d}\}_{g,d})
    \right] 
    \\&+
    \ee_m^\ell\left[
    \eta_\mathrm{pd}(O;\{f^\ell_{\mu,g,d}\}_{g,d}, \{f^\ell_{\pi,g,d}\}_{g,d})
    \right] -
    \pp\left[
    \eta_\mathrm{pd}(O;\{f^\ell_{\mu,g,d}\}_{g,d}, \{f^\ell_{\pi,g,d}\}_{g,d})
    \right]
    \\&+
    \pp\left[
    \eta_\mathrm{pd}(O;\{f^\ell_{\mu,g,d}\}_{g,d}, \{f^\ell_{\pi,g,d}\}_{g,d})
    \right]
    -
    \frac{\psi_\mathrm{pd}}{e},
\end{align*}
\endgroup
where the first term is $o_p(1)$ because of $L^1$ convergence and an application of Chebyshev's inequality,
the second term is $o_p(1)$ since $\eta_\mathrm{pd}(O;\{f^\ell_{\mu,g,d}\}_{g,d}, \{f^\ell_{\pi,g,d}\}_{g,d})$ is integrable and the weak law of large numbers applies, and the third term is $0$ by \Cref{eq:proofdrexpec}.
Therefore,
\[\ee_m^\ell\left[\eta_\mathrm{pd}(O;\{\hat{\mu}^\ell_{g,d,\Delta}\}_{g,d}, \{\hat{\pi}^\ell_{r,g,d}\}_{g,d})\right]\overset{p}{\to}\frac{\psi_\mathrm{pd}}{e},\]
and by \Cref{lem:ratioestimator}, $\hat{e}^\ell\overset{p}{\to}e$.
Using Theorem 2.7 of \cite{van2000asymptotic}, we get
\[\hat{e}^\ell\cdot\ee_m^\ell\left[\eta_\mathrm{pd}(O;\{\hat{\mu}^\ell_{g,d,\Delta}\}_{g,d}, \{\hat{\pi}^\ell_{r,g,d}\}_{g,d})\right]\overset{p}{\to}\psi_\mathrm{pd},\]
and since the latter holds for every fold $\ell\in\{1,\dots,L\}$, $\hat{\psi}_\mathrm{pd}^\mathrm{dr}$ is consistent for $\psi_\mathrm{pd}$.
\end{proof}

\thmcanp*
\begin{proof}
Define $e=\frac{1}{\ex{}[G\cdot D]}$ and $\pi_{r,g,d}(X)=\frac{\pi_{1,1}(X)}{\pi_{g,d}(X)}$.
Suppose that the data is split into $L$ folds of size $m$, so that $n=mL$. 
We drop the subscript of $\hat{e}_\mathrm{pd}^\ell$ in the remainder of this proof.

\begingroup
\allowdisplaybreaks
    \begin{align}
        \sqrt{n}&(\hat{\psi}_\mathrm{pd}^\mathrm{dr} - \psi_\mathrm{pd}) = \notag
        \\&
        \sqrt{n}
            \big(\frac{1}{L}
            \sum_{\ell=1}^{L}
                \hat{e}^\ell\cdot\ee_m^\ell\big[\eta_\mathrm{pd}(O;\{\hat{\mu}^\ell_{g,d,\Delta}\}_{g,d}, \{\hat{\pi}^\ell_{r,g,d}\}_{g,d})\big]
            -\psi_\mathrm{pd}\big)\notag
        \\&
            =\frac{\sqrt{mL}}{L}\cdot
            \sum_{\ell=1}^{L}
                \frac{\hat{e}^\ell}{e}\cdot\big(e\cdot\ee_m^\ell\big[\eta_\mathrm{pd}(O;\{\hat{\mu}^\ell_{g,d,\Delta}\}_{g,d}, \{\hat{\pi}^\ell_{r,g,d}\}_{g,d})\big]
            -e\cdot\psi_\mathrm{pd}\cdot\ee_m^\ell[G\cdot D]\big)\notag
        \\&
        =\frac{1}{\sqrt{L}}
            \sum_{\ell=1}^{L}
                (1+\frac{\hat{e}^\ell-e}{e})\cdot\sqrt{m}\big(e\cdot\ee_m^\ell\big[\eta_\mathrm{pd}(O;\{\hat{\mu}^\ell_{g,d,\Delta}\}_{g,d}, \{\hat{\pi}^\ell_{r,g,d}\}_{g,d})\big]
            -e\cdot\psi_\mathrm{pd}\cdot\ee_m^\ell[G\cdot D]\big)\notag
        \\&
        \begin{aligned}
            &=
            \frac{1}{\sqrt{L}}\sum_{\ell=1}^{L}
            (1+\frac{\hat{e}^\ell-e}{e})\cdot
            \sqrt{m}(\ee_m^\ell-\pp)
                \Big[e\cdot\eta_\mathrm{pd}(O;\{\hat{\mu}^\ell_{g,d,\Delta}\}_{g,d}, \{\hat{\pi}^\ell_{r,g,d}\}_{g,d})
                \\&\hspace{20em}
                -e\cdot\eta_\mathrm{pd}(O;\{\mu_{g,d,\Delta}\}_{g,d}, \{\pi_{r,g,d}\}_{g,d})\Big]
        \end{aligned}\label{eq:t1}\tag{$T_{1}$}
        \\&
        \begin{aligned}
            &+
            \frac{1}{\sqrt{L}}\sum_{\ell=1}^{L}
            (1+\frac{\hat{e}^\ell-e}{e})\cdot
            \sqrt{m}(\ee_m^\ell-\pp)
                \Big[e\cdot\eta_\mathrm{pd}(O;\{\mu_{g,d,\Delta}\}_{g,d}, \{\pi_{r,g,d}\}_{g,d})-e\cdot G\cdot D\cdot \psi_\mathrm{pd}\Big]
        \end{aligned}\label{eq:t2}\tag{$T_{2}$}
        \\&
        \begin{aligned}
            &+
            \frac{1}{\sqrt{L}}\sum_{\ell=1}^{L}
            (1+\frac{\hat{e}^\ell-e}{e})\cdot
            \sqrt{m}\big(\pp
                \big[e\cdot\eta_\mathrm{pd}(O;\{\hat{\mu}^\ell_{g,d,\Delta}\}_{g,d}, \{\hat{\pi}^\ell_{r,g,d}\}_{g,d})
                \big]-\psi_\mathrm{pd}\big).
        \end{aligned}\label{eq:t3}\tag{$T_{3}$}
    \end{align}
\endgroup

We analyze each of the terms above separately.
In the remainder of the proof, we use $O_n^{-\ell}$ to represent the data in all but the $\ell$-th fold.

\textbf{Term \ref{eq:t1}:}

Define $A_m^\ell = \sqrt{m}(\ee_m^\ell-\pp)
                \big[e\cdot\eta_\mathrm{pd}(O;\{\hat{\mu}^\ell_{g,d,\Delta}\}_{g,d}\allowbreak, \allowbreak\{\hat{\pi}^\ell_{r,g,d}\}_{g,d})\allowbreak-e\cdot\eta_\mathrm{pd}(O;\{\mu_{g,d,\Delta}\}_{g,d}, \{\pi_{r,g,d}\}_{g,d})\big]$.
Note that $\ex{}[A_m^\ell\mid O^{-\ell}]=0$. 
Next, we bound the conditional variance as
\begingroup\allowdisplaybreaks
\begin{align*}
    \mathrm{var}&(A_m^\ell\mid O^{-\ell})
    \\&= 
    m\cdot e^2\cdot\mathrm{var}(\ee_m^\ell
        [
            \eta_\mathrm{pd}(O;\{\hat{\mu}^\ell_{g,d,\Delta}\}_{g,d}, \{\hat{\pi}^\ell_{r,g,d}\}_{g,d})-\eta_\mathrm{pd}(O;\{\mu_{g,d,\Delta}\}_{g,d}, \{\pi_{r,g,d}\}_{g,d})
        ]
        \mid O^{-\ell}
    )
    \\&=
    e^2\cdot\mathrm{var}(
            \eta_\mathrm{pd}(O;\{\hat{\mu}^\ell_{g,d,\Delta}\}_{g,d}, \{\hat{\pi}^\ell_{r,g,d}\}_{g,d})-\eta_\mathrm{pd}(O;\{\mu_{g,d,\Delta}\}_{g,d}, \{\pi_{r,g,d}\}_{g,d})
        \mid O^{-\ell}
    )
    \\&\leq e^2\cdot\pp\big[\big(
            \eta_\mathrm{pd}(O;e,\{\hat{\mu}^\ell_{g,d,\Delta}\}_{g,d}, \{\hat{\pi}^\ell_{r,g,d}\}_{g,d})-\eta_\mathrm{pd}(O;e,\{\mu_{g,d,\Delta}\}_{g,d}, \{\pi_{r,g,d}\}_{g,d})\big)^2
        \mid O^{-\ell}
    \big]
    \\&=e^2\cdot
    \pp\Big[\Big(\sum_{g,d\in\{0,1\}}(-1)^{(g+d+1)}\hat{w}^\ell_{g,d}(X,G,D)\cdot
            \big(
                Y_1-Y_0-\hat{\mu}^\ell_{g,d,\Delta}(X)
            \big)
        \\&\hspace{5em}-
            \sum_{g,d\in\{0,1\}}(-1)^{(g+d+1)}w_{g,d}(X,G,D)\cdot
            \big(
                Y_1-Y_0-\mu_{g,d,\Delta}(X)
            \big)
            \Big)^2
        \:\Big\vert\: O^{-\ell}
    \Big]
    \\&\overset{(a)}{\leq}4e^2\sum_{g,d\in\{0,1\}}
    \pp\Big[\Big(\hat{w}^\ell_{g,d}(X,G,D)\cdot
            \big(
                Y_1-Y_0-\hat{\mu}^\ell_{g,d,\Delta}(X)
            \big)
        \\&\hspace{10em}-
            w_{g,d}(X,G,D)\cdot
            \big(
                Y_1-Y_0-\mu_{g,d,\Delta}(X)
            \big)
            \Big)^2
        \:\Big\vert\: O^{-\ell}
    \Big]
    \\&\overset{(b)}{\leq}12e^2\sum_{g,d\in\{0,1\}}
    \pp\Big[\Big(G\cdot D\cdot\big(\hat{\mu}^\ell_{g,d}(X)-\mu_{g,d}(X)\big)\Big)^2
        \:\Big\vert\: O^{-\ell}
    \Big]
        \\&\hspace{5em}+12e^2\sum_{g,d\in\{0,1\}}
        \pp\Big[\Big(\big(\hat{\pi}^\ell_{r,g,d}(X)-\pi_{r,g,d}(X)\big)\cdot\big(Y_1-Y_0-\mu_{g,d}(X)\big)\Big)^2
            \:\Big\vert\: O^{-\ell}
        \Big]
        \\&\hspace{5em}+12e^2\sum_{g,d\in\{0,1\}}
        \pp\Big[\Big(\hat{\pi}^\ell_{r,g,d}(X)\cdot\big(\hat{\mu}^\ell_{g,d}(X)-\mu_{g,d}(X)\big)\Big)^2
            \:\Big\vert\: O^{-\ell}
        \Big]
    \\&\overset{(c)}{\leq}
        72e^2\sum_{g,d\in\{0,1\}}\Big(\bignorm{
            \hat{\mu}^\ell_{g,d} - \mu_{g,d}
        }^2
        +
        \bignorm{
            (\hat{\pi}^\ell_{r,g,d})^2-\pi_{r,g,d}^2
        }\cdot
        \big(\norm{
            Y_1^2
        }
        +
        \norm{Y_0^2}
        +
        \norm{\mu_{g,d}^2}\big)
        \\&\hspace{5em}+
        \bignorm{
            (\hat{\mu}^\ell_{g,d})^2 - \mu_{g,d}^2
        }\cdot
        \big(\bignorm{
            (\hat{\pi}^\ell_{r,g,d})^2-\pi_{r,g,d}^2
        }
        +\bignorm{\pi_{r,g,d}^2}\big)
        \Big),
\end{align*}
\endgroup
where $(a)$ and $(b)$ are due to Cauchy-Schwartz, and $(c)$ is an application of Hölder, Cauchy-Schwartz, and Minkowski inequalities.
Under \Cref{as:infp} and strict positivity, the conditional variance above is $o_p(1)$.
By the law of iterated expectations and Chebyshev's inequality, 
we conclude $A_m^\ell\overset{p}{\to}0$.
$(\hat{e}^\ell-e)/e$ is $o_p(1)$ by \Cref{lem:ratioestimator}, and using Slutsky's theorem and Theorem 2.7 of \citep{van2000asymptotic}, $(1+(\hat{e}^\ell-e)/e)\cdot A_m^\ell\overset{p}{\to}0$ for every $\ell$.
Finally, since the number of folds ($L$) is finite, \ref{eq:t1} converges to $0$ in probability.

\textbf{Term \ref{eq:t2}:}
\begin{align*}
    T_2&=\frac{1}{\sqrt{L}}\sum_{\ell=1}^{L}
            (1+\frac{\hat{e}^\ell-e}{e})\cdot\sqrt{m}(\ee_m^\ell-\pp)
                \big[e\cdot\eta_\mathrm{pd}(O;\{\mu_{g,d,\Delta}\}_{g,d}, \{\pi_{g,d}\}_{g,d})-e\cdot G\cdot D\cdot \psi_\mathrm{pd}\big]
    \\&=
    \frac{1}{\sqrt{L}}\sum_{\ell=1}^{L}
            (1+\frac{\hat{e}^\ell-e}{e})\cdot\sqrt{m}(\ee_m^\ell-\pp)
                \big[\psi^1_\mathrm{pd}(O)\big]
    \\&\overset{(a)}{=}\sqrt{n}(\ee_n-\pp)\big[\psi^1_\mathrm{pd}(O)\big] + o_p(1),
\end{align*}
where $(a)$ is because (i) $\sqrt{m}(\ee_m^\ell-\pp)\big[\psi^1_\mathrm{pd}(O)\big]$ is $O_p(1)$ due to the central limit theorem, (ii) $(\hat{e}^\ell-e)/e$ is $O_p(m^{-1/2})$ due to \Cref{lem:ratioestimator}, so their product converges to $0$ in probability due to Slutsky's theorem and Theorem 2.7 of \citep{van2000asymptotic}, and (iii) the number of folds ($L$) is finite.
By the central limit theorem, the term $\sqrt{n}(\ee_n-\pp)\big[\psi^1_\mathrm{pd}(O)\big]$ converges in distribution to $\mathcal{N}\Big(0, \mathrm{var}\big(\psi^1_\mathrm{pd}(O)\big)\Big)$, as long as the variance of $\psi^1_\mathrm{pd}(\cdot)$ is bounded.
This variance can be bounded as follows.
\begin{align*}
    \mathrm{var}&\big(e\cdot\eta_\mathrm{pd}(O; \allowbreak\{\mu_{g,d,\Delta}\}_{g,d}, \{\pi_{g,d}\}_{g,d})-e\cdot G\cdot D\cdot \psi_\mathrm{pd}\big)
    \\&\leq
    e^2\cdot\ex{}\big[\big(\eta_\mathrm{pd}(O; \allowbreak\{\mu_{g,d,\Delta}\}_{g,d}, \{\pi_{g,d}\}_{g,d})- G\cdot D\cdot \psi_\mathrm{pd}\big)^2\big]
    \\&\leq
    2e^2\cdot\ex{}\big[\big(\eta_\mathrm{pd}(O; \allowbreak\{\mu_{g,d,\Delta}\}_{g,d}, \{\pi_{g,d}\}_{g,d})\big)^2 + \big(G\cdot D\cdot \psi_\mathrm{pd}\big)^2\big]
    \\&\leq
    8e^2\sum_{g,d\in\{0,1\}}\ex{}\big[w_{g,d}^2\cdot\big(Y_1-Y_0-\mu_{g,d}(X)\big)^2\big] + 2e^2\cdot\psi_\mathrm{pd}^2
    \\&\leq
    24e^2\sum_{g,d\in\{0,1\}}\bignorm{w_{g,d}^2}\cdot\Big(\bignorm{Y_1^2}+\bignorm{Y_0^2}+\bignorm{\mu_{g,d}^2}
    \Big)+ 2e^2\cdot\psi_\mathrm{pd}^2.
\end{align*}
which is bounded under \Cref{as:infp} and strict positivity.

\textbf{Term \ref{eq:t3}:}

Define $B_m^\ell\coloneqq \sqrt{m}\big(\pp
                \big[e\cdot\eta_\mathrm{pd}(O;\{\hat{\mu}^\ell_{g,d,\Delta}\}_{g,d}, \{\hat{\pi}^\ell_{r,g,d}\}_{g,d})
                \big]-\psi_\mathrm{pd}\big)$.
\begingroup
\allowdisplaybreaks
\begin{align*}
    B_m^\ell&=
    \sqrt{m}\cdot e\cdot
            \big(\pp
                \big[\eta_\mathrm{pd}(O;\{\hat{\mu}^\ell_{g,d,\Delta}\}_{g,d}, \{\hat{\pi}^\ell_{r,g,d}\}_{g,d})\big]-\pp
                \big[\eta_\mathrm{pd}(O;\{{\mu}_{g,d,\Delta}\}_{g,d}, \{{\pi}_{r,g,d}\}_{g,d})\big]\big)
    \\&=\sqrt{m}\cdot
        e\cdot \pp\big[
            \sum_{g,d\in\{0,1\}}(-1)^{(g+d+1)}\hat{w}^\ell_{g,d}(X,G,D)
            \cdot
            \big(
                Y_1-Y_0-\hat{\mu}^\ell_{g,d,\Delta}(X)
            \big)
        \\&\hspace{9em}-
        \sum_{g,d\in\{0,1\}}(-1)^{(g+d+1)}w_{g,d}(X,G,D)
            \cdot
            \big(
                Y_1-Y_0-\mu_{g,d,\Delta}(X)
            \big)
        \big]
        \\&\leq
        \sqrt{m}\cdot{e}
        \sum_{g,d\in\{0,1\}}\big\vert
        \pp\big[
        \hat{w}^\ell_{g,d}(X,G,D)\cdot \big(Y_1-Y_0-\hat{\mu}^\ell_{g,d,\Delta}(X))
        \\&\hspace{12em}-
        w_{g,d}(X,G,D)\cdot \big(Y_1-Y_0-\mu_{g,d,\Delta}(X))
        \big]
        \Big\vert
        \\&=
        \sqrt{m}\cdot{e}
        \sum_{g,d\in\{0,1\}}\Big\vert
        \pp\Big[\pp[
        \hat{w}^\ell_{g,d}(X,G,D)\cdot \big(Y_1-Y_0-\hat{\mu}^\ell_{g,d,\Delta}(X))
        \\&\hspace{12em}-
        w_{g,d}(X,G,D)\cdot \big(Y_1-Y_0-\mu_{g,d,\Delta}(X))
        \mid X,G,D\big]\Big]
        \Big\vert
        \\&=
        \sqrt{m}\cdot{e}
        \sum_{g,d\in\{0,1\}}\Big\vert
        \pp[-
        \hat{\pi}^\ell_{r,g,d}(X)\cdot\mathbbm{1}\{G=g,D=d\} \big(\mu_{g,d,\Delta}(X)-\hat{\mu}^\ell_{g,d,\Delta}(X))
        \\&\hspace{8em}+
        p(G=1,D=1\mid  X)\cdot\frac{\mathbbm{1}\{G=g,D=d\}}{p(G=g,D=d\mid X)}\cdot \big(\mu_{g,d,\Delta}(X)-\mu_{g,d,\Delta}(X))\big]
        \\&\hspace{12em}+\pp\big[
        G\cdot D\cdot \big(\mu_{g,d,\Delta}(X)-\hat{\mu}^\ell_{g,d,\Delta}(X)\big)
        \big]
        \Big\vert
        \\&=
        \sqrt{m}\cdot{e}
        \sum_{g,d\in\{0,1\}}
        \big\vert
            \pp\big[\big(
                G\cdot D-\hat{\pi}^\ell_{r,g,d}(X)\cdot\mathbbm{1}\{G=g,D=d\}
            \big)\cdot\big(
                \mu_{g,d,\Delta}(X)-\hat{\mu}^\ell_{g,d,\Delta}(X)
            \big)\big]
        \big\vert
        \\&=
        \sqrt{m}\cdot{e}
        \sum_{g,d\in\{0,1\}}
        \big\vert
            \pp\big[\big(
                \pi_{1,1}(X)-\hat{\pi}^\ell_{r,g,d}(X)\cdot\pi_{g,d}(X)
            \big)\cdot\big(
                \mu_{g,d,\Delta}(X)-\hat{\mu}^\ell_{g,d,\Delta}(X)
            \big)\big]
        \big\vert
        \\&=
        \sqrt{m}\cdot{e}
        \sum_{g,d\in\{0,1\}}
        \big\vert
            \pp\big[\pi_{g,d}(X)\cdot\big(
                \frac{\pi_{1,1}(X)}{\pi_{g,d}(X)}-\hat{\pi}^\ell_{r,g,d}(X)
            \big)\cdot\big(
                \mu_{g,d,\Delta}(X)-\hat{\mu}^\ell_{g,d,\Delta}(X)
            \big)\big]
        \big\vert
        \\&\overset{}{\leq}
        \sqrt{m}\cdot{e}
        \sum_{g,d\in\{0,1\}}
        \bignorm{\pi_{g,d}\cdot\big(
                \frac{\pi_{1,1}}{\pi_{g,d}}-\hat{\pi}^\ell_{r,g,d}
            \big)}\cdot
        \bignorm{
                \mu_{g,d,\Delta}-\hat{\mu}^\ell_{g,d,\Delta}
        }
        \\&\leq
        \sqrt{m}\cdot{e}
        \sum_{g,d\in\{0,1\}}
        \bignorm{
                \frac{\pi_{1,1}}{\pi_{g,d}}-\hat{\pi}^\ell_{r,g,d}
            }\cdot
        \bignorm{
                \mu_{g,d,\Delta}-\hat{\mu}^\ell_{g,d,\Delta}
        },
\end{align*}
\endgroup
where each summand is $o_p(n^{-1/2})$ under \Cref{as:infp}, and since $\sqrt{m} = \sqrt{n}/\sqrt{L}$ where $L$ is a fixed value, $B_m^\ell$ is $o_p(1)$.
Based on the inequality above, 
\begingroup
\allowdisplaybreaks
\begin{align*}
    T_3 &= \frac{1}{\sqrt{L}}\sum_{\ell=1}^{L}(1+\frac{\hat{e}^\ell-e}{e})\cdot B_m^\ell
\end{align*}
\endgroup
is also $o_p(1)$. Note that $(\hat{e}^\ell-e)/e$ is $o_p(1)$ due to \Cref{lem:ratioestimator}, and the result follows from Slutsky's theorem and Theorem 2.7 of \citep{van2000asymptotic}.
\end{proof}

\thmdrrc*
\begin{proof}
Define $e=\frac{1}{\ex{}[G\cdot D\cdot T]}$ and $\pi_{r,g,d,t}(X)=\frac{\pi_{1,1,1}(X)}{\pi_{g,d,t}(X)}$.
Suppose that the data is split into $L$ folds of size $m$, so that $n=mL$. 
We drop the subscript of $\hat{e}_\mathrm{rc}^\ell$ in the remainder of this proof.
We first show that $\ex{}\left[
\eta_\mathrm{rc,1}(O;\{f^\ell_{\mu,g,d,t}\}_{g,d,t}, \{f^\ell_{\pi,g,d,t}\}_{g,d,t})
\right]=\psi_\mathrm{rc}/e$.
\begingroup
\allowdisplaybreaks
\begin{align*}
    &\hspace{-.5em}\ex{}\left[\eta_\mathrm{rc,1}(O;\{f^\ell_{\mu,g,d,t}\}_{g,d,t}, \{f^\ell_{\pi,g,d,t}\}_{g,d,t})
    -
    \eta_\mathrm{rc,1}(O;\{\mu_{g,d,t}\}_{g,d,t}, \{\pi_{r,g,d,t}\}_{g,d,t})
    \right] 
    =
    \\
    &\sum_{g,d,t\in\{0,1\}}(-1)^{(g+d+t)}
    \ex{}\Big[
        -G\cdot D\cdot T\cdot(f^\ell_{\mu,g,d,t}-\mu_{g,d,t})
        \\&-\ind{G=g,D=d,T=t}\cdot Y\cdot(f^\ell_{\pi,g,d,t}-\pi_{r,g,d,t})
        \\&+\ind{G=g,D=d,T=t}\cdot(f^\ell_{\mu,g,d,t}f^\ell_{\pi,g,d,t}-\mu_{g,d,t}\pi_{r,g,d,t})
    \Big]
    \\&=
    -\sum_{g,d,t\in\{0,1\}}(-1)^{(g+d+t)}
    \ex{}\left[
        \big(G\cdot D\cdot T - \ind{G=g,D=d,T=t}\cdot\pi_{r,g,d,t}\big)\cdot(f^\ell_{\mu,g,d,t}-\mu_{g,d,t})
        \right]
        \\&
        -\sum_{g,d,t\in\{0,1\}}(-1)^{(g+d+t)}\ex{}\left[
        \ind{G=g,D=d,T=t}\cdot(Y-\mu_{g,d,t})\cdot(f^\ell_{\pi,g,d,t}-\pi_{r,g,d,t})
        \right]
        \\&
        +\sum_{g,d,t\in\{0,1\}}(-1)^{(g+d+t)}\ex{}\left[
        \ind{G=g,D=d,T=t}\cdot(f^\ell_{\mu,g,d,t}-\mu_{g,d,t})\cdot(f^\ell_{\pi,g,d,t}-\pi_{r,g,d,t})
    \right]
    \\&\overset{(a)}{=}
    -\sum_{g,d,t\in\{0,1\}}(-1)^{(g+d+t)}
    \ex{}\left[
        \big(\pi_{1,1,1} - \pi_{g,d,t}\cdot\pi_{r,g,d,t}\big)\cdot(f^\ell_{\mu,g,d,t}-\mu_{g,d,t})
        \right]
        \\&
        -\sum_{g,d,t\in\{0,1\}}(-1)^{(g+d+t)}\ex{}\left[
        \ind{G=g,D=d,T=t}\cdot(Y-\mu_{g,d,t})\cdot(f^\ell_{\pi,g,d,t}-\pi_{r,g,d,t})
        \right]
        \\&
        +\sum_{g,d,t\in\{0,1\}}(-1)^{(g+d+t)}\ex{}\left[
        \ind{G=g,D=d,T=t}\cdot(f^\ell_{\mu,g,d,t}-\mu_{g,d,t})\cdot(f^\ell_{\pi,g,d,t}-\pi_{r,g,d,t})
    \right]
    \\&\overset{(b)}{=}
        -\sum_{g,d,t\in\{0,1\}}(-1)^{(g+d+t)}\ex{}\left[
        \ind{G=g,D=d,T=t}\cdot(Y-\mu_{g,d,t})\cdot(f^\ell_{\pi,g,d,t}-\pi_{r,g,d,t})
        \right]
    \\&\overset{(c)}{=}
        -\sum_{g,d,t\in\{0,1\}}(-1)^{(g+d+t)}\ex{}\left[
        (Y-\mu_{g,d,t})\cdot(f^\ell_{\pi,g,d,t}-\pi_{r,g,d,t})
        \big| G=g,D=d,T=t
        \right]\\&\cdot p(G=g,D=d,T=t)
        \\&\overset{(d)}{=}
        -\sum_{g,d,t\in\{0,1\}}(-1)^{(g+d+t)}\ex{}\bigg[
        \ex{}\left[
        Y-\mu_{g,d,t}|X,G=g,D=d,T=t\right]\cdot(f^\ell_{\pi,g,d,t}-\pi_{r,g,d,t})
        \big|\\& G=g,D=d,T=t
        \bigg]\cdot p(G=g,D=d,T=t)
        \\&\overset{(e)}{=}
        -\sum_{g,d,t\in\{0,1\}}(-1)^{(g+d+t)}\ex{}\left[
        (\mu_{g,d,t}-\mu_{g,d,t})\cdot(f^\ell_{\pi,g,d,t}-\pi_{r,g,d,t})
        \big| G=g,D=d,T=t
        \right]\\&\cdot p(G=g,D=d,T=t)
        \\&=0,
\end{align*}
where $(a)$ and $(d)$ are due to the law of iterated expectations, $(b)$ follows since $\pi_{1,1,1}-\pi_{g,d,t}\pi_{r,g,d,t}=\pi_{1,1,1}-\pi_{1,1,1}=0$, 
and the third term is $0$ because either $f^\ell_{\mu,g,d,t}-\mu_{g,d,t}=0$ or $f^\ell_{\pi,g,d,t}-\pi_{r,g,d,t}=0$,
$(c)$ is an application of the total probability law, and $(e)$ is by definition of $\mu_{g,d,t}$.
We get
\begin{equation}\label{eq:proofdrexpecrc}
    \ex{}\left[
\eta_\mathrm{rc,1}(O;\{f^\ell_{\mu,g,d,t}\}_{g,d,t}, \{f^\ell_{\pi,g,d,t}\}_{g,d,t})
\right]
=
\ex{}\left[
    \eta_\mathrm{rc,1}(O;\{\mu_{g,d,t}\}_{g,d,t}, \{\pi_{r,g,d,t}\}_{g,d,t})
\right]
=\frac{\psi_\mathrm{rc}}{e}.
\end{equation}
\endgroup
We also show that $\eta_\mathrm{rc,1}(O;\{f^\ell_{\mu,g,d,t}\}_{g,d,t}, \{f^\ell_{\pi,g,d,t}\}_{g,d,t})$ is integrable:
\begingroup
\allowdisplaybreaks
\begin{align*}
    \ex{}\big[\big\vert\eta_\mathrm{rc,1}(O;&\{f^\ell_{\mu,g,d,t}\}_{g,d,t}, \{f^\ell_{\pi,g,d,t}\}_{g,d,t})\big\vert\big] 
    \\&\overset{(a)}{\leq}
    \sum_{g,d,t\in\{0,1\}}\Big[
    \pp\big[\vert G\cdot D\cdot T\cdot Y\vert\big]
    +\pp\big[\vert Y\cdot f^\ell_{\pi,g,d,t}(X)\vert\big]
    \\&\hspace{3em}+\pp\big[\vert G\cdot D\cdot T\cdot f^\ell_{\mu,g,d,t}(X)\vert\big]
    +\pp\big[\vert
    f^\ell_{\mu,g,d,t}(X)\cdot f^\ell_{\pi,g,d,t}(X)
    \vert\big]
            \Big]
    \\&\overset{(b)}{\leq}
    \sum_{g,d,t\in\{0,1\}}\Big[
    \norm{Y}
    \cdot(1+ \norm{f^\ell_{\pi,g,d,t}})
    \\&\hspace{7em}+\norm{f^\ell_{\mu,g,d,t}}
    +\norm{
    f^\ell_{\mu,g,d,t}}\cdot \norm{f^\ell_{\pi,g,d,t}}
            \Big]
            \overset{(c)}{<}\infty,
\end{align*}
\endgroup
where $(a)$, $(b)$, and $(c)$ are due to triangle inequality, Cauchy-Schwarz, and the bounded moment assumptions of the proposition.
Moreover,
\begingroup
\allowdisplaybreaks
\begin{align*}
    &\hspace{-8em}\pp\left[\big\vert\eta_\mathrm{rc,1}(O;\{\hat{\mu}^\ell_{g,d,t}\}_{g,d,t}, \{\hat{\pi}^\ell_{r,g,d,t}\}_{g,d,t})
    -
    \eta_\mathrm{rc,1}(O;\{f^\ell_{\mu,g,d,t}\}_{g,d,t}, \{f^\ell_{\pi,g,d,t}\}_{g,d,t})\big\vert\right]
    \\&\overset{(a)}{\leq}
    \sum_{g,d,t\in\{0,1\}}\Big[
    \pp\big[\vert G\cdot D\cdot T\cdot\big(\hat{\mu}^\ell_{g,d,t}(X) - f^\ell_{\mu,g,d,t}(X)\big)\vert\big]
    \\&+\pp\big[\vert Y\cdot\big(
    \hat{\pi}^\ell_{r,g,d,t}(X)
    - f^\ell_{\pi,g,d,t}(X)\big)\vert\big]
    \\&+
    \pp\big[\vert
    f^\ell_{\mu,g,d,t}(X)\big(
    \hat{\pi}^\ell_{r,g,d,t}(X)
    - f^\ell_{\pi,g,d,t}(X)\big)
    \vert\big]
    \\&+
    \pp\big[\vert
    f^\ell_{\pi,g,d,t}(X)\big(
    \hat{\mu}^\ell_{g,d,t}(X)
    - f^\ell_{\mu,g,d,t}(X)\big)
    \vert\big]
    \\&+
    \pp\big[\vert
    \big(
    \hat{\pi}^\ell_{r,g,d,t}(X)
    - f^\ell_{\pi,g,d,t}(X)\big)
    \big(
    \hat{\mu}^\ell_{g,d,t}(X)
    - f^\ell_{\mu,g,d,t}(X)\big)\vert\big]
            \Big]
    \\
    &\overset{(b)}{\leq}\sum_{g,d,t\in\{0,1\}}\Big[
    \norm{\big(\hat{\mu}^\ell_{g,d,t} - f^\ell_{\mu,g,d,t}\big)}
    \\&+\norm{Y}\cdot\norm{\big(
    \hat{\pi}^\ell_{r,g,d,t}
    - f^\ell_{\pi,g,d,t}\big)}
    \\&+
    \norm{
    f^\ell_{\mu,g,d,t}}\cdot\norm{\big(
    \hat{\pi}^\ell_{r,g,d,t}
    - f^\ell_{\pi,g,d,t}\big)
    }
    \\&+
    \norm{
    f^\ell_{\pi,g,d,t}}\cdot\norm{\big(
    \hat{\mu}^\ell_{g,d,t}
    - f^\ell_{\mu,g,d,t}\big)}
    \\&+
    \norm{
    \big(
    \hat{\pi}^\ell_{r,g,d,t}
    - f^\ell_{\pi,g,d,t}\big)}\cdot\norm{
    \big(
    \hat{\mu}^\ell_{g,d,t}
    - f^\ell_{\mu,g,d,t}\big)}
            \Big]
    \\&\overset{(c)}{=}o_p(1),
\end{align*}
\endgroup
where $(a)$ is due to triangle inequality, $(b)$ is due to Cauchy-Schwartz, and $(c)$ holds because of the assumptions of the proposition.
Therefore, $\eta_\mathrm{rc,1}(O;\{\hat{\mu}^\ell_{g,d,t}\}_{g,d,t}, \{\hat{\pi}^\ell_{r,g,d,t}\}_{g,d,t})$ converges in $L_1$ to $\eta_\mathrm{rc,1}(O;\{f^\ell_{\mu,g,d,t}\}_{g,d,t}, \{f^\ell_{\pi,g,d,t}\}_{g,d,t})$.
Now consider 
\begingroup
\allowdisplaybreaks
\begin{align*}
    &\hspace{-2em}\ee_m^\ell\left[\eta_\mathrm{rc,1}(O;\{\hat{\mu}^\ell_{g,d,t}\}_{g,d,t}, \{\hat{\pi}^\ell_{r,g,d,t}\}_{g,d,t})\right] - \frac{\psi_\mathrm{rc}}{e}
    \\
    &=
    \ee_m^\ell\left[\eta_\mathrm{rc,1}(O;\{\hat{\mu}^\ell_{g,d,t}\}_{g,d,t}, \{\hat{\pi}^\ell_{r,g,d,t}\}_{g,d,t})
    -
    \eta_\mathrm{rc,1}(O;\{f^\ell_{\mu,g,d,t}\}_{g,d,t}, \{f^\ell_{\pi,g,d,t}\}_{g,d,t})
    \right] 
    \\&+
    \ee_m^\ell\left[
    \eta_\mathrm{rc,1}(O;\{f^\ell_{\mu,g,d,t}\}_{g,d,t}, \{f^\ell_{\pi,g,d,t}\}_{g,d,t})
    \right] -
    \pp\left[
    \eta_\mathrm{rc,1}(O;\{f^\ell_{\mu,g,d,t}\}_{g,d,t}, \{f^\ell_{\pi,g,d,t}\}_{g,d,t})
    \right]
    \\&+
    \pp\left[
    \eta_\mathrm{rc,1}(O;\{f^\ell_{\mu,g,d,t}\}_{g,d,t}, \{f^\ell_{\pi,g,d,t}\}_{g,d,t})
    \right]
    -
    \frac{\psi_\mathrm{rc}}{e},
\end{align*}
\endgroup
where the first term is $o_p(1)$ because of $L^1$ convergence and an application of Chebyshev's inequality,
the second term is $o_p(1)$ since $\eta_\mathrm{rc,1}(O;\{f^\ell_{\mu,g,d,t}\}_{g,d,t}, \{f^\ell_{\pi,g,d,t}\}_{g,d,t})$ is integrable and the weak law of large numbers applies, and the third term is $0$ by \Cref{eq:proofdrexpecrc}.
Therefore,
\[\ee_m^\ell\left[\eta_\mathrm{rc,1}(O;\{\hat{\mu}^\ell_{g,d,t}\}_{g,d,t}, \{\hat{\pi}^\ell_{r,g,d,t}\}_{g,d,t})\right]\overset{p}{\to}\frac{\psi_\mathrm{rc}}{e},\]
and by \Cref{lem:ratioestimator}, $\hat{e}^\ell\overset{p}{\to}e$.
Using Theorem 2.7 of \cite{van2000asymptotic}, we get
\[\hat{e}^\ell\cdot\ee_m^\ell\left[\eta_\mathrm{rc,1}(O;\{\hat{\mu}^\ell_{g,d,t}\}_{g,d,t}, \{\hat{\pi}^\ell_{r,g,d,t}\}_{g,d,t})\right]\overset{p}{\to}\psi_\mathrm{rc},\]
and since the latter holds for every fold $\ell\in\{1,\dots,L\}$, $\hat{\psi}_\mathrm{rc,1}^\mathrm{dr}$ is consistent.
\end{proof}

\thmcanrc*
\begin{proof}
The proof is similar to that of \Cref{thm:canp}:
define $e=\frac{1}{\ex{}[G\cdot D \cdot T]}$ and $\pi_{r,g,d,t}(X)=\frac{\pi_{1,1,1}(X)}{\pi_{g,d,t}(X)}$.
Suppose that the data is split into $L$ folds of size $m$, so that $n=mL$. 
We drop the subscript of $\hat{e}_\mathrm{rc}^\ell$ in the remainder of this proof.

\begingroup
\allowdisplaybreaks
    \begin{align}
        \sqrt{n}&(\hat{\psi}_\mathrm{rc,1}^\mathrm{dr} - \psi_\mathrm{rc}) = \notag
        \\&
        \sqrt{n}
            \big(\frac{1}{L}
            \sum_{\ell=1}^{L}
                \hat{e}^\ell\cdot\ee_m^\ell\big[\eta_\mathrm{rc,1}(O;\{\hat{\mu}^\ell_{g,d,t}\}_{g,d,t}, \{\hat{\pi}^\ell_{r,g,d,t}\}_{g,d,t})\big]
            -\psi_\mathrm{rc}\big)\notag
        \\&
            =\frac{\sqrt{mL}}{L}\cdot
            \sum_{\ell=1}^{L}
                \frac{\hat{e}^\ell}{e}\cdot\big(e\cdot\ee_m^\ell\big[\eta_\mathrm{rc,1}(O;\{\hat{\mu}^\ell_{g,d,t}\}_{g,d,t}, \{\hat{\pi}^\ell_{r,g,d,t}\}_{g,d,t})\big]
            -e\cdot\psi_\mathrm{rc}\cdot\ee_m^\ell[G\cdot D\cdot T]\big)\notag
        \\&
        =\frac{1}{\sqrt{L}}
            \sum_{\ell=1}^{L}
                (1+\frac{\hat{e}^\ell-e}{e})\cdot\sqrt{m}\big(e\cdot\ee_m^\ell\big[\eta_\mathrm{rc,1}(O;\{\hat{\mu}^\ell_{g,d,t}\}_{g,d,t}, \{\hat{\pi}^\ell_{r,g,d,t}\}_{g,d,t})\big]
            -e\cdot\psi_\mathrm{rc}\cdot\ee_m^\ell[G\cdot D\cdot T]\big)\notag
        \\&
        \begin{aligned}
            &=
            \frac{1}{\sqrt{L}}\sum_{\ell=1}^{L}
            (1+\frac{\hat{e}^\ell-e}{e})\cdot
            \sqrt{m}(\ee_m^\ell-\pp)
                \Big[e\cdot\eta_\mathrm{rc,1}(O;\{\hat{\mu}^\ell_{g,d,t}\}_{g,d,t}, \{\hat{\pi}^\ell_{r,g,d,t}\}_{g,d,t})
                \\&\hspace{20em}
                -e\cdot\eta_\mathrm{rc,1}(O;\{\mu_{g,d,t}\}_{g,d,t}, \{\pi_{r,g,d,t}\}_{g,d,t})\Big]
        \end{aligned}\label{eq:rct1}\tag{$T_{1}$}
        \\&
        \begin{aligned}
            &+
            \frac{1}{\sqrt{L}}\sum_{\ell=1}^{L}
            (1+\frac{\hat{e}^\ell-e}{e})\cdot
            \sqrt{m}(\ee_m^\ell-\pp)
                \Big[e\cdot\eta_\mathrm{rc,1}(O;\{\mu_{g,d,t}\}_{g,d,t}, \{\pi_{r,g,d,t}\}_{g,d,t})\\&\hspace{25em}-e\cdot G\cdot D\cdot T\cdot \psi_\mathrm{rc}\Big]
        \end{aligned}\label{eq:rct2}\tag{$T_{2}$}
        \\&
        \begin{aligned}
            &+
            \frac{1}{\sqrt{L}}\sum_{\ell=1}^{L}
            (1+\frac{\hat{e}^\ell-e}{e})\cdot
            \sqrt{m}\big(\pp
                \big[e\cdot\eta_\mathrm{rc,1}(O;\{\hat{\mu}^\ell_{g,d,t}\}_{g,d,t}, \{\hat{\pi}^\ell_{r,g,d,t}\}_{g,d,t})
                \big]-\psi_\mathrm{rc}\big).
        \end{aligned}\label{eq:rct3}\tag{$T_{3}$}
    \end{align}
\endgroup
We analyze each of the terms above separately.
In the remainder of the proof, we use $O_n^{-\ell}$ to represent the data in all but the $\ell$-th fold.

\textbf{Term \ref{eq:rct1}:}

Define $A_m^\ell = \sqrt{m}(\ee_m^\ell-\pp)
                \big[e\cdot\eta_\mathrm{rc,1}(O;\{\hat{\mu}^\ell_{g,d,t}\}_{g,d,t}\allowbreak, \allowbreak\{\hat{\pi}^\ell_{r,g,d,t}\}_{g,d,t})\allowbreak-e\cdot\eta_\mathrm{rc,1}(O;\{\mu_{g,d,t}\}_{g,d,t}, \{\pi_{r,g,d,t}\}_{g,d,t})\big]$.
Note that $\ex{}[A_m^\ell\mid O^{-\ell}]=0$. 
Next, we bound the conditional variance as
\begingroup\allowdisplaybreaks
\begin{align*}
    \mathrm{var}&(A_m^\ell\mid O^{-\ell})
    \\&= 
    m\cdot e^2\cdot\mathrm{var}(\ee_m^\ell
        [
            \eta_\mathrm{rc,1}(O;\{\hat{\mu}^\ell_{g,d,t}\}_{g,d,t}, \{\hat{\pi}^\ell_{r,g,d,t}\}_{g,d,t})-\eta_\mathrm{rc,1}(O;\{\mu_{g,d,t}\}_{g,d,t}, \{\pi_{r,g,d,t}\}_{g,d,t})
        ]
        \mid O^{-\ell}
    )
    \\&=
    e^2\cdot\mathrm{var}(
            \eta_\mathrm{rc,1}(O;\{\hat{\mu}^\ell_{g,d,t}\}_{g,d,t}, \{\hat{\pi}^\ell_{r,g,d,t}\}_{g,d,t})-\eta_\mathrm{rc,1}(O;\{\mu_{g,d,t}\}_{g,d,t}, \{\pi_{r,g,d,t}\}_{g,d,t})
        \mid O^{-\ell}
    )
    \\&\leq e^2\cdot\pp\big[\big(
            \eta_\mathrm{rc,1}(O;e,\{\hat{\mu}^\ell_{g,d,t}\}_{g,d,t}, \{\hat{\pi}^\ell_{r,g,d,t}\}_{g,d,t})-\eta_\mathrm{rc,1}(O;e,\{\mu_{g,d,t}\}_{g,d,t}, \{\pi_{r,g,d,t}\}_{g,d,t})\big)^2
        \mid O^{-\ell}
    \big]
    \\&=e^2\cdot
    \pp\Big[\Big(\sum_{g,d,t\in\{0,1\}}(-1)^{(g+d+t)}\hat{\omega}^\ell_{g,d,t}(X,G,D,T)\cdot
            \big(
                Y-\hat{\mu}^\ell_{g,d,t}(X)
            \big)
        \\&\hspace{5em}-
            \sum_{g,d,t\in\{0,1\}}(-1)^{(g+d+t)}\omega_{g,d,t}(X,G,D,T)\cdot
            \big(
                Y-\mu_{g,d,t}(X)
            \big)
            \Big)^2
        \:\Big\vert\: O^{-\ell}
    \Big]
    \\&\overset{(a)}{\leq}8e^2\sum_{g,d,t\in\{0,1\}}
    \pp\Big[\Big(\hat{\omega}^\ell_{g,d,t}(X,G,D,T)\cdot
            \big(
                Y-\hat{\mu}^\ell_{g,d,t}(X)
            \big)
        \\&\hspace{10em}-
            \omega_{g,d,t}(X,G,D,T)\cdot
            \big(
                Y-\mu_{g,d,t}(X)
            \big)
            \Big)^2
        \:\Big\vert\: O^{-\ell}
    \Big]
    \\&\overset{(b)}{\leq}24e^2\sum_{g,d,t\in\{0,1\}}
    \pp\Big[\Big(G\cdot D\cdot T\cdot\big(\hat{\mu}^\ell_{g,d,t}(X)-\mu_{g,d,t}(X)\big)\Big)^2
        \:\Big\vert\: O^{-\ell}
    \Big]
        \\&\hspace{5em}+24e^2\sum_{g,d,t\in\{0,1\}}
        \pp\Big[\Big(\big(\hat{\pi}^\ell_{r,g,d,t}(X)-\pi_{r,g,d,t}(X)\big)\cdot\big(Y-\mu_{g,d,t}(X)\big)\Big)^2
            \:\Big\vert\: O^{-\ell}
        \Big]
        \\&\hspace{5em}+24e^2\sum_{g,d,t\in\{0,1\}}
        \pp\Big[\Big(\hat{\pi}^\ell_{r,g,d,t}(X)\cdot\big(\hat{\mu}^\ell_{g,d,t}(X)-\mu_{g,d,t}(X)\big)\Big)^2
            \:\Big\vert\: O^{-\ell}
        \Big]
    \\&\overset{(c)}{\leq}
        96e^2\sum_{g,d,t\in\{0,1\}}\Big(\bignorm{
            \hat{\mu}^\ell_{g,d,t} - \mu_{g,d,t}
        }^2
        +
        \bignorm{
            (\hat{\pi}^\ell_{r,g,d,t})^2-\pi_{r,g,d,t}^2
        }\cdot
        \big(\norm{
            Y^2
        }
        +
        \norm{\mu_{g,d,t}^2}\big)
        \\&\hspace{5em}+
        \bignorm{
            (\hat{\mu}^\ell_{g,d,t})^2 - \mu_{g,d,t}^2
        }\cdot
        \big(\bignorm{
            (\hat{\pi}^\ell_{r,g,d,t})^2-\pi_{r,g,d,t}^2
        }
        +\bignorm{\pi_{r,g,d,t}^2}\big)
        \Big),
\end{align*}
\endgroup
where $(a)$ and $(b)$ are due to Cauchy-Schwartz, and $(c)$ is an application of Hölder, Cauchy-Schwartz, and Minkowski inequalities.
Under \Cref{as:infrc} and strict positivity, the conditional variance above is $o_p(1)$.
By the law of iterated expectations and Chebyshev's inequality, 
we conclude $A_m^\ell\overset{p}{\to}0$.
$(\hat{e}^\ell-e)/e$ is $o_p(1)$ by \Cref{lem:ratioestimator}, and using Slutsky's theorem and Theorem 2.7 of \citep{van2000asymptotic}, $(1+(\hat{e}^\ell-e)/e)\cdot A_m^\ell\overset{p}{\to}0$ for every $\ell$.
Finally, since the number of folds ($L$) is finite, \ref{eq:rct1} converges to $0$ in probability.

\textbf{Term \ref{eq:rct2}:}
\begin{align*}
    T_2&=\frac{1}{\sqrt{L}}\sum_{\ell=1}^{L}
            (1+\frac{\hat{e}^\ell-e}{e})\cdot\sqrt{m}(\ee_m^\ell-\pp)
                \big[e\cdot\eta_\mathrm{rc,1}(O;\{\mu_{g,d,t}\}_{g,d,t}, \{\pi_{g,d,t}\}_{g,d,t})-e\cdot G\cdot D\cdot T\cdot \psi_\mathrm{rc}\big]
    \\&=
    \frac{1}{\sqrt{L}}\sum_{\ell=1}^{L}
            (1+\frac{\hat{e}^\ell-e}{e})\cdot\sqrt{m}(\ee_m^\ell-\pp)
                \big[\psi^1_\mathrm{rc,1}(O)\big]
    \\&\overset{(a)}{=}\sqrt{n}(\ee_n-\pp)\big[\psi^1_\mathrm{rc,1}(O)\big] + o_p(1),
\end{align*}
where $(a)$ is because (i) $\sqrt{m}(\ee_m^\ell-\pp)\big[\psi^1_\mathrm{rc,1}(O)\big]$ is $O_p(1)$ due to the central limit theorem, (ii) $(\hat{e}^\ell-e)/e$ is $O_p(m^{-1/2})$ due to \Cref{lem:ratioestimator}, so their product converges to $0$ in probability due to Slutsky's theorem and Theorem 2.7 of \citep{van2000asymptotic}, and (iii) the number of folds ($L$) is finite.
By the central limit theorem, the term $\sqrt{n}(\ee_n-\pp)\big[\psi^1_\mathrm{rc,1}(O)\big]$ converges in distribution to $\mathcal{N}\Big(0, \mathrm{var}\big(\psi^1_\mathrm{rc,1}(O)\big)\Big)$, as long as the variance of $\psi^1_\mathrm{rc,1}(\cdot)$ is bounded.
This variance can be bounded as follows.
\begin{align*}
    \mathrm{var}&\big(e\cdot\eta_\mathrm{rc,1}(O; \allowbreak\{\mu_{g,d,t}\}_{g,d,t}, \{\pi_{g,d,t}\}_{g,d,t})-e\cdot G\cdot D\cdot T\cdot \psi_\mathrm{rc}\big)
    \\&\leq
    e^2\cdot\ex{}\big[\big(\eta_\mathrm{rc,1}(O; \allowbreak\{\mu_{g,d,t}\}_{g,d,t}, \{\pi_{g,d,t}\}_{g,d,t})- G\cdot D\cdot T\cdot \psi_\mathrm{rc}\big)^2\big]
    \\&\leq
    2e^2\cdot\ex{}\big[\big(\eta_\mathrm{rc,1}(O; \allowbreak\{\mu_{g,d,t}\}_{g,d,t}, \{\pi_{g,d,t}\}_{g,d,t})\big)^2 + \big(G\cdot D\cdot T\cdot \psi_\mathrm{rc}\big)^2\big]
    \\&\leq
    16e^2\sum_{g,d,t\in\{0,1\}}\ex{}\big[\omega_{g,d,t}^2\cdot\big(Y-\mu_{g,d,t}(X)\big)^2\big] + 2e^2\cdot\psi_\mathrm{rc}^2
    \\&\leq
    32e^2\sum_{g,d,t\in\{0,1\}}\bignorm{\omega_{g,d,t}^2}\cdot\Big(\bignorm{Y^2}+\bignorm{\mu_{g,d,t}^2}
    \Big)+ 2e^2\cdot\psi_\mathrm{rc}^2.
\end{align*}
which is bounded under \Cref{as:infrc} and strict positivity.

\textbf{Term \ref{eq:rct3}:}

Define $B_m^\ell\coloneqq \sqrt{m}\big(\pp
                \big[e\cdot\eta_\mathrm{rc,1}(O;\{\hat{\mu}^\ell_{g,d,t}\}_{g,d,t}, \{\hat{\pi}^\ell_{r,g,d,t}\}_{g,d,t})
                \big]-\psi_\mathrm{rc}\big)$.
\begingroup
\allowdisplaybreaks
\begin{align*}
    B_m^\ell&=
    \sqrt{m}\cdot e\cdot
            \big(\pp
                \big[\eta_\mathrm{rc,1}(O;\{\hat{\mu}^\ell_{g,d,t}\}_{g,d,t}, \{\hat{\pi}^\ell_{r,g,d,t}\}_{g,d,t})\big]-\pp
                \big[\eta_\mathrm{rc,1}(O;\{{\mu}_{g,d,t}\}_{g,d,t}, \{{\pi}_{r,g,d,t}\}_{g,d,t})\big]\big)
    \\&=\sqrt{m}\cdot
        e\cdot \pp\big[
            \sum_{g,d,t\in\{0,1\}}(-1)^{(g+d+t)}\hat{\omega}^\ell_{g,d,t}(X,G,D,T)
            \cdot
            \big(
                Y-\hat{\mu}^\ell_{g,d,t}(X)
            \big)
        \\&\hspace{9em}-
        \sum_{g,d,t\in\{0,1\}}(-1)^{(g+d+t)}\omega_{g,d,t}(X,G,D,T)
            \cdot
            \big(
                Y-\mu_{g,d,t}(X)
            \big)
        \big]
        \\&\leq
        \sqrt{m}\cdot{e}
        \sum_{g,d,t\in\{0,1\}}\big\vert
        \pp\big[
        \hat{\omega}^\ell_{g,d,t}(X,G,D,T)\cdot \big(Y-\hat{\mu}^\ell_{g,d,t}(X))
        \\&\hspace{12em}-
        \omega_{g,d,t}(X,G,D,T)\cdot \big(Y-\mu_{g,d,t}(X))
        \big]
        \Big\vert
        \\&=
        \sqrt{m}\cdot{e}
        \sum_{g,d,t\in\{0,1\}}\Big\vert
        \pp\Big[\pp[
        \hat{\omega}^\ell_{g,d,t}(X,G,D,T)\cdot \big(Y-\hat{\mu}^\ell_{g,d,t}(X))
        \\&\hspace{12em}-
        \omega_{g,d,t}(X,G,D,T)\cdot \big(Y-\mu_{g,d,t}(X))
        \mid X,G,D,T\big]\Big]
        \Big\vert
        \\&=
        \sqrt{m}\cdot{e}
        \sum_{g,d,t\in\{0,1\}}\Big\vert
        \pp[-
        \hat{\pi}^\ell_{r,g,d,t}(X)\cdot\mathbbm{1}\{G=g,D=d, T=t\} \big(\mu_{g,d,t}(X)-\hat{\mu}^\ell_{g,d,t}(X))
        \\&\hspace{3em}+
        p(G=1,D=1, T=1\mid  X)\cdot\frac{\mathbbm{1}\{G=g,D=d, T=t\}}{p(G=g,D=d, T=t\mid X)}\cdot \big(\mu_{g,d,t}(X)-\mu_{g,d,t}(X))\big]
        \\&\hspace{12em}+\pp\big[
        G\cdot D\cdot T\cdot \big(\mu_{g,d,t}(X)-\hat{\mu}^\ell_{g,d,t}(X)\big)
        \big]
        \Big\vert
        \\&=
        \sqrt{m}\cdot{e}
        \sum_{g,d,t\in\{0,1\}}
        \big\vert
            \pp\big[\big(
                G\cdot D\cdot T-\hat{\pi}^\ell_{r,g,d,t}(X)\cdot\mathbbm{1}\{G=g,D=d, T=t\}
            \big)\cdot\\&\hspace{25em}\big(
                \mu_{g,d,t}(X)-\hat{\mu}^\ell_{g,d,t}(X)
            \big)\big]
        \big\vert
        \\&=
        \sqrt{m}\cdot{e}
        \sum_{g,d,t\in\{0,1\}}
        \big\vert
            \pp\big[\big(
                \pi_{1,1,1}(X)-\hat{\pi}^\ell_{r,g,d,t}(X)\cdot\pi_{g,d,t}(X)
            \big)\cdot\big(
                \mu_{g,d,t}(X)-\hat{\mu}^\ell_{g,d,t}(X)
            \big)\big]
        \big\vert
        \\&=
        \sqrt{m}\cdot{e}
        \sum_{g,d,t\in\{0,1\}}
        \big\vert
            \pp\big[\pi_{g,d,t}(X)\cdot\big(
                \frac{\pi_{1,1,1}(X)}{\pi_{g,d,t}(X)}-\hat{\pi}^\ell_{r,g,d,t}(X)
            \big)\cdot\big(
                \mu_{g,d,t}(X)-\hat{\mu}^\ell_{g,d,t}(X)
            \big)\big]
        \big\vert
        \\&\overset{}{\leq}
        \sqrt{m}\cdot{e}
        \sum_{g,d,t\in\{0,1\}}
        \bignorm{\pi_{g,d,t}\cdot\big(
                \frac{\pi_{1,1,1}}{\pi_{g,d,t}}-\hat{\pi}^\ell_{r,g,d,t}
            \big)}\cdot
        \bignorm{
                \mu_{g,d,t}-\hat{\mu}^\ell_{g,d,t}
        }
        \\&\leq
        \sqrt{m}\cdot{e}
        \sum_{g,d,t\in\{0,1\}}
        \bignorm{
                \frac{\pi_{1,1,1}}{\pi_{g,d,t}}-\hat{\pi}^\ell_{r,g,d,t}
            }\cdot
        \bignorm{
                \mu_{g,d,t}-\hat{\mu}^\ell_{g,d,t}
        },
\end{align*}
\endgroup
where each summand is $o_p(n^{-1/2})$ under \Cref{as:infrc}, and since $\sqrt{m} = \sqrt{n}/\sqrt{L}$ where $L$ is a fixed value, $B_m^\ell$ is $o_p(1)$.
Based on the inequality above, 
\begingroup
\allowdisplaybreaks
\begin{align*}
    T_3 &= \frac{1}{\sqrt{L}}\sum_{\ell=1}^{L}(1+\frac{\hat{e}^\ell-e}{e})\cdot B_m^\ell
\end{align*}
\endgroup
is also $o_p(1)$. Note that $(\hat{e}^\ell-e)/e$ is $o_p(1)$ due to \Cref{lem:ratioestimator}, and the result follows from Slutsky's theorem and Theorem 2.7 of \citep{van2000asymptotic}.
\end{proof}

\thmindep*
\begin{proof}
Define $e=\frac{1}{\ex{}[G\cdot D]}$ and $\pi_{r,g,d}(X)=\frac{\pi_{1,1}(X)}{\pi_{g,d}(X)}$.
Suppose that the data is split into $L$ folds of size $m$, so that $n=mL$. 
We drop the subscript of $\hat{e}_\mathrm{rc,2}^\ell$ in the remainder of this proof.
We first show that $\ex{}\left[
\eta_\mathrm{rc,2}(O;\{f^\ell_{\mu,g,d,t}\}_{g,d,t}, \{f^\ell_{\pi,g,d}\}_{g,d})
\right]=\psi_\mathrm{rc}/e$.
\begingroup
\allowdisplaybreaks
\begin{align*}
    &\hspace{-.5em}\ex{}\left[\eta_\mathrm{rc,2}(O;\{f^\ell_{\mu,g,d,t}\}_{g,d,t}, \{f^\ell_{\pi,g,d}\}_{g,d})
    -
    \eta_\mathrm{rc,2}(O;\{\mu_{g,d,t}\}_{g,d,t}, \{\pi_{r,g,d}\}_{g,d})
    \right] 
    =
    \\
    &\sum_{g,d,t\in\{0,1\}}(-1)^{(g+d+t)}
    \ex{}\Big[
        -G\cdot D\cdot(f^\ell_{\mu,g,d,t}-\mu_{g,d,t})
        \\&-\frac{\ind{T=t}}{p(T=t)}\cdot\ind{G=g,D=d}\cdot Y\cdot(f^\ell_{\pi,g,d}-\pi_{r,g,d})
        \\&+\frac{\ind{T=t}}{p(T=t)}\cdot\ind{G=g,D=d}\cdot(f^\ell_{\mu,g,d,t}f^\ell_{\pi,g,d}-\mu_{g,d,t}\pi_{r,g,d})
    \Big]
    \\&=
    -\sum_{g,d,t\in\{0,1\}}(-1)^{(g+d+t)}
    \ex{}\left[
        \big(G\cdot D - \frac{\ind{T=t}}{p(T=t)}\cdot\ind{G=g,D=d}\cdot\pi_{r,g,d}\big)\cdot(f^\ell_{\mu,g,d,t}-\mu_{g,d,t})
        \right]
        \\&
        -\sum_{g,d,t\in\{0,1\}}(-1)^{(g+d+t)}\ex{}\left[
        \frac{\ind{T=t}}{p(T=t)}\cdot\ind{G=g,D=d}\cdot(Y-\mu_{g,d,t})\cdot(f^\ell_{\pi,g,d}-\pi_{r,g,d})
        \right]
        \\&
        +\sum_{g,d,t\in\{0,1\}}(-1)^{(g+d+t)}\ex{}\left[
        \frac{\ind{T=t}}{p(T=t)}\cdot\ind{G=g,D=d}\cdot(f^\ell_{\mu,g,d,t}-\mu_{g,d,t})\cdot(f^\ell_{\pi,g,d}-\pi_{r,g,d})
    \right]
    \\&\overset{(a)}{=}
    -\sum_{g,d,t\in\{0,1\}}(-1)^{(g+d+t)}
    \ex{}\left[
        \big(\pi_{1,1} - \pi_{g,d}\cdot\pi_{r,g,d}\big)\cdot(f^\ell_{\mu,g,d,t}-\mu_{g,d,t})
        \right]
        \\&
        -\sum_{g,d,t\in\{0,1\}}(-1)^{(g+d+t)}\ex{}\left[
        \frac{\ind{T=t}}{p(T=t)}\cdot\ind{G=g,D=d}\cdot(Y-\mu_{g,d,t})\cdot(f^\ell_{\pi,g,d}-\pi_{r,g,d})
        \right]
        \\&
        +\sum_{g,d,t\in\{0,1\}}(-1)^{(g+d+t)}\ex{}\left[
        \ind{G=g,D=d}\cdot(f^\ell_{\mu,g,d,t}-\mu_{g,d,t})\cdot(f^\ell_{\pi,g,d}-\pi_{r,g,d})
    \right]
    \\&\overset{(b)}{=}
        -\sum_{g,d,t\in\{0,1\}}(-1)^{(g+d+t)}\ex{}\left[
        \frac{\ind{T=t}}{p(T=t)}\cdot\ind{G=g,D=d}\cdot(Y-\mu_{g,d,t})\cdot(f^\ell_{\pi,g,d}-\pi_{r,g,d})
        \right]
    \\&
        +\sum_{g,d\in\{0,1\}}(-1)^{(g+d+1)}\ex{}\Big[
        \ind{G=g,D=d}\cdot\big((f^\ell_{\mu,g,d,1}-f^\ell_{\mu,g,d,0})-(\mu_{g,d,1}-\mu_{g,d,0})\big)\\&\cdot(f^\ell_{\pi,g,d}-\pi_{r,g,d})
        \Big]
    \\&\overset{(c)}{=}
        -\sum_{g,d,t\in\{0,1\}}(-1)^{(g+d+t)}\ex{}\left[
        \frac{\ind{T=t}}{p(T=t)}\cdot\ind{G=g,D=d}\cdot(Y-\mu_{g,d,t})\cdot(f^\ell_{\pi,g,d}-\pi_{r,g,d})
        \right]
    \\&\overset{(d)}{=}
        -\sum_{g,d,t\in\{0,1\}}(-1)^{(g+d+t)}\ex{}\left[
        (Y-\mu_{g,d,t})\cdot(f^\ell_{\pi,g,d}-\pi_{r,g,d})
        \big| G=g,D=d,T=t
        \right]\cdot p(G=g,D=d)
        \\&\overset{(e)}{=}
        -\sum_{g,d,t\in\{0,1\}}(-1)^{(g+d+t)}\ex{}\Big[
        \ex{}\left[
        Y-\mu_{g,d,t}|X,G=g,D=d,T=t\right]\\&\cdot(f^\ell_{\pi,g,d}-\pi_{r,g,d})
        \big| G=g,D=d,T=t
        \Big]\cdot p(G=g,D=d)
        \\&\overset{(f)}{=}
        -\sum_{g,d,t\in\{0,1\}}(-1)^{(g+d+t)}\ex{}\left[
        (\mu_{g,d,t}-\mu_{g,d,t})\cdot(f^\ell_{\pi,g,d}-\pi_{r,g,d})
        \big| G=g,D=d,T=t
        \right]\\&\cdot p(G=g,D=d)
        \\&=0,
\end{align*}
where $(a)$ and $(e)$ are due to the law of iterated expectations, $(b)$ follows since $\pi_{1,1}-\pi_{g,d}\pi_{r,g,d}=\pi_{1,1}-\pi_{1,1}=0$, 
$(c)$ holds because either $(f^\ell_{\mu,g,d,1}-f^\ell_{\mu,g,d,0})-(\mu_{g,d,1}-\mu_{g,d,0})=0$ or $f^\ell_{\pi,g,d}-\pi_{r,g,d}=0$,
$(d)$ is an application of the total probability law, and $(f)$ is by definition of $\mu_{g,d,t}$.
We get
\begin{equation}\label{eq:proofdrexpecrc2}
    \ex{}\left[
\eta_\mathrm{rc,2}(O;\{f^\ell_{\mu,g,d,t}\}_{g,d,t}, \{f^\ell_{\pi,g,d}\}_{g,d})
\right]
=
\ex{}\left[
    \eta_\mathrm{rc,2}(O;\{\mu_{g,d,t}\}_{g,d,t}, \{\pi_{r,g,d}\}_{g,d})
\right]
=\frac{\psi_\mathrm{rc}}{e}.
\end{equation}
\endgroup
We also show that $\eta_\mathrm{rc,2}(O;\{f^\ell_{\mu,g,d,t}\}_{g,d,t}, \{f^\ell_{\pi,g,d}\}_{g,d})$ is integrable:
\begingroup
\allowdisplaybreaks
\begin{align*}
    \ex{}\big[\big\vert\eta_\mathrm{rc,2}(O;&\{f^\ell_{\mu,g,d,t}\}_{g,d,t}, \{f^\ell_{\pi,g,d}\}_{g,d})\big\vert\big] 
    \\&\overset{(a)}{\leq}
    \sum_{g,d,t\in\{0,1\}}\Big[
    \pp\big[\vert G\cdot D\cdot Y\vert\big]
    +\frac{1}{p(T=t)}\cdot\pp\big[\vert Y\cdot f^\ell_{\pi,g,d}(X)\vert\big]
    \\&\hspace{3em}+\pp\big[\vert G\cdot D\cdot f^\ell_{\mu,g,d,t}(X)\vert\big]
    +\frac{1}{p(T=t)}\cdot\pp\big[\vert
    f^\ell_{\mu,g,d,t}(X)\cdot f^\ell_{\pi,g,d}(X)
    \vert\big]
            \Big]
    \\&\overset{(b)}{\leq}
    \sum_{g,d,t\in\{0,1\}}\Big[
    \norm{Y}
    \cdot(1+ \frac{\norm{f^\ell_{\pi,g,d}}}{p(T=t)})
    \\&\hspace{7em}+\norm{f^\ell_{\mu,g,d,t}}
    +\frac{\norm{
    f^\ell_{\mu,g,d,t}}\cdot \norm{f^\ell_{\pi,g,d}}}{p(T=t)}
            \Big]
            \overset{(c)}{<}\infty,
\end{align*}
\endgroup
where $(a)$, $(b)$, and $(c)$ are due to triangle inequality, Cauchy-Schwarz, and the bounded moment assumptions of the proposition.
Moreover,
\begingroup
\allowdisplaybreaks
\begin{align*}
    &\hspace{-8em}\pp\left[\big\vert\eta_\mathrm{rc,2}(O;\{\hat{\mu}^\ell_{g,d,t}\}_{g,d,t}, \{\hat{\pi}^\ell_{r,g,d}\}_{g,d})
    -
    \eta_\mathrm{rc,2}(O;\{f^\ell_{\mu,g,d,t}\}_{g,d,t}, \{f^\ell_{\pi,g,d}\}_{g,d})\big\vert\right]
    \\&\overset{(a)}{\leq}
    \sum_{g,d,t\in\{0,1\}}\Big[
    \pp\big[\vert G\cdot D\cdot\big(\hat{\mu}^\ell_{g,d,t}(X) - f^\ell_{\mu,g,d,t}(X)\big)\vert\big]
    \\&+\frac{1}{p(T=t)}\cdot\pp\big[\vert Y\cdot\big(
    \hat{\pi}^\ell_{r,g,d}(X)
    - f^\ell_{\pi,g,d}(X)\big)\vert\big]
    \\&+\frac{1}{p(T=t)}\cdot
    \pp\big[\vert
    f^\ell_{\mu,g,d,t}(X)\big(
    \hat{\pi}^\ell_{r,g,d}(X)
    - f^\ell_{\pi,g,d}(X)\big)
    \vert\big]
    \\&+\frac{1}{p(T=t)}\cdot
    \pp\big[\vert
    f^\ell_{\pi,g,d}(X)\big(
    \hat{\mu}^\ell_{g,d,t}(X)
    - f^\ell_{\mu,g,d,t}(X)\big)
    \vert\big]
    \\&+\frac{1}{p(T=t)}\cdot
    \pp\big[\vert
    \big(
    \hat{\pi}^\ell_{r,g,d}(X)
    - f^\ell_{\pi,g,d}(X)\big)
    \big(
    \hat{\mu}^\ell_{g,d,t}(X)
    - f^\ell_{\mu,g,d,t}(X)\big)\vert\big]
            \Big]
    \\
    &\overset{(b)}{\leq}\sum_{g,d,t\in\{0,1\}}\Big[
    \norm{\big(\hat{\mu}^\ell_{g,d,t} - f^\ell_{\mu,g,d,t}\big)}
    \\&+\frac{1}{p(T=t)}\cdot\norm{Y}\cdot\norm{\big(
    \hat{\pi}^\ell_{r,g,d}
    - f^\ell_{\pi,g,d}\big)}
    \\&+\frac{1}{p(T=t)}\cdot
    \norm{
    f^\ell_{\mu,g,d,t}}\cdot\norm{\big(
    \hat{\pi}^\ell_{r,g,d}
    - f^\ell_{\pi,g,d}\big)
    }
    \\&+\frac{1}{p(T=t)}\cdot
    \norm{
    f^\ell_{\pi,g,d}}\cdot\norm{\big(
    \hat{\mu}^\ell_{g,d,t}
    - f^\ell_{\mu,g,d,t}\big)}
    \\&+\frac{1}{p(T=t)}\cdot
    \norm{
    \big(
    \hat{\pi}^\ell_{r,g,d}
    - f^\ell_{\pi,g,d}\big)}\cdot\norm{
    \big(
    \hat{\mu}^\ell_{g,d,t}
    - f^\ell_{\mu,g,d,t}\big)}
            \Big]
    \\&\overset{(c)}{=}o_p(1),
\end{align*}
\endgroup
where $(a)$ is due to triangle inequality, $(b)$ is due to Cauchy-Schwartz, and $(c)$ holds because of the assumptions of the proposition.
Therefore, $\eta_\mathrm{rc,2}(O;\{\hat{\mu}^\ell_{g,d,t}\}_{g,d,t}, \{\hat{\pi}^\ell_{r,g,d}\}_{g,d})$ converges in $L_1$ to $\eta_\mathrm{rc,2}(O;\{f^\ell_{\mu,g,d,t}\}_{g,d,t}, \{f^\ell_{\pi,g,d}\}_{g,d})$.
Now consider 
\begingroup
\allowdisplaybreaks
\begin{align*}
    &\hspace{-2em}\ee_m^\ell\left[\eta_\mathrm{rc,2}(O;\{\hat{\mu}^\ell_{g,d,t}\}_{g,d,t}, \{\hat{\pi}^\ell_{r,g,d}\}_{g,d})\right] - \frac{\psi_\mathrm{rc}}{e}
    \\
    &=
    \ee_m^\ell\left[\eta_\mathrm{rc,2}(O;\{\hat{\mu}^\ell_{g,d,t}\}_{g,d,t}, \{\hat{\pi}^\ell_{r,g,d}\}_{g,d})
    -
    \eta_\mathrm{rc,2}(O;\{f^\ell_{\mu,g,d,t}\}_{g,d,t}, \{f^\ell_{\pi,g,d}\}_{g,d})
    \right] 
    \\&+
    \ee_m^\ell\left[
    \eta_\mathrm{rc,2}(O;\{f^\ell_{\mu,g,d,t}\}_{g,d,t}, \{f^\ell_{\pi,g,d}\}_{g,d})
    \right] -
    \pp\left[
    \eta_\mathrm{rc,2}(O;\{f^\ell_{\mu,g,d,t}\}_{g,d,t}, \{f^\ell_{\pi,g,d}\}_{g,d})
    \right]
    \\&+
    \pp\left[
    \eta_\mathrm{rc,2}(O;\{f^\ell_{\mu,g,d,t}\}_{g,d,t}, \{f^\ell_{\pi,g,d}\}_{g,d})
    \right]
    -
    \frac{\psi_\mathrm{rc}}{e},
\end{align*}
\endgroup
where the first term is $o_p(1)$ because of $L^1$ convergence and an application of Chebyshev's inequality,
the second term is $o_p(1)$ since $\eta_\mathrm{rc,2}(O;\{f^\ell_{\mu,g,d,t}\}_{g,d,t}, \{f^\ell_{\pi,g,d}\}_{g,d})$ is integrable and the weak law of large numbers applies, and the third term is $0$ by \Cref{eq:proofdrexpecrc2}.
Therefore,
\[\ee_m^\ell\left[\eta_\mathrm{rc,2}(O;\{\hat{\mu}^\ell_{g,d,t}\}_{g,d,t}, \{\hat{\pi}^\ell_{r,g,d}\}_{g,d})\right]\overset{p}{\to}\frac{\psi_\mathrm{rc}}{e},\]
and by \Cref{lem:ratioestimator}, $\hat{e}^\ell\overset{p}{\to}e$.
Using Theorem 2.7 of \cite{van2000asymptotic}, we get
\[\hat{e}^\ell\cdot\ee_m^\ell\left[\eta_\mathrm{rc,2}(O;\{\hat{\mu}^\ell_{g,d,t}\}_{g,d,t}, \{\hat{\pi}^\ell_{r,g,d}\}_{g,d})\right]\overset{p}{\to}\psi_\mathrm{rc},\]
and since the latter holds for every fold $\ell\in\{1,\dots,L\}$, $\hat{\psi}_\mathrm{rc,2}^\mathrm{dr}$ is consistent for $\psi_\mathrm{rc}$.
\end{proof}

\thmcanrctwo*
\begin{proof}
The proof is similar to those of \Cref{thm:canp} and \Cref{thm:canrc}:
define $e=\frac{1}{\ex{}[G\cdot D]}$ and $\pi_{r,g,d}(X)=\frac{\pi_{1,1}(X)}{\pi_{g,d}(X)}$.
Suppose that the data is split into $L$ folds of size $m$, so that $n=mL$. 
We drop the subscript of $\hat{e}_\mathrm{rc,2}^\ell$ in the remainder of this proof.

\begingroup
\allowdisplaybreaks
    \begin{align}
        \sqrt{n}&(\hat{\psi}_\mathrm{rc,2}^\mathrm{dr} - \psi_\mathrm{rc}) = \notag
        \\&
        \sqrt{n}
            \big(\frac{1}{L}
            \sum_{\ell=1}^{L}
                \hat{e}^\ell\cdot\ee_m^\ell\big[\eta_\mathrm{rc,2}(O;\{\hat{\mu}^\ell_{g,d,t}\}_{g,d,t}, \{\hat{\pi}^\ell_{r,g,d}\}_{g,d})\big]
            -\psi_\mathrm{rc}\big)\notag
        \\&
            =\frac{\sqrt{mL}}{L}\cdot
            \sum_{\ell=1}^{L}
                \frac{\hat{e}^\ell}{e}\cdot\big(e\cdot\ee_m^\ell\big[\eta_\mathrm{rc,2}(O;\{\hat{\mu}^\ell_{g,d,t}\}_{g,d,t}, \{\hat{\pi}^\ell_{r,g,d}\}_{g,d})\big]
            -e\cdot\psi_\mathrm{rc}\cdot\ee_m^\ell[G\cdot D]\big)\notag
        \\&
        =\frac{1}{\sqrt{L}}
            \sum_{\ell=1}^{L}
                (1+\frac{\hat{e}^\ell-e}{e})\cdot\sqrt{m}\big(e\cdot\ee_m^\ell\big[\eta_\mathrm{rc,2}(O;\{\hat{\mu}^\ell_{g,d,t}\}_{g,d,t}, \{\hat{\pi}^\ell_{r,g,d}\}_{g,d})\big]
            -e\cdot\psi_\mathrm{rc}\cdot\ee_m^\ell[G\cdot D]\big)\notag
        \\&
        \begin{aligned}
            &=
            \frac{1}{\sqrt{L}}\sum_{\ell=1}^{L}
            (1+\frac{\hat{e}^\ell-e}{e})\cdot
            \sqrt{m}(\ee_m^\ell-\pp)
                \Big[e\cdot\eta_\mathrm{rc,2}(O;\{\hat{\mu}^\ell_{g,d,t}\}_{g,d,t}, \{\hat{\pi}^\ell_{r,g,d}\}_{g,d})
                \\&\hspace{20em}
                -e\cdot\eta_\mathrm{rc,2}(O;\{\mu_{g,d,t}\}_{g,d,}, \{\pi_{r,g,d}\}_{g,d})\Big]
        \end{aligned}\label{eq:rc2t1}\tag{$T_{1}$}
        \\&
        \begin{aligned}
            &+
            \frac{1}{\sqrt{L}}\sum_{\ell=1}^{L}
            (1+\frac{\hat{e}^\ell-e}{e})\cdot
            \sqrt{m}(\ee_m^\ell-\pp)
                \Big[e\cdot\eta_\mathrm{rc,2}(O;\{\mu_{g,d,t}\}_{g,d,t}, \{\pi_{r,g,d}\}_{g,d})\\&\hspace{25em}-e\cdot G\cdot D\cdot \psi_\mathrm{rc}\Big]
        \end{aligned}\label{eq:rc2t2}\tag{$T_{2}$}
        \\&
        \begin{aligned}
            &+
            \frac{1}{\sqrt{L}}\sum_{\ell=1}^{L}
            (1+\frac{\hat{e}^\ell-e}{e})\cdot
            \sqrt{m}\big(\pp
                \big[e\cdot\eta_\mathrm{rc,2}(O;\{\hat{\mu}^\ell_{g,d,t}\}_{g,d,t}, \{\hat{\pi}^\ell_{r,g,d}\}_{g,d})
                \big]-\psi_\mathrm{rc}\big).
        \end{aligned}\label{eq:rc2t3}\tag{$T_{3}$}
    \end{align}
\endgroup
We analyze each of the terms above separately.
In the remainder of the proof, we use $O_n^{-\ell}$ to represent the data in all but the $\ell$-th fold.

\textbf{Term \ref{eq:rc2t1}:}

Define $A_m^\ell = \sqrt{m}(\ee_m^\ell-\pp)
                \big[e\cdot\eta_\mathrm{rc,2}(O;\{\hat{\mu}^\ell_{g,d,t}\}_{g,d,t}\allowbreak, \allowbreak\{\hat{\pi}^\ell_{r,g,d}\}_{g,d})\allowbreak-e\cdot\eta_\mathrm{rc,2}(O;\{\mu_{g,d,t}\}_{g,d,t}, \{\pi_{r,g,d}\}_{g,d})\big]$.
Note that $\ex{}[A_m^\ell\mid O^{-\ell}]=0$. 
Next, we bound the conditional variance as
\begingroup\allowdisplaybreaks
\begin{align*}
    \mathrm{var}&(A_m^\ell\mid O^{-\ell})
    \\&= 
    m\cdot e^2\cdot\mathrm{var}(\ee_m^\ell
        [
            \eta_\mathrm{rc,2}(O;\{\hat{\mu}^\ell_{g,d,t}\}_{g,d,t}, \{\hat{\pi}^\ell_{r,g,d}\}_{g,d})-\eta_\mathrm{rc,2}(O;\{\mu_{g,d,t}\}_{g,d,t}, \{\pi_{r,g,d}\}_{g,d})
        ]
        \mid O^{-\ell}
    )
    \\&=
    e^2\cdot\mathrm{var}(
            \eta_\mathrm{rc,2}(O;\{\hat{\mu}^\ell_{g,d,t}\}_{g,d,t}, \{\hat{\pi}^\ell_{r,g,d}\}_{g,d})-\eta_\mathrm{rc,2}(O;\{\mu_{g,d,t}\}_{g,d,t}, \{\pi_{r,g,d}\}_{g,d})
        \mid O^{-\ell}
    )
    \\&\leq e^2\cdot\pp\big[\big(
            \eta_\mathrm{rc,2}(O;e,\{\hat{\mu}^\ell_{g,d,t}\}_{g,d,t}, \{\hat{\pi}^\ell_{r,g,d,t}\}_{g,d,t})-\eta_\mathrm{rc,2}(O;e,\{\mu_{g,d,t}\}_{g,d,t}, \{\pi_{r,g,d,t}\}_{g,d,t})\big)^2
        \mid O^{-\ell}
    \big]
    \\&=e^2\cdot
    \pp\Big[\Big(\sum_{g,d,t\in\{0,1\}}(-1)^{(g+d+t)}\hat{\alpha}^\ell_{g,d,t}(X,G,D,T)\cdot
            \big(
                Y-\hat{\mu}^\ell_{g,d,t}(X)
            \big)
        \\&\hspace{5em}-
            \sum_{g,d,t\in\{0,1\}}(-1)^{(g+d+t)}\alpha_{g,d,t}(X,G,D,T)\cdot
            \big(
                Y-\mu_{g,d,t}(X)
            \big)
            \Big)^2
        \:\Big\vert\: O^{-\ell}
    \Big]
    \\&\overset{(a)}{\leq}8e^2\sum_{g,d,t\in\{0,1\}}
    \pp\Big[\Big(\hat{\alpha}^\ell_{g,d,t}(X,G,D,T)\cdot
            \big(
                Y-\hat{\mu}^\ell_{g,d,t}(X)
            \big)
        \\&\hspace{10em}-
            \alpha_{g,d,t}(X,G,D,T)\cdot
            \big(
                Y-\mu_{g,d,t}(X)
            \big)
            \Big)^2
        \:\Big\vert\: O^{-\ell}
    \Big]
    \\&\overset{(b)}{\leq}24e^2\sum_{g,d,t\in\{0,1\}}
    \pp\Big[\Big(G\cdot D\cdot\big(\hat{\mu}^\ell_{g,d,t}(X)-\mu_{g,d,t}(X)\big)\Big)^2
        \:\Big\vert\: O^{-\ell}
    \Big]
        \\&\hspace{5em}+24e^2\sum_{g,d,t\in\{0,1\}}
        \frac{1}{p(T=t)^2}\cdot\pp\Big[\Big(\big(\hat{\pi}^\ell_{r,g,d}(X)-\pi_{r,g,d}(X)\big)\cdot\big(Y-\mu_{g,d,t}(X)\big)\Big)^2
            \:\Big\vert\: O^{-\ell}
        \Big]
        \\&\hspace{5em}+24e^2\sum_{g,d,t\in\{0,1\}}
        \frac{1}{p(T=t)^2}\cdot\pp\Big[\Big(\hat{\pi}^\ell_{r,g,d}(X)\cdot\big(\hat{\mu}^\ell_{g,d,t}(X)-\mu_{g,d,t}(X)\big)\Big)^2
            \:\Big\vert\: O^{-\ell}
        \Big]
    \\&\overset{(c)}{\leq}
        96e^2\sum_{g,d,t\in\{0,1\}}\Big(\bignorm{
            \hat{\mu}^\ell_{g,d,t} - \mu_{g,d,t}
        }^2
        +\frac{1}{p(T=t)^2}\cdot
        \bignorm{
            (\hat{\pi}^\ell_{r,g,d,t})^2-\pi_{r,g,d,t}^2
        }\cdot
        \big(\norm{
            Y^2
        }
        +
        \norm{\mu_{g,d,t}^2}\big)
        \\&\hspace{5em}+\frac{1}{p(T=t)^2}\cdot
        \bignorm{
            (\hat{\mu}^\ell_{g,d,t})^2 - \mu_{g,d,t}^2
        }\cdot
        \big(\bignorm{
            (\hat{\pi}^\ell_{r,g,d,t})^2-\pi_{r,g,d,t}^2
        }
        +\bignorm{\pi_{r,g,d,t}^2}\big)
        \Big),
\end{align*}
\endgroup
where $(a)$ and $(b)$ are due to Cauchy-Schwartz, and $(c)$ is an application of Hölder, Cauchy-Schwartz, and Minkowski inequalities.
Under \Cref{as:infrc2} and strict positivity, the conditional variance above is $o_p(1)$.
By the law of iterated expectations and Chebyshev's inequality, 
we conclude $A_m^\ell\overset{p}{\to}0$.
$(\hat{e}^\ell-e)/e$ is $o_p(1)$ by \Cref{lem:ratioestimator}, and using Slutsky's theorem and Theorem 2.7 of \citep{van2000asymptotic}, $(1+(\hat{e}^\ell-e)/e)\cdot A_m^\ell\overset{p}{\to}0$ for every $\ell$.
Finally, since the number of folds ($L$) is finite, \ref{eq:rc2t1} converges to $0$ in probability.

\textbf{Term \ref{eq:rc2t2}:}
\begin{align*}
    T_2&=\frac{1}{\sqrt{L}}\sum_{\ell=1}^{L}
            (1+\frac{\hat{e}^\ell-e}{e})\cdot\sqrt{m}(\ee_m^\ell-\pp)
                \big[e\cdot\eta_\mathrm{rc,2}(O;\{\mu_{g,d,t}\}_{g,d,t}, \{\pi_{g,d}\}_{g,d})-e\cdot G\cdot D\cdot \psi_\mathrm{rc}\big]
    \\&=
    \frac{1}{\sqrt{L}}\sum_{\ell=1}^{L}
            (1+\frac{\hat{e}^\ell-e}{e})\cdot\sqrt{m}(\ee_m^\ell-\pp)
                \big[\psi^1_\mathrm{rc,2}(O)\big]
    \\&\overset{(a)}{=}\sqrt{n}(\ee_n-\pp)\big[\psi^1_\mathrm{rc,2}(O)\big] + o_p(1),
\end{align*}
where $(a)$ is because (i) $\sqrt{m}(\ee_m^\ell-\pp)\big[\psi^1_\mathrm{rc,2}(O)\big]$ is $O_p(1)$ due to the central limit theorem, (ii) $(\hat{e}^\ell-e)/e$ is $O_p(m^{-1/2})$ due to \Cref{lem:ratioestimator}, so their product converges to $0$ in probability due to Slutsky's theorem and Theorem 2.7 of \citep{van2000asymptotic}, and (iii) the number of folds ($L$) is finite.
By the central limit theorem, the term $\sqrt{n}(\ee_n-\pp)\big[\psi^1_\mathrm{rc,2}(O)\big]$ converges in distribution to $\mathcal{N}\Big(0, \mathrm{var}\big(\psi^1_\mathrm{rc,2}(O)\big)\Big)$, as long as the variance of $\psi^1_\mathrm{rc,2}(\cdot)$ is bounded.
This variance can be bounded as follows.
\begin{align*}
    \mathrm{var}&\big(e\cdot\eta_\mathrm{rc,2}(O; \allowbreak\{\mu_{g,d,t}\}_{g,d,t}, \{\pi_{g,d}\}_{g,d})-e\cdot G\cdot D\cdot \psi_\mathrm{rc}\big)
    \\&\leq
    e^2\cdot\ex{}\big[\big(\eta_\mathrm{rc,2}(O; \allowbreak\{\mu_{g,d,t}\}_{g,d,t}, \{\pi_{g,d}\}_{g,d})- G\cdot D\cdot \psi_\mathrm{rc}\big)^2\big]
    \\&\leq
    2e^2\cdot\ex{}\big[\big(\eta_\mathrm{rc,2}(O; \allowbreak\{\mu_{g,d,t}\}_{g,d,t}, \{\pi_{g,d}\}_{g,d})\big)^2 + \big(G\cdot D\cdot \psi_\mathrm{rc}\big)^2\big]
    \\&\leq
    16e^2\sum_{g,d,t\in\{0,1\}}\ex{}\big[\alpha_{g,d,t}^2\cdot\big(Y-\mu_{g,d,t}(X)\big)^2\big] + 2e^2\cdot\psi_\mathrm{rc}^2
    \\&\leq
    32e^2\sum_{g,d,t\in\{0,1\}}\bignorm{\alpha_{g,d,t}^2}\cdot\Big(\bignorm{Y^2}+\bignorm{\mu_{g,d,t}^2}
    \Big)+ 2e^2\cdot\psi_\mathrm{rc}^2.
\end{align*}
which is bounded under \Cref{as:infrc2} and strict positivity.

\textbf{Term \ref{eq:rc2t3}:}

Define $B_m^\ell\coloneqq \sqrt{m}\big(\pp
                \big[e\cdot\eta_\mathrm{rc,2}(O;\{\hat{\mu}^\ell_{g,d,t}\}_{g,d,t}, \{\hat{\pi}^\ell_{r,g,d}\}_{g,d})
                \big]-\psi_\mathrm{rc}\big)$.
We have
\begingroup
\allowdisplaybreaks
\begin{align*}
    B_m^\ell&=
    \sqrt{m}\cdot e\cdot
            \big(\pp
                \big[\eta_\mathrm{rc,2}(O;\{\hat{\mu}^\ell_{g,d,t}\}_{g,d,t}, \{\hat{\pi}^\ell_{r,g,d}\}_{g,d})\big]-\pp
                \big[\eta_\mathrm{rc,2}(O;\{{\mu}_{g,d,t}\}_{g,d,t}, \{{\pi}_{r,g,d}\}_{g,d})\big]\big)
    \\&=\sqrt{m}\cdot
        e\cdot \pp\big[
            \sum_{g,d,t\in\{0,1\}}(-1)^{(g+d+t)}\hat{\alpha}^\ell_{g,d,t}(X,G,D,T)
            \cdot
            \big(
                Y-\hat{\mu}^\ell_{g,d,t}(X)
            \big)
        \\&\hspace{9em}-
        \sum_{g,d,t\in\{0,1\}}(-1)^{(g+d+t)}\alpha_{g,d,t}(X,G,D,T)
            \cdot
            \big(
                Y-\mu_{g,d,t}(X)
            \big)
        \big]
        \\&\leq
        \sqrt{m}\cdot{e}
        \sum_{g,d\in\{0,1\}}\Big\vert
        \pp\big[
        \hat{\alpha}^\ell_{g,d,1}(X,G,D,T)\cdot \big(Y-\hat{\mu}^\ell_{g,d,1}(X))
        \\&\hspace{12em}-
        \hat{\alpha}^\ell_{g,d,0}(X,G,D,T)\cdot \big(Y-\hat{\mu}^\ell_{g,d,0}(X))
        \\&\hspace{12em}-
        \alpha_{g,d,1}(X,G,D,T)\cdot \big(Y-\mu_{g,d,1}(X))
        \\&\hspace{12em}+
        \alpha_{g,d,0}(X,G,D,T)\cdot \big(Y-\mu_{g,d,0}(X))
        \big]
        \Big\vert
        \\&=
        \sqrt{m}\cdot{e}
        \sum_{g,d\in\{0,1\}}\Big\vert
        \pp\Big[\pp\big[
        \hat{\alpha}^\ell_{g,d,1}(X,G,D,T)\cdot \big(Y-\hat{\mu}^\ell_{g,d,1}(X))
        \\&\hspace{12em}-
        \hat{\alpha}^\ell_{g,d,0}(X,G,D,T)\cdot \big(Y-\hat{\mu}^\ell_{g,d,0}(X))
        \\&\hspace{12em}-
        \alpha_{g,d,1}(X,G,D,T)\cdot \big(Y-\mu_{g,d,1}(X))
        \\&\hspace{12em}+
        \alpha_{g,d,0}(X,G,D,T)\cdot \big(Y-\mu_{g,d,0}(X))
        \mid X,G,D,T\big]\Big]
        \Big\vert
        \\&=
        \sqrt{m}\cdot{e}
        \sum_{g,d\in\{0,1\}}\Big\vert
        \pp[-
        \frac{\hat{\pi}^\ell_{r,g,d}(X)}{p(T=1)}\cdot\mathbbm{1}\{G=g,D=d, T=1\} \big(\mu_{g,d,1}(X)-\hat{\mu}^\ell_{g,d,1}(X))
        \\&\hspace{3em}+\frac{\hat{\pi}^\ell_{r,g,d}(X)}{p(T=0)}\cdot\mathbbm{1}\{G=g,D=d, T=0\} \big(\mu_{g,d,0}(X)-\hat{\mu}^\ell_{g,d,0}(X))
        \\&\hspace{3em}+
        \pi_{r,g,d}(X)\cdot\frac{\mathbbm{1}\{G=g,D=d, T=1\}}{p(T=1)}\cdot \big(\mu_{g,d,1}(X)-\mu_{g,d,1}(X))
        \\&\hspace{3em}-
        \pi_{r,g,d}(X)\cdot\frac{\mathbbm{1}\{G=g,D=d, T=0\}}{p(T=0)}\cdot \big(\mu_{g,d,0}(X)-\mu_{g,d,0}(X))\big]
        \\&\hspace{12em}+\pp\big[
        G\cdot D\cdot\big(\mu_{g,d,1}(X)-\mu_{g,d,0}(X)-\hat{\mu}^\ell_{g,d,1}(X)+\hat{\mu}^\ell_{g,d,0}(X)\big)
        \big]
        \Big\vert
        \\&=
        \sqrt{m}\cdot{e}
        \sum_{g,d\in\{0,1\}}\Big\vert
        \pp[-
        \hat{\pi}^\ell_{r,g,d}(X)\cdot\pi_{g,d}(X)\cdot \big(\mu_{g,d,1}(X)-\hat{\mu}^\ell_{g,d,1}(X))
        \\&\hspace{3em}+\hat{\pi}^\ell_{r,g,d}(X)\cdot\pi_{g,d}(X)\cdot \big(\mu_{g,d,0}(X)-\hat{\mu}^\ell_{g,d,0}(X))
        \\&\hspace{12em}+\pp\big[
        \pi_{1,1}(X)\cdot\big(\mu_{g,d,1}(X)-\mu_{g,d,0}(X)-\hat{\mu}^\ell_{g,d,1}(X)+\hat{\mu}^\ell_{g,d,0}(X)\big)
        \big]
        \Big\vert
        \\&=
        \sqrt{m}\cdot{e}
        \sum_{g,d\in\{0,1\}}
        \Big\vert
            \pp\big[\big(
                \pi_{1,1}(X)-\hat{\pi}^\ell_{r,g,d}(X)\cdot\pi_{g,d}(X)
            \big)\\&\hspace{9em}\cdot\big(
                \mu_{g,d,1}(X)-\mu_{g,d,0}(X)-\hat{\mu}^\ell_{g,d,1}(X)+\hat{\mu}^\ell_{g,d,0}(X)
            \big)\big]
        \Big\vert
        \\&=
        \sqrt{m}\cdot{e}
        \sum_{g,d\in\{0,1\}}
        \Big\vert
            \pp\big[\pi_{g,d}(X)\cdot\big(
                \frac{\pi_{1,1}(X)}{\pi_{g,d}(X)}-\hat{\pi}^\ell_{r,g,d,t}(X)
            \big)\\&\hspace{9em}\cdot\big(
                \mu_{g,d,1}(X)-\mu_{g,d,0}(X)-\hat{\mu}^\ell_{g,d,1}(X)+\hat{\mu}^\ell_{g,d,0}(X)
            \big)\big]
        \Big\vert
        \\&\overset{}{\leq}
        \sqrt{m}\cdot{e}
        \sum_{g,d\in\{0,1\}}
        \bignorm{\pi_{g,d}\cdot\big(
                \pi_{r,g,d}-\hat{\pi}^\ell_{r,g,d}
            \big)}\cdot
        \bignorm{
                \mu_{g,d,1}-\mu_{g,d,0}-\hat{\mu}^\ell_{g,d,1}+\hat{\mu}^\ell_{g,d,0}
        }
        \\&\leq
        \sqrt{m}\cdot{e}
        \sum_{g,d\in\{0,1\}}
        \bignorm{
                \pi_{r,g,d}-\hat{\pi}^\ell_{r,g,d,t}
            }\cdot
        \bignorm{
                (\mu_{g,d,1}-\mu_{g,d,0})-(\hat{\mu}^\ell_{g,d,1}-\hat{\mu}^\ell_{g,d,0})
        },
\end{align*}
\endgroup
where each summand is $o_p(n^{-1/2})$ under \Cref{as:infrc2}, and since $\sqrt{m} = \sqrt{n}/\sqrt{L}$ where $L$ is a fixed value, $B_m^\ell$ is $o_p(1)$.
Based on the inequality above, 
\begingroup
\allowdisplaybreaks
\begin{align*}
    T_3 &= \frac{1}{\sqrt{L}}\sum_{\ell=1}^{L}(1+\frac{\hat{e}^\ell-e}{e})\cdot B_m^\ell
\end{align*}
\endgroup
is also $o_p(1)$. Note that $(\hat{e}^\ell-e)/e$ is $o_p(1)$ due to \Cref{lem:ratioestimator}, and the result follows from Slutsky's theorem and Theorem 2.7 of \citep{van2000asymptotic}.
\end{proof}

\section{Details of the Synthetic Data Generation Mechanism}\label{apx:sim}
We explain the details of the data generating mechanism for simulations of \Cref{sec:sim}.
We begin by generating five latent features $(Z_1, Z_2, Z_3, \allowbreak Z_4, U)$ from a standard normal distribution, and nonlinear transforms are applied to them as:
\[
\begin{cases}
    \tilde{Z}_1 = \exp(Z_1/2),\\
    \tilde{Z}_2 = 10 + Z_2/\exp(Z_1),\\
    \tilde{Z}_3 = (0.6+Z_1*Z_3/25)^3,\\
    \tilde{Z}_4 = (20+Z_2+Z_4)^2.
\end{cases}
\]
$U$ is treated as an unobserved covariate, while observed covariates $X_1,\dots, X_4$ are defined as $X_i = clip(\bar{Z}_i, -5, 5)$, where $\bar{Z}$ is the centered and normalized version of $Z_i$, and $clip(z, -5, 5)$ is the operation that limits the value of $z$ between $-5$ and $5$, i.e.,
\[
    clip(z, -5, 5) = \max(\min(z,5),-5).
\]
The first four variables $(X_0, X_1, X_2, X_3)$ are treated as observed covariates, while the fifth variable is reserved as an unobserved confounder $U$.
Here, our data generating mechanism follows that of \citep{kang2007demystifying}, except for the introduction of the unobserved confounder $U$.
This is a crucial difference, often overlooked by similar papers evaluating their difference-in-differences estimators using this model, because the triple difference (or even the classic difference-in-differences) framework is reasonable only when hidden confounding is present.
Otherwise, one can estimate the parameter of interest using covariate adjustment and simpler estimators.

Treatment and domain assignments are generated simultaneously through a multinomial logistic model over the four possible combinations $(G, D) \in \{0,1\}^2$. 
The logits are constructed as linear functions of both the observed covariates and the unobserved confounder $U$ with the following coefficients followed by softmax normalization to produce valid assignment probabilities.
\[
\begin{cases}
    p_{00} = -3X_1^2+0.6U,\\
    p_{01} = 0.75*(1 -0.5X_1 +0.5X_2 -0.25X_3 -0.1X_4 -4X_1^2 -2X_4^2)+0.6U,\\
    p_{10} = 0.75*(-0.5 +0.5X_1 -0.5X_2 +0.5X_3 +0.5X_4 -4X_1^2 +2X_4^2)+0.6U,\\
    p_{11} = 0.75*(-0.2X_1 +0.5X_2 -0.2X_3 -0.1X_4 -4X_1^2 -0.8X_4^2)+0.6U,\\
\end{cases}
\]
In the repeated cross-sections setting, the time indicator $T$ is sampled from an expit model with parameter $0.1X_4 + 0.15(D-\ex{}[D]) + 0.1(G-\ex{}[G])$.
The latter is also an important deviation from what is commonly done in the literature, in that it violates \Cref{as:nocompch}.

The outcome models are defined based on the following function:
\[\begin{split}
    k_t(X_1,X_2,X_3,X_4, G,D)\coloneqq&0.1*\alpha_t(G,D)*h(X_1, X_2, X_3, X_4, U) \\&+ 2*(-1)^{G+D}*p(G,D\mid X_1,X_2,X_3,X_4)*U+N,
\end{split}\]
where the production function $h(\cdot)$ is defined similarly to \citep{kang2007demystifying} with added nonlinearity through clipped quadratic terms:
\[\begin{split}
    h(X_1, X_2, X_3, X_4, U)\coloneqq 210 + 27.4X_1 &+ 13.7(X_2 + X_3 + X4) \\&-13.7* clip(X_2^2, -10,10) + 20*clip(X_4^2, -10, 10),
\end{split}
\]
$\alpha_t(G,D)$ for $t\in\{0,1\}$ is a coefficient defined as 
\[
 \alpha_0(G,D)\coloneqq 1+D, \quad\alpha_1(G,D)\coloneqq 1+G+2D,
\]
and $N$ is an independent normally distributed exogenous variable.
In the panel data setting, the potential outcomes are generated as
\[\begin{cases}
    \Y{0}=Y^{(1)}_0 = k_0(X_1,X_2,X_3,X_4, G,D),\\
    \Y{1}= k_1(X_1,X_2,X_3,X_4, G,D),\\
    Y^{(1)}_1 = \ex{}[\Y{1}\mid G=1, D=1] + \tilde{N} + 10,
\end{cases}\]
where $\tilde{N}$ is an independent normally distributed exogenous variable.
In the repeated cross-sections setting, the potential outcomes are generated as
\[
\begin{cases}
    \Y{}=Tk_1(X_1,X_2,X_3,X_4, G,D) + (1-T)k_0(X_1,X_2,X_3,X_4, G,D),\\
    Y^{(1)}= T(\ex{}[\Y{1}\mid G=1, D=1] + \tilde{N} + 10)+ (1-T)k_0(X_1,X_2,X_3,X_4, G,D).
\end{cases}
\]
Note that the functions $k_t$ are defined in a way to satisfy the conditional parallel difference in trends assumptions, and the potential outcomes are generated in such a way that the true parameter of interest is always equal to $10$.
Finally, the observed outcome variables are generated to satisfy the consistency assumptions in each setting.
\end{document}